\documentclass[seceq]{ptptex}

\usepackage{epsfig,multirow,booktabs}
\newcommand{\aap}{Astron. Astrophys.}
\newcommand{\apj}{Astrophys. J.}
\newcommand{\physrep}{Phys. Rep.}
\newcommand{\apjl}{Astrophys. J. Lett.}
\newcommand{\apjs}{Astrophys. J. Suppl.}

\newcommand{\pasj}{Pub. Astro. Soc. Japan}
\newcommand{\mnras}{MNRAS}
\newcommand{\beq}{\begin{equation}}
\newcommand{\eeq}{\end{equation}}
\newcommand{\beqn}{\begin{eqnarray}}
\newcommand{\eeqn}{\end{eqnarray}}

\title{
Core-Collapse Supernovae as Supercomputing Science: a status report
 toward 6D simulations with exact Boltzmann neutrino transport 
in full general relativity}

\author{
Kei \textsc{Kotake}$^{1,2}$, %
Kohsuke \textsc{Sumiyoshi}$^{3}$, %
Shoichi \textsc{Yamada}$^{4,5}$, 
\\
Tomoya \textsc{Takiwaki}$^{2}$,  Takami \textsc{Kuroda}$^{1}$,
 Yudai \textsc{Suwa}$^6$, and Hiroki \textsc{Nagakura}$^{4,6}$\\
}

\inst{
$^1$Division of Theoretical Astronomy, National Astronomical Observatory of Japan, 
2-21-1, Osawa, Mitaka, Tokyo, 181-8588, Japan,\\
$^2$Center for Computational Astrophysics, National Astronomical Observatory of Japan, 2-21-1, Osawa, Mitaka, Tokyo, 181-8588, Japan \\
$^3$  Numazu College of Technology, 
       Ooka 3600, Numazu, Shizuoka 410-8501, Japan \\
       \& \\
       Theory Center, 
       High Energy Accelerator Research Organization (KEK), \\
       Oho 1-1, Tsukuba, Ibaraki 305-0801, Japan,\\
$^4$Science \& Engineering, Waseda University,
3-4-1 Okubo, Shinjuku, Tokyo, 169-8555, Japan\\
$^5$Advanced Research Institute for Science and Engineering,
Waseda University,
3-4-1 Okubo, Shinjuku, Tokyo, 169-8555, Japan\\

$^{6}$Yukawa Institute for Theoretical Physics, Kyoto
  University, Oiwake-cho, Kitashirakawa, Sakyo-ku, Kyoto, 606-8502,
  Japan
}
\abst{
This is a status report on our endeavor to reveal the mechanism of core-collapse
supernovae (CCSNe) by large-scale numerical simulations. 
Multi-dimensionality of the supernova engine, general relativistic 
magnetohydrodynamics, energy and lepton number transport by neutrinos
 emitted from the forming neutron star as well as nuclear interactions there,
 are all believed to 
play crucial roles in repelling infalling matter and producing energetic explosions. 
These ingredients are nonlinearly coupled with one another in the dynamics of
 core-collapse, bounce, and shock expansion. 
Serious quantitative studies of CCSNe hence make extensive numerical 
computations mandatory. Since neutrinos are neither in thermal nor in chemical 
equilibrium in general, their distributions in the phase space should be computed.
This is a six dimensional (6D) neutrino transport problem
 and quite a challenge even for those with an access to the most 
advanced numerical resources such as the ``K computer''. To tackle this problem, we have 
embarked on multi-front efforts. In particular we report in this paper our recent progresses in 
the treatments of multi-dimensional (multi-D) radiation-hydrodynamics. We are currently proceeding on two different paths 
to the ultimate goal; in one approach we employ an approximate but highly efficient
scheme for neutrino transport and treat 3D hydrodynamics and/or general relativity rigorously; 
some neutrino-driven explosions will be presented and comparisons will be made between 2D and 3D models 
quantitatively; in the second approach, on the other hand, exact but so far Newtonian Boltzmann 
equations are solved in two and three spatial dimensions; we will show some demonstrative test simulations.
We will also address the perspectives of exa-scale computations on the next generation supercomputers. 
}

\begin{document}

\maketitle

\section{Explorations of CCSNe by multi-D simulations}

From the very beginning~\cite{Colgate66}, CCSN research has been one of the 
greatest challenges in computational astrophysics. This is simply due to the 
following facts~(see, e.g., \citen{kota06,Kotake11} and references therein): 3D 
macroscopic magnetohydrodynamics on stellar scales is largely 
dictated by weak/strong interactions on femto-meter scales; not only
dynamical time scales of milliseconds but neutrino-diffusion time scales
of $\gtrsim 100$ms become also important in the dynamics of massive star cores; 
compactness of proto-neutron stars (PNSs) and high velocities of collapsing 
material make
it simply impossible to neglect special/general relativity; (magneto)hydrodynamical 
instabilities render the core dynamics intrinsically non-spherical;
phase transitions in dense hadronic matter and/or neutrino oscillations may have a
critical impact. These diverse physical elements involved in CCSNe, although they are the reasons 
why many researchers from different disciplines have been attracted so much, are
 the very sources of difficulties in numerical simulations.

Energetics also suggests the subtlety in the explosion dynamics. 
 Recently advancing ability of the HST has enabled the direct 
observation of the progenitors of nearby CCSNe from pre-explosion images
 (see \cite{smartt09} for review). The accumulating observations suggest
  that the majority of
 CCSNe (the so-called type II-plateau (II-P)) comes 
predominantly from stars in the range about of 8 - 16 
$M_{\odot}$ \cite{smartt09}. A generic explosion energy for the 
 progenitors in the mass range is roughly on the order 
of $10^{51}$ erg 
\cite{utrobin11}, which is much smaller than the available energy, 
$\sim \!\! 10^{53}$erg, i.e., the gravitational binding energy of PNSs. 
This large amount of energy is stored as the internal energy of PNSs 
and tapped slowly by neutrinos. After the successful detections of 
neutrinos from SN1987A followed by
 detailed analyses of these events (e.g., \citen{Sato-and-Suzuki}, and 
see also \citen{raffelt12} for a recent review), 
CCSN researchers have a confidence that the explosion energy is indeed 
supplied by this reservoir. Present state-of-the-art supernova 
simulations require more than $10^{18}$ operations and typically more than several months for a single model
on currently available best supercomputing platforms.
 The energy conservation should be satisfied within 
an error of at least less than 
$\sim 1$\% $(=10^{51}{\rm erg}/10^{53}{\rm erg})$ in such very long-term 
simulations to obtain reliable results. It may not be difficult then to imagine how demanding the numerical simulations of CCSNe are. 

It is interesting that after +45 years of intensive and extensive theoretical 
 studies we are still working on the same scenario that Colgate and White (1966) 
envisaged in their seminal paper~\cite{Colgate66}, in which they reported
 the first CCSN simulation. The abstract of their paper ended
 with the sentence, {\it "The energy release corresponds to the change in gravitational potential of 
the unstable imploding core; the transfer of energy takes place by the emission and deposition of neutrinos"}. 
As is now well known, this is exactly the essence of the so-called neutrino-heating mechanism (e.g., \citen{jank07} 
for review). Although they proposed originally that the mechanism works promptly after bounce, 
Bethe and Wilson later amended it~\cite{Wilson85,Bethe85} to the currently prevailing form,
in which a bounce-generated shock wave is stalled first, but revived later by the deposition of neutrinos and 
an explosion follows in several hundred milliseconds after bounce. 
 The scenario was investigated intensively for the first 5 years in the new millennium under the
assumption of spherically symmetry but with the full Boltzmann treatment of neutrino transfer~\cite{Rampp00,Liebendorfer01,thom03,Sumiyoshi05}. 
General relativity~\cite{Liebendorfer01,Sumiyoshi05} and/or various neutrino reactions, some of which are supposed to be of minor
importance~\cite{horowitz02,buras03,burrows06npa}, were also implemented.  

These explorations made clear, however, that the neutrino-heating mechanism fails to produce explosions
in 1D spherical symmetry except for super-AGB stars at the low-mass end~\cite{Kitaura06}.
Not deterred by this failure, researchers changed gear to multi-D modeling. By this time there had already been mounting observational 
evidence that supernova explosions are indeed aspherical in general (see, e.g., \citen{wang01,Maeda08,Tanaka09} and references therein). 
Numerical experiments also suggested that breaking spherical symmetry holds 
a key to success of the
neutrino-heating mechanism; convective motions (e.g., \citen{Herant92,shimizu93,Burrows95,Janka96,fryer04a}) 
and/or the so-called standing accretion shock instability, or SASI, (e.g., 
\citen{Blondin03,scheck06,Ohnishi06,ohnishi07,Foglizzo06,Iwakami08,iwakami2,Murphy08,rodrigo09_2} and
 references therein) help the onset of neutrino-driven explosions.

In the following years, we have indeed witnessed some exploding models by the neutrino-heating mechanism in 
axisymmetric 2D simulations (see, e.g., table~1 in \citen{Kotake11}).  Employing 
one of the best approximations for 
2D neutrino transfer, Buras et al.~\cite{Buras06a} reported 
explosions firstly for a non-rotating low-mass 
($11.2 M_\odot$) progenitor~\cite{woos02} and then for a $15 M_{\odot}$ progenitor \cite{WW95} with a moderate rotation being imposed.\cite{Marek09} Implementing a multi-group flux-limited diffusion algorithm to their CHIMERA code 
in a ray-by-ray manner, on the other hand, Bruenn et al.~\cite{bruenn} obtained explosions for non-rotating 
progenitors~\cite{woos02} in the mass range from $12 M_{\odot}$ to 25$M_{\odot}$.  
Implementing the ray-by-ray isotropic diffusion source approximation (IDSA)~\cite{idsa} in the ZEUS code with a reduced set of 
weak interactions, Suwa~et~al.~\cite{Suwa10} pointed out that 
a rapidly rotating $13M_\odot$ progenitor produced a stronger explosion than the 
non-rotating counterpart did~\cite{Kotake03}. 

Accompanying these successes are new questions, however. In addition to the apparent contradictions among the groups 
(see Table~1 in \citen{Kotake11}), 
the models mentioned above produced generically under-energetic explosions at the end of simulations, with the diagnostic 
explosion energy being smaller by one or two orders of magnitude than the canonical kinetic energy of supernova ejecta 
($\sim 10^{51}$erg). Hence, it is a legitimate concern whether
 we can obtain energetic explosions comparable to observations by the neutrino-heating mechanism with appropriate 
nucleosynthetic yield, which is one of the most important observables~\cite{yamamoto12}. In the above-mentioned 
computations, the softest version of Lattimer \& Swesty's (LS) equation of state (EOS) ~\cite{latt91} with an incompressibility at nuclear
density, $K$, of 180 MeV, was commonly employed. In addition to the fact that recent experiments suggest a stiffer EOS with 
$K=240 \pm 20$ MeV \cite{shlo06}, it is now thought to be a serious flaw that the LS180 EOS cannot support a 
$2 M_{\odot}$ cold neutron star that is certainly existent in the universe~\cite{demo10}\footnote{The maximum mass for 
the LS180 EOS is about $1.8 M_{\odot}$ (see, e.g., \citen{oconnor,kiuc08}).}. Employing a stiffer EOS with $K=263$MeV 
based on the Hartree-Fock approximation~\cite{wolff}, Marek et al.~\cite{Marek09} found no explosion for the same progenitor 
model, whereas they indeed obtained an explosion for the Shen's EOS that is even stiffer with $K=281$MeV\cite{Janka-pr}. 
Suwa et al. also pointed out that not only the incompressibility but the symmetry energy also matters for the success of 
neutrino-driven explosions~\cite{suwa12}. 
Impacts of more detailed properties of nuclear EOSs (such as the density dependence of 
symmetry energy and the skewness of compressibility~\cite{steiner,lattimer12})
 on the multi-D neutrino-heating mechanism
 are remaining to be understood. 

The paramount interest of supernova researchers at present is 3D effects, however. We know in fact that SASI
is qualitatively different between 2D and 3D ~\cite{blo07,iwakami2} and it is naturally expected that 
this may have some consequences to success of the neutrino-heating mechanism in 3D. So far experimental simulations
are contradicting each other: Nordhaus et al.~\cite{Nordhaus10} claimed that 3D dynamics will make shock revival easier
than 2D. The assertion was challenged later by Hanke et al.~\cite{Hanke11}, who found little difference in the critical neutrino 
luminosity for shock revival between their 2D and 3D simulations. In both of the experimental computations, the neutrino 
transfer was not solved and the controversy will be settled only by detailed self-consistent 3D simulations. 
In the next section, we present our recent findings that illuminate 3D effects on the  neutrino-driven mechanism. 
The former part of \S\ref{3d} is devoted to our 3D Newtonian hydrodynamical simulations with spectral neutrino transport 
whereas in the latter half we show the latest results of our fully general relativistic (GR) 3D simulations that employ a 
more approximate neutrino transport scheme. 

Although we will focus on the neutrino-heating mechanism in this paper, it should be mentioned here that there are some other 
viable mechanisms. In the so-called acoustic mechanism~\cite{Burrows06}, oscillations 
of PNSs are supposed to 
produce pressure perturbations and send acoustic powers to the stalled shock until it revives and produces an explosion. 
The mechanism will then be particularly
important in the later postbounce phase, when the neutrino luminosity has 
already declined and the neutrino-heating mechanism has no
chance of success. The merit of this mechanism is that matter accretion, the source of acoustic powers, will last long,
possibly until the shock is revived. 
The scenario was challenged by Quataert et al.~\cite{weinberg}, however, who demonstrated that the amplitudes of g-mode
oscillations of PNSs will not be so large (see also \citen{yoshida}). Although the additional energy input by acoustic 
waves is very appealing, it still remains an issue under vivid debates and has yet to be confirmed by other groups~\cite{Marek09}. 
The magnetohydrodynamical (MHD) mechanism taps rotational energies of stellar cores (e.g., 
\citen{leblanc,yamasawa,kota04b,taki04,kota05,kotake_prd,suwa07,suwa07a,taki09,burr07,fogli_B,martin11,taki_kota,endeve12}. See also  
\citen{kota06} for collective references). Magnetic fields are expected to be amplified spontaneously by 
the magneto-rotational instability (MRI) even if they are tiny prior to collapse~\cite{aki}. This mechanism requires rapid 
rotation of stellar cores at the onset of core collapse~\cite{aki}. Recent stellar evolution models predict that such a condition 
can be realized only in the special case that experiences the so-called chemically homogeneous evolution~\cite{woos06,yoon} and
applies to just a fraction ($\sim 1$\%) of massive stars. Further investigations are currently hampered by the fact that high numerical 
resolutions are required to accurately compute the growth of the MRI~\cite{shapiro,
martin_mri}.
 We finally list other possibilities proposed in the literature:
exotic physics in the proto-neutron star~\cite{takahara88,sage09,fischer}, viscous heating by MRI\cite{thomp05,masa11}, and 
dissipation of Alfv\'en waves~\cite{suzu08}.

Improvements of input physics are another important ingredient in the CCSN simulations. One of the authors provided an EOS table 
to the society~\cite{Shen98,Shen98b,Shen11}, which is referred to as Shen's EOS and is based on the relativistic mean field theory 
and Thomas-Fermi approximation and is now a standard choice for core-collapse simulations. 
Besides this and another representative EOS by Lattimer \& Swesty~\cite{latt91}, new sets of 
 EOS's have been reported recently~\cite{Hempel11,Furu11,GShen11}. We have joined this effort to expand the inventory of EOS's 
both above and below nuclear density for the last few years. Above saturation density, we have later included 
hyperons in the same framework~\cite{ish08}. We have also combined the original table with an EOS for quark matter, adopting 
the MIT bag model and the Gibbs condition for the first order phase transition~\cite{nak10b}. These modifications manifest themselves
at high densities and will be more important for CCSNe with black hole formations than 
those with neutron star formations. It is emphasized that
the EOS at high densities can be probed by neutrino and/or gravitational wave signals from core-collapse 
events~\cite{sum06,nak10a,oconnor,fischer09}. The EOS below 
nuclear saturation density is no less important. In the conventional EOSs~\cite{latt91,Shen98,Shen98b,Shen11}, the so-called single nucleus approximation 
was employed, in which thermally populated heavy nuclei are represented by a single, supposedly the most abundant nucleus. 
Combining experimental nuclear mass data and a mass formula, 
we have solved Saha-like equations to obtain populations of various
nuclei in constructing the EOS \cite{Furu11}. In doing so, not only the excluded-volume effect but the emergence of pasta phases as well as the modifications of 
bulk, Coulomb and surface energies by surrounding nucleons and nuclei are also taken into account phenomenologically. 
We are currently preparing an EOS table including the electron capture rates according to the obtained 
populations~\cite{Furu2}. Detailed comparisons with other tables~\cite{Hempel11,Botvina} will be also published soon.  

The ultimate goal of CCSN simulations is 
3D neutrino-radiation-(magneto-) hydrodynamics in full GR, 
in which the exact Boltzmann 
equations are solved and all the relevant weak interactions are included with sufficient realism. It is worth mentioning that more massive 
stars than the SN II-p mentioned earlier 
are expected to lose much of their mass and explode as 
hydrogen-stripped SNe (Ib/c and IIb). Among them, the type Ic-BL SNe, 
which are associated with long gamma-ray bursts
 \cite{woos_blom}, all show much broader lines than SNe Ic. 
Due to the large kinetic energies of $2 - 5 \times 10^{52}$ erg,
 they have been referred to as "hypernovae"  (e.g., Ref.\cite{modjaz11}
 for a recent review), the central engine of which is likely to 
  be associated with massive stellar core-collapse with black-hole 
formation (e.g., collapsar, see collective references in
 Refs. \cite{macfadyen99,Woosley2011}). In such a system, the 
implementation of full GR in numerical simulations is 
essential. In one of our approaches mentioned above, 3D hydrodynamics in full GR is first addressed in \S2 with neutrino transport being approximated one way or another. On the other front, we are pursuing the Boltzmann transport in 3D
first. We present the current status of our efforts along this path in \S\ref{KS-Boltzmann}. Our method is based on an implicit
finite differencing of the Boltzmann equations and the inversion of large matrices in a very efficient way is one of the 
major challenges. As discussed in the final section, the 3D version of the code may work only on the next generation, exa-scale
platforms. In this sense, what we provide in the following sections is just a snapshot of a long, on-going documentary film that will record our 
struggles to make the "dream simulation" come true. 

\section{3D hydrodynamical simulations with approximate neutrino transport}\label{3d}
\subsection{3D Newtonian simulations with spectral neutrino transport}\label{3d_1}
It is generally very computationally expensive to solve neutrino
 transport in 3D and a light-bulb scheme\cite{jankamueller96} has been 
widely used so far, in which neutrino heating and cooling are treated 
in a parametric manner to trigger 3D explosions.  Using this prescription, Nordhaus 
et al.~\cite{Nordhaus10}
 was the first to argue that 
the critical neutrino luminosity for producing neutrino-driven explosions 
becomes significantly smaller in 3D than in 2D (see, however, \citen{Hanke11}). 
They employed the CASTRO code with an adaptive mesh refinement technique, by which unprecedentedly 
high resolution 3D calculations were made possible. 

Since the light-bulb scheme can capture fundamental properties of 
 neutrino-driven explosions (albeit on a qualitative basis), it is one of 
the most prevailing approximations adopted in recent 3D models
(e.g., \citen{Iwakami08,iwakami2,annop}).
 A number of important findings have been reported recently in these simulations,
such as a potential role of non-axisymmetric SASI flows in generating spins 
(see \citen{annop,rantsiou} as well as \citen{blo07,fern}) and magnetic 
fields~\cite{endeve} of pulsars, stochastic nature of 
 gravitational-wave (e.g., \citen{kotake09a,Kotake11,ewald11}) and 
 neutrino emission (see \citen{kneller} for recent review).

To go up the ladders beyond the light-bulb scheme, 
we studied 3D effects on the supernova mechanism by performing the first 3D, multi-energy-group, 
radiation-hydrodynamical core-collapse simulations \cite{Takiwaki11}.
For the spectral transport, the IDSA scheme is implemented, 
which can be done rather in a 
 straightforward manner by extending our 2D modules \cite{Suwa10,suwa12} to 3D.
This can be made possible because we apply the so-called
ray-by-ray approach (e.g., \citen{Buras06a}) 
in which the neutrino transport is solved along a
given radial direction by assuming the angle-averaged matter 
distribution in a spherically symmetric manner. 
From a technical point of view, it is worth
 mentioning that the ray-by-ray treatment is highly efficient in paralellization\footnote{along each radial ray}
 on present supercomputers, most of which 
 employ the message-passing-interface (MPI) routines.
The IDSA scheme splits the neutrino distribution
 into two components, each of which is solved with different numerical techniques
 (see \citen{idsa} for more details). 
 A drawback in the current
 version of the IDSA scheme is that heavy lepton neutrinos
 ($\nu_x$, i.e., $\nu_{\mu}$, $\nu_{\tau}$ and their 
anti-particles) as well as the energy-coupling weak interactions 
 have yet to be implemented.
The approximation level of the IDSA
 scheme is basically the same as the one of the Multi-Group Flux-Limited Diffusion 
MGFLD scheme. 
 The main advantage of the IDSA scheme is that the fluxes in the transparent
 region can be determined by the non-local distribution of sources
 rather than the gradient of the local intensity like in MGFLD.
 In the following, we briefly summarize the main results on 
 our 3D simulations, in which we obtained a first 3D explosion for an 11.2 $M_{\odot}$ 
star\cite{woos02}.

\subsubsection{3D dynamics from core-collapse through postbounce turbulence till 
 explosion}
\begin{figure}[htbp]
    \centering
    \includegraphics[width=.42\linewidth]{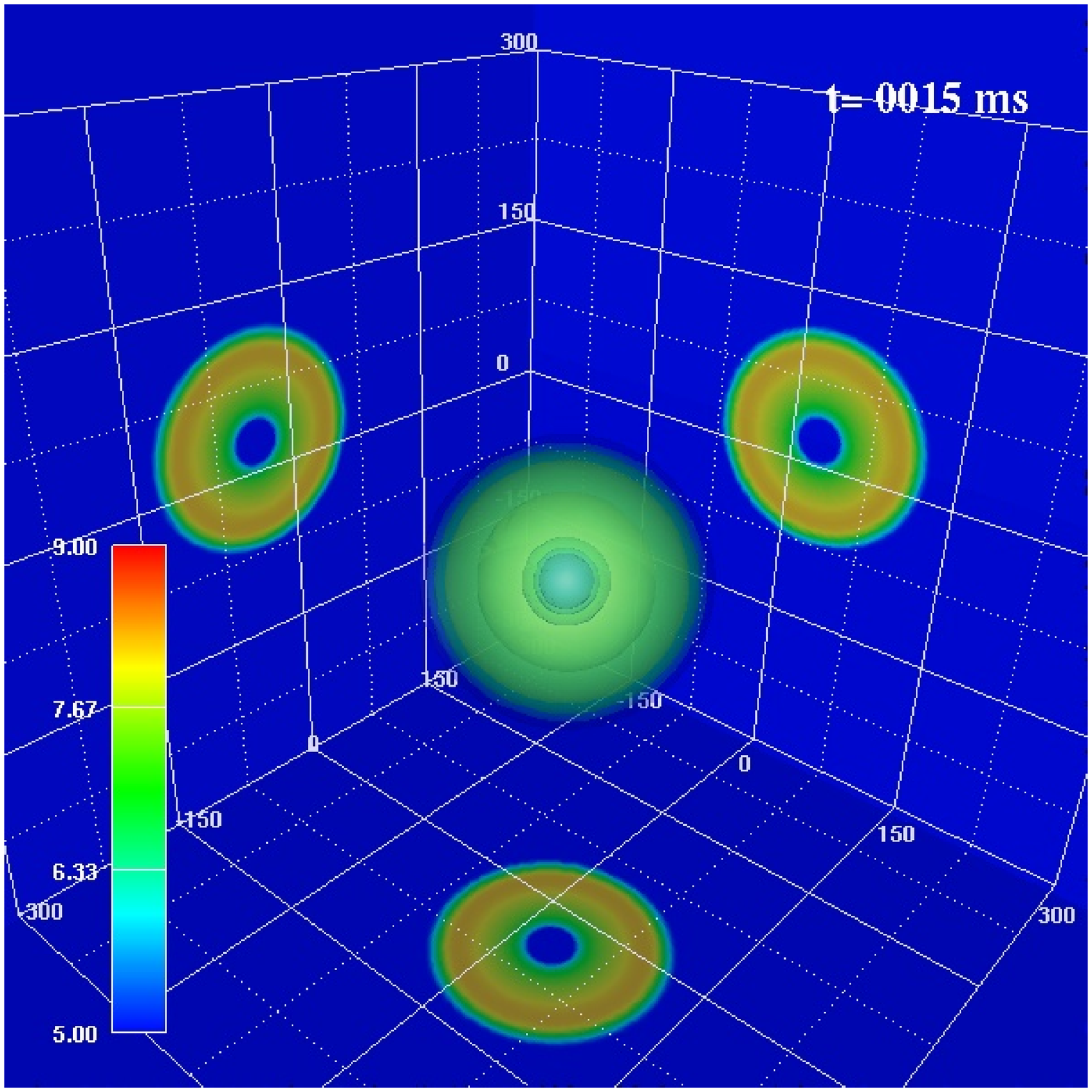}
    \includegraphics[width=.42\linewidth]{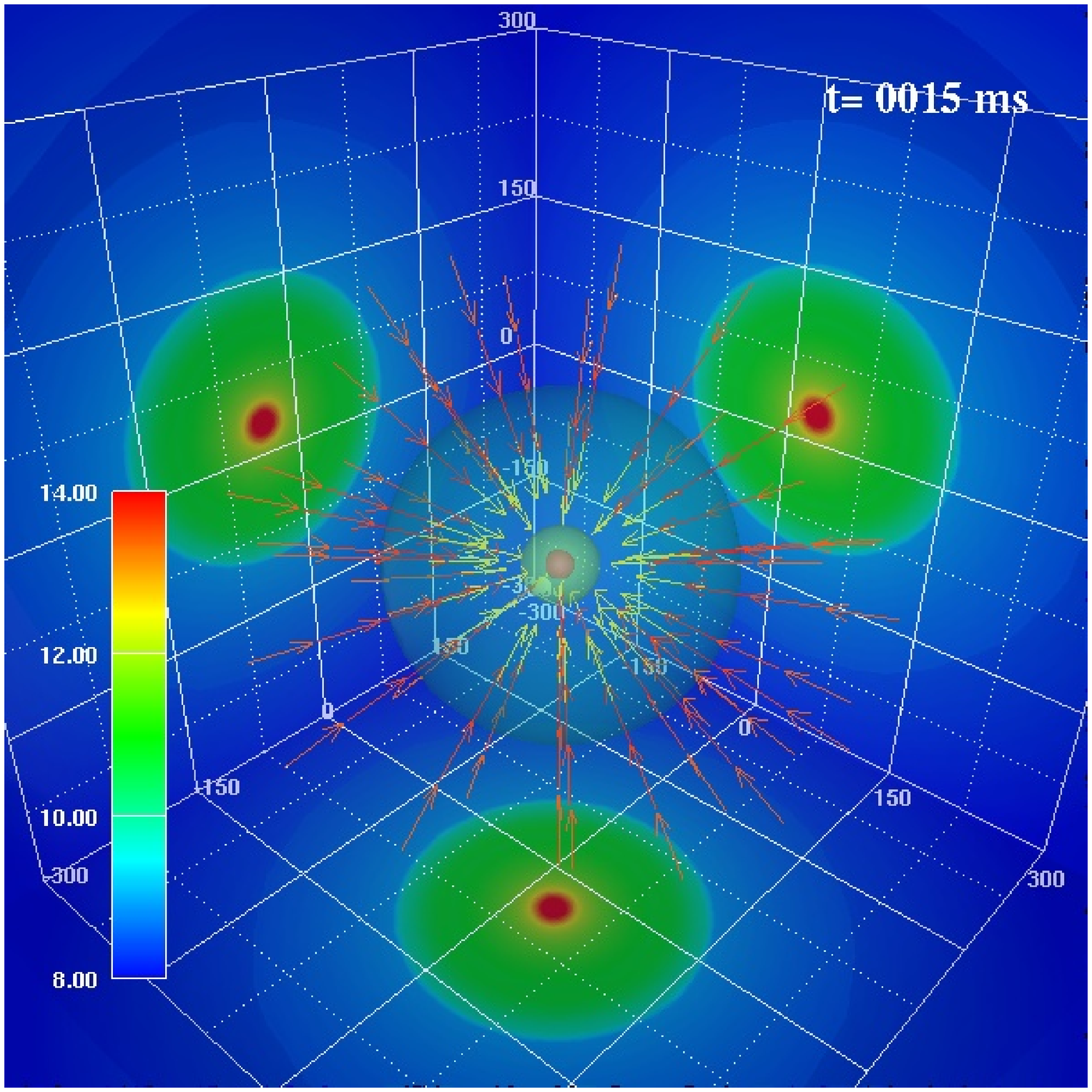}\\
    \includegraphics[width=.42\linewidth]{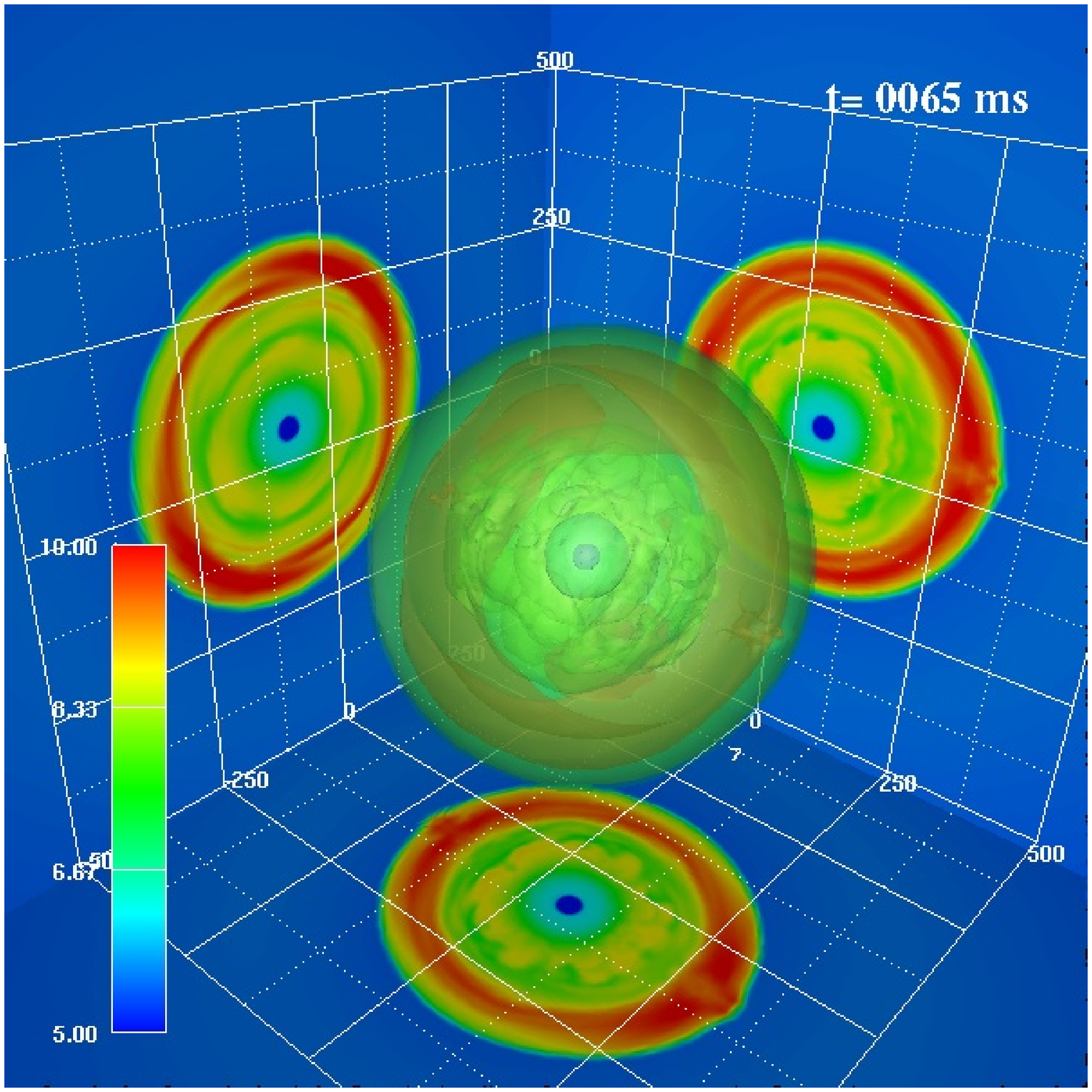}
    \includegraphics[width=.42\linewidth]{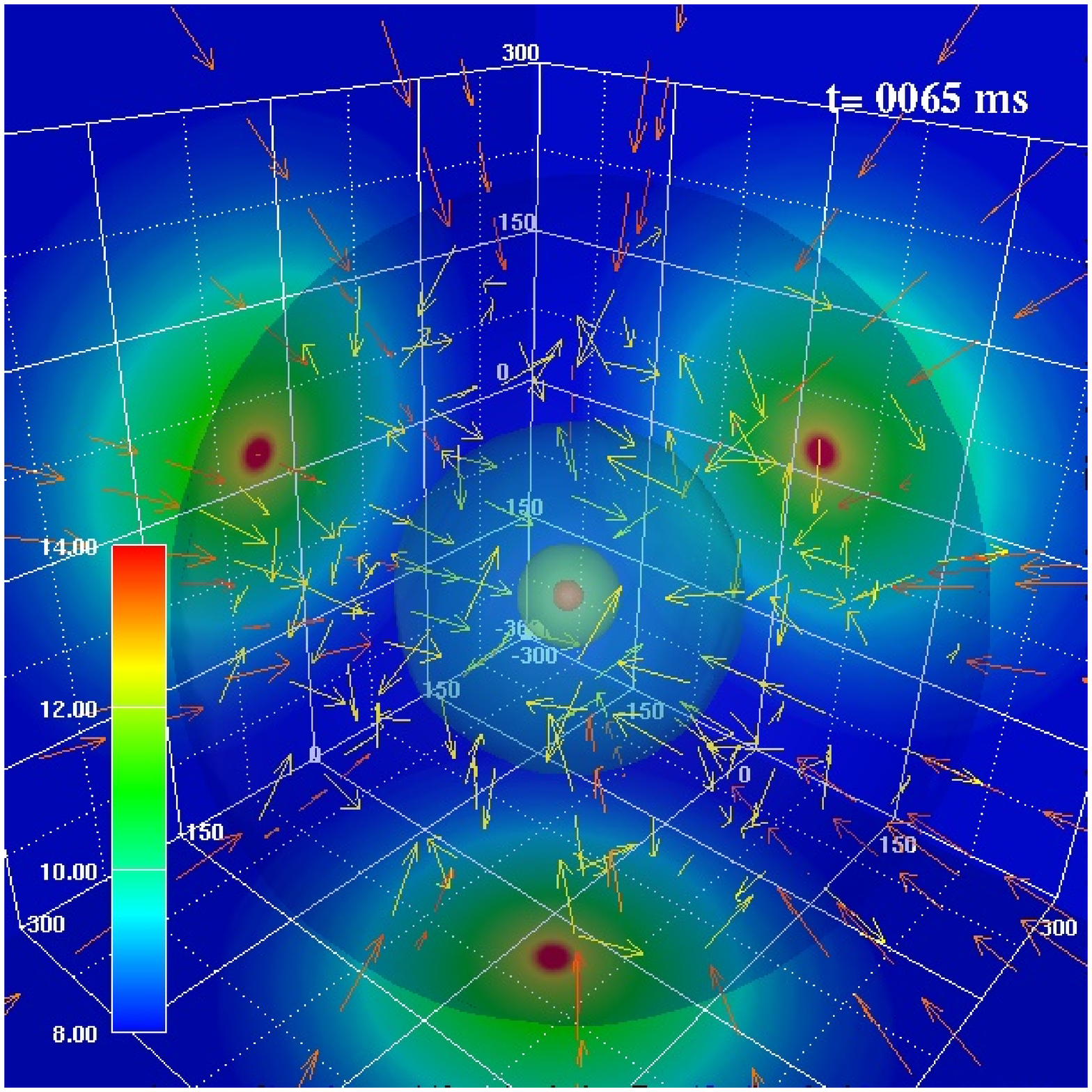}\\
    \includegraphics[width=.42\linewidth]{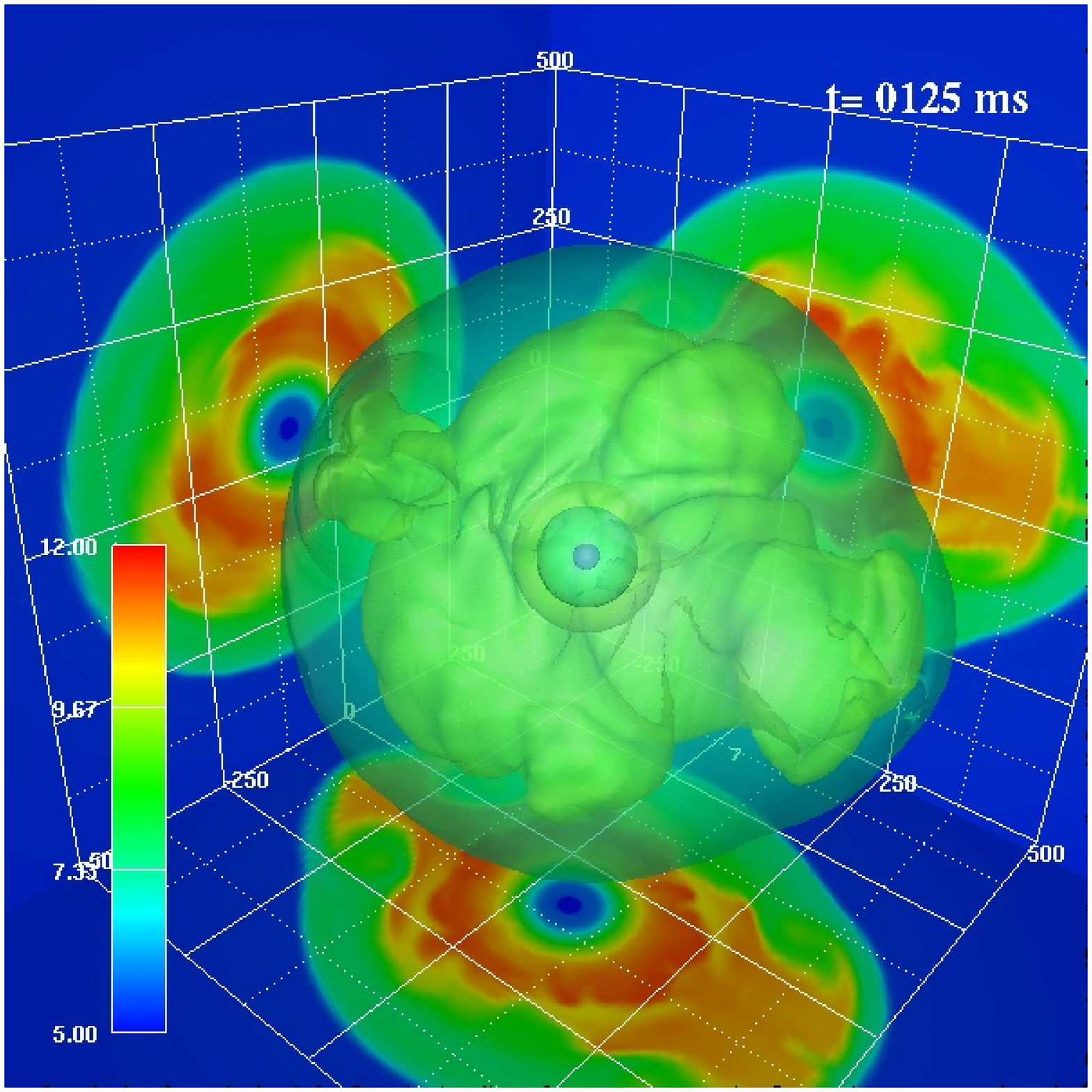}
    \includegraphics[width=.42\linewidth]{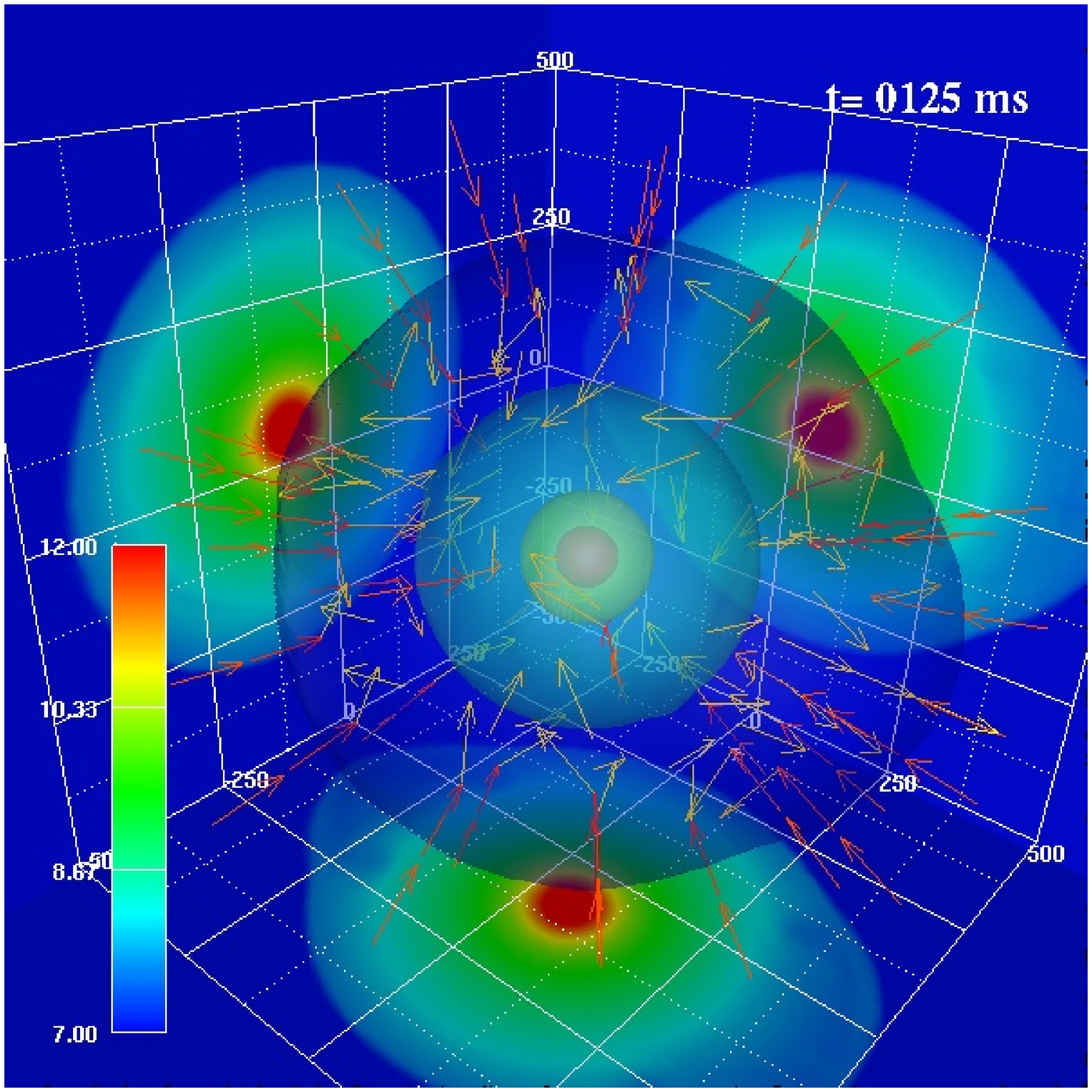}\\
 \caption{3D plots of entropy per baryon (left panels) 
and logarithmic density
 (right panels, in unit of ${\rm g}/{\rm cm}^3$) for three snapshots (top; $t = 15$ ms, middle; $t=65$ ms, and 
 bottom; $t=125$ ms after bounce ($t\equiv 0$)) during
 the evolution of a (non-rotating) exploding 3D model of a 11.2 $M_{\odot}$ star (figures
 taken from \citen{Takiwaki11}).
 In the right panels,  velocities are indicated by arrows.
The color contours in the 
$x=0$ (back right), $y=0$ (back bottom), and $z=0$ (back left) planes are
projected on the sidewalls of the graphs. For each snapshot, 
 the linear scale is shown along the axis in unit of km
 (figures taken from \citen{Takiwaki11}, reproduced by permissions
 of the AAS).}
\label{f1}
\end{figure}

Figure~\ref{f1} shows three snapshots, which 
are helpful to characterize hydrodynamic features in 3D simulations.
The top panel corresponds to $t=15$ ms after bounce, showing that the bounce shock stalls 
(indicated by inward arrows in the top right panel) at a radius of $150$ km.  
Note that the colors of velocity arrows are chosen so that they would change from yellow to red as 
 the absolute values become larger. Looking carefully at the top right panel, 
 we find that matter flows supersonically (indicated by reddish arrows) 
into the standing shock (the central transparent sphere), and then advects
 subsonically (indicated by yellowish arrows) onto the proto-neutron star (PNS, the central bluish region 
in the top left panel).
For the non-rotating progenitor, the dynamics of collapsing iron core 
 proceeds perfectly spherically till the stall of the bounce shock. This is the reason
 why multi-D effects are invisible in the entropy (top left panel) and density (top right panel) distributions 
right after bounce.

 The middle panels show the epoch ($t=65$ ms)
 when the neutrino-driven convection is already active.
 From the right panel, turbulent motions can be 
 seen (arrows in random directions) inside the standing shock, which is
 indicated by the boundary between red and yellow arrows.
 The entropy behind the standing shock becomes high by the neutrino-heating 
 (reddish regions in the left panel). The size of neutrino-heated hot 
 bubbles becomes larger in a non-axisymmetric way later on, which is indicated by  
 smaller structures encompassed by the stalled shock (i.e., inside the central 
greenish sphere in the left panel). 

 The bottom panels ($t=125$ ms) show the epoch when the revived shock is expanding 
 aspherically, which is indicated by the outgoing yellowish arrows 
in the right panel. The asphericity of 
 the expanding shocks could be more clearly visible by the sidewall panels.
From the entropy distribution (left panel), the expanding shock is 
shown to touch a radius of $\sim 500$ km (the projected back bottom panel). 
 Inside the expanding shock (enclosed by the greenish membrane in the 
 left panel), the bumpy structures of the hot bubbles are seen. In contrast to
 these smaller asphericities, the deformation of the shock surface is mild.
 This is a consequence of SASI, leading to the shock deformation dominated 
 by low spherical-harmonics modes ($\ell = 1, 2$).

\begin{figure}[htbp]
    \centering
    \includegraphics[width=.49\linewidth]{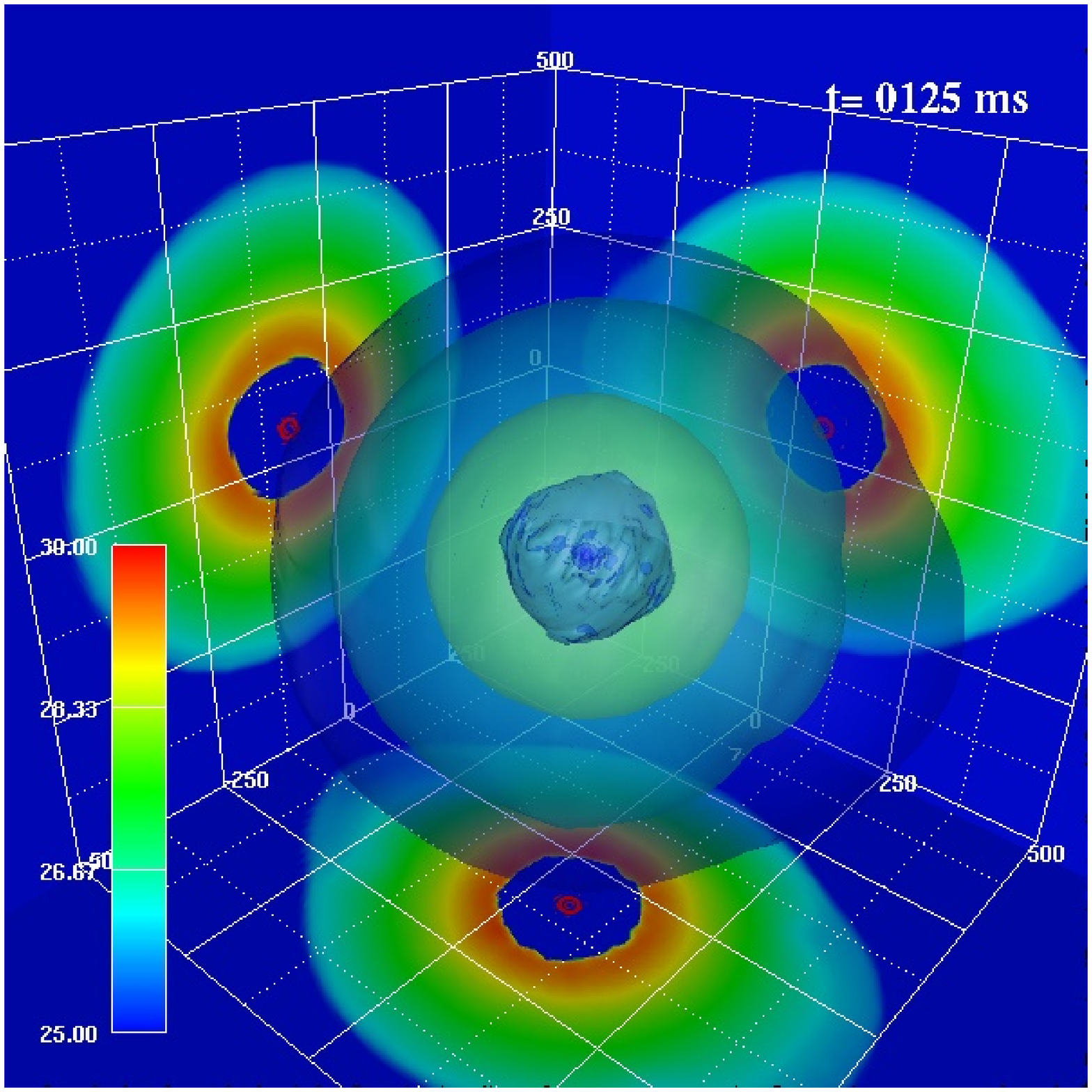}
    \includegraphics[width=.49\linewidth]{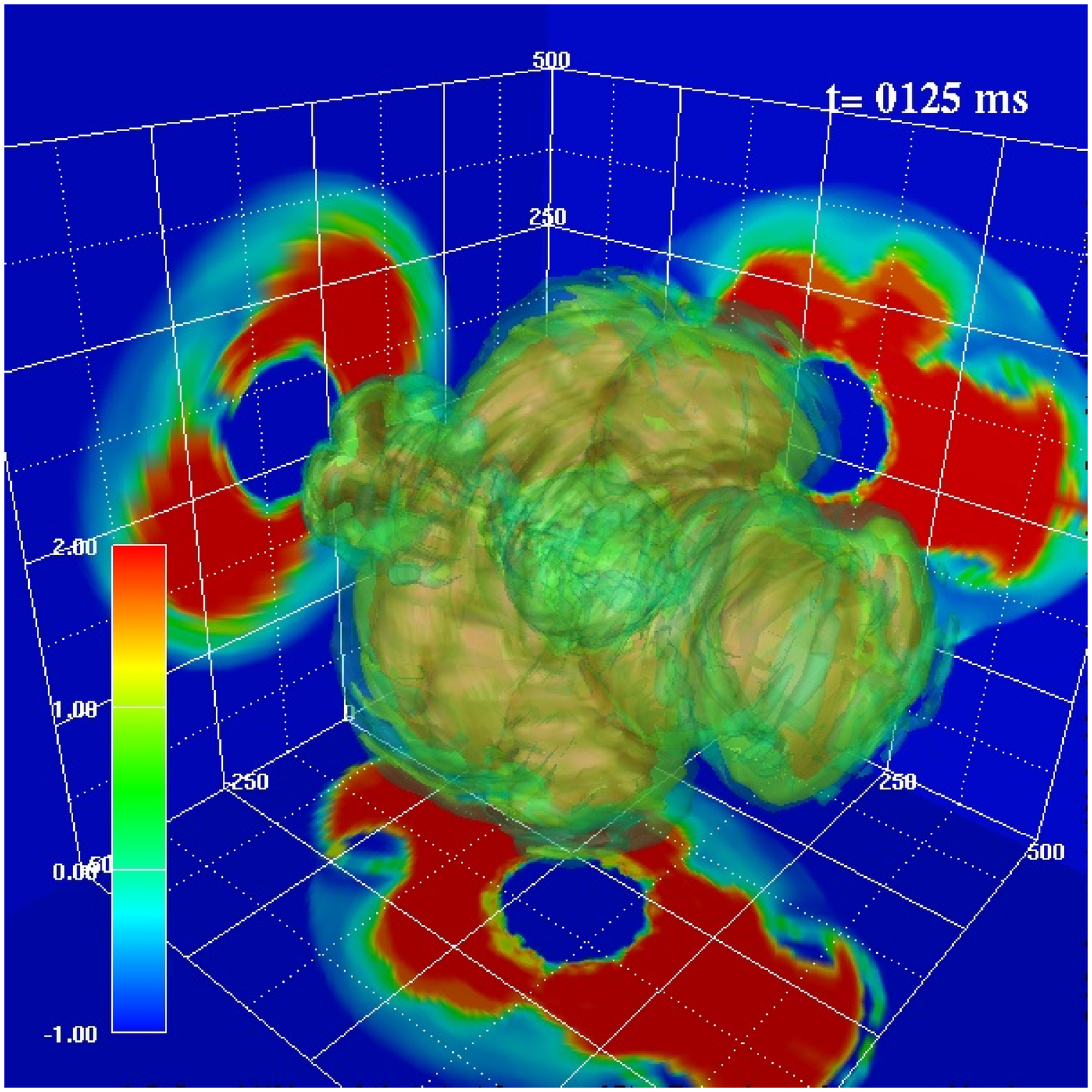}\\
 \caption{Same as Figure 1 but for the net neutrino heating rate (left panel,
 logarithmic in unit of erg/cm$^3$/s) and $\tau_\mathrm{res}/\tau_\mathrm{heat}$
 (right panel), which is the ratio of the residency to the neutrino heating time scales (see the text for details). The gain region (colored 
 in red in left panel), in which neutrino heating dominates over cooling, 
is shown to be formed. The right panel shows that 
the condition of $\tau_\mathrm{res}/\tau_\mathrm{heat}\gtrsim1$ is satisfied 
behind the globally aspherical shock, which is a characteristics of  
  SASI (figures taken from \citen{Takiwaki11}, reproduced by permissions
 of the AAS).}
\label{f2}
\end{figure}

Figure \ref{f2} shows the net neutrino heating rate (left panel) and 
$\tau_\mathrm{res}/\tau_\mathrm{heat}$: the 
 ratio of the residency time scale to the neutrino-heating time scale (right panel) 
for the snapshot of $t=125$ ms in Fig.~\ref{f1}. 
 This time scale ratio is known to be a useful quantity to diagnose 
the success ($\tau_\mathrm{res}/\tau_\mathrm{heat}\gtrsim 
1$, i.e., the neutrino-heating timescale is shorter than the dwell 
time scale of material in the gain region\footnote{ 
 where the neutrino heating dominates over the neutrino cooling
 (see Ref \cite{Janka01} for more detail).}  or failure 
($\tau_\mathrm{res}/\tau_\mathrm{heat}\lesssim 1$) of the 
neutrino-driven explosion (e.g., \citen{goshy,Janka01,thomp05,Murphy08}).
 The left panel of Fig.~\ref{f2} shows that there forms the so-called gain 
 region, in which the neutrino heating dominates over cooling (seen 
 as reddish regions in the wall panels).
 As seen in the right panel, the time scale ratio reaches $\sim2$ in the gain region, the 
  evidence that the shock-revival is driven by the neutrino-heating mechanism.

\begin{figure}[htbp]
    \centering
    \includegraphics[width=.49\linewidth]{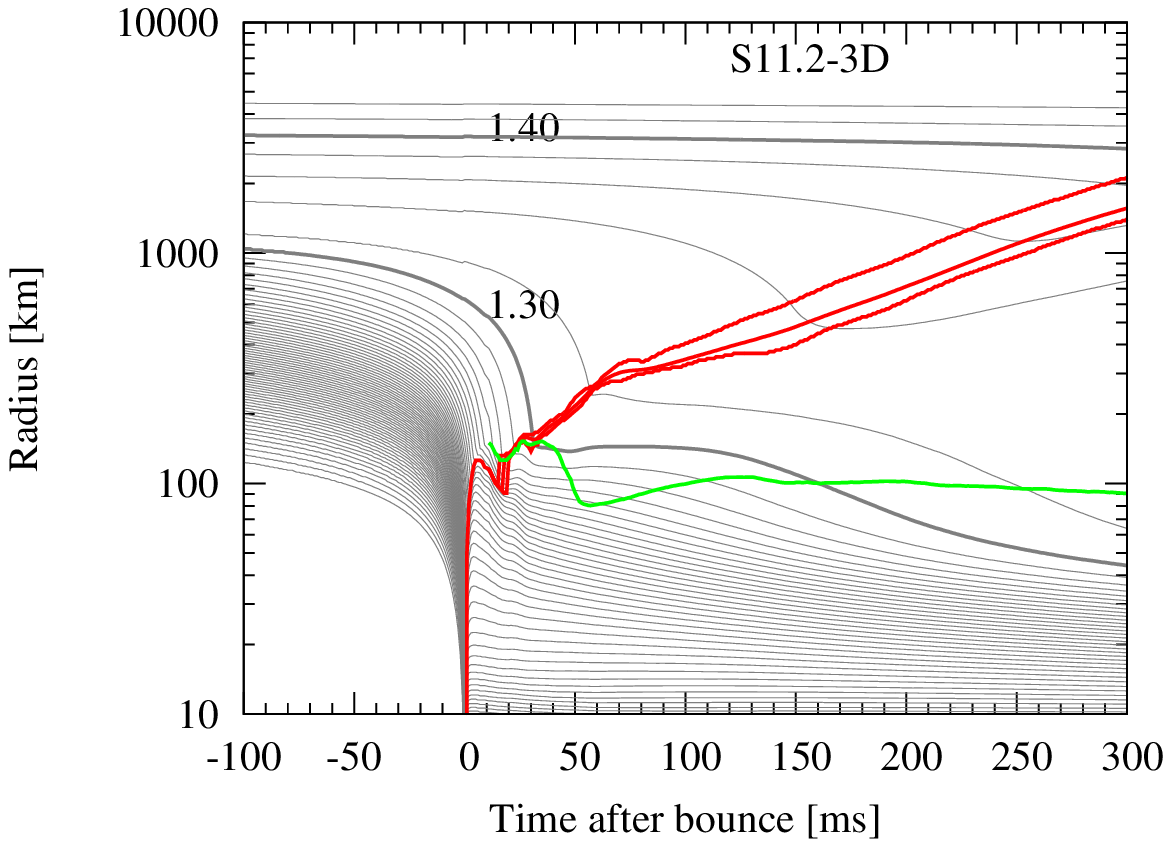}
    \includegraphics[width=.49\linewidth]{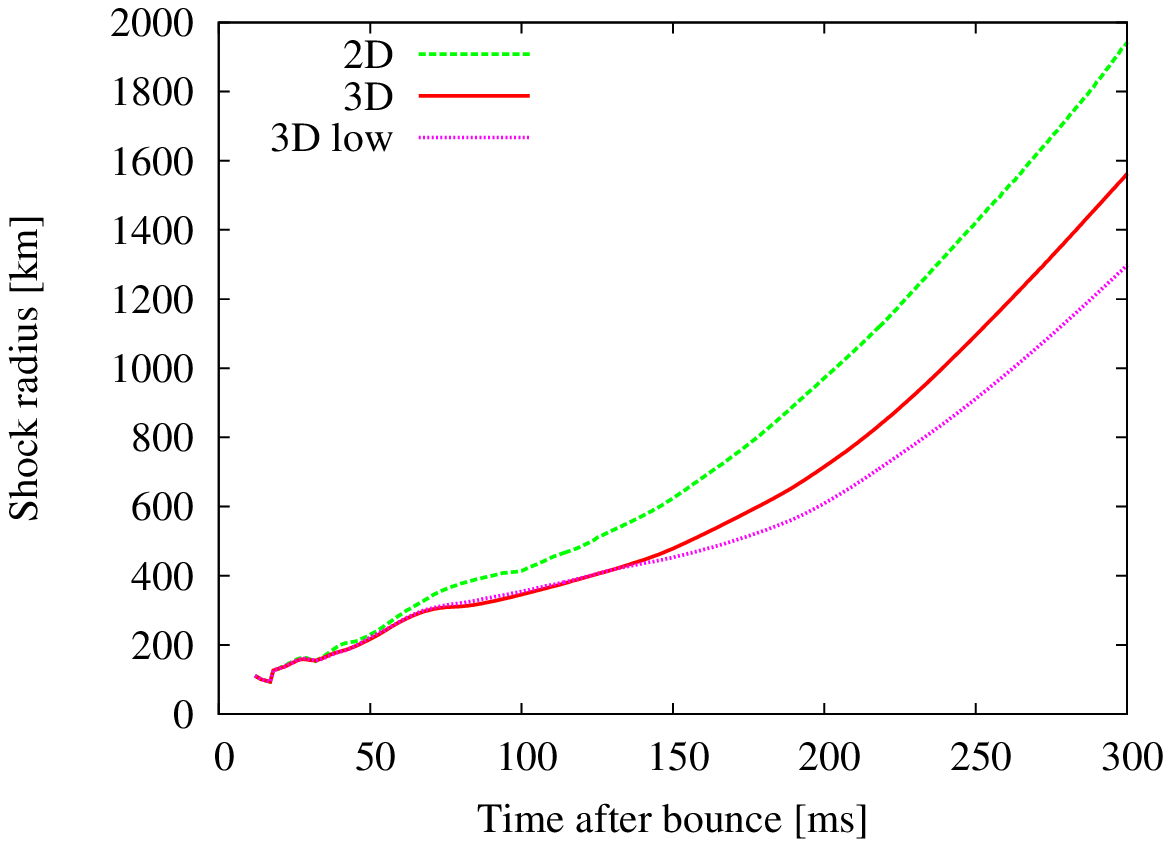}
 \caption{Left panel: the time evolution of 3D model, visualized by mass shell trajectories 
in thin gray lines. Thick red lines show the positions of shock: the maximum (top),
 average (middle) and minimum (bottom) radii. The green line represents the shock position for
the 1D model. The numbers "1.30" and "1.40" indicate the masses in unit of $M_{\odot}$ 
that are enclosed inside each mass-shell. Right panel: the evolution of average shock radii 
for the 2D (green line)  and 3D (red line) models. The pink line presents the low 
resolution 3D model, in which the azimuthal grid number is reduced to half the number for 
the standard model (see the body). Figures are taken 
from \citen{Takiwaki11} (reproduced by permissions
 of the AAS).}
\label{f3}
\end{figure}

\subsubsection{2D vs. 3D: which is more advantageous for the neutrino-driven 
 explosion?}

 The left panel of Fig.~\ref{f3} shows the comparison of mass-shell trajectories 
 between the 3D (red lines) and corresponding 1D models (green line).
 At around 300ms after bounce, the average shock radius for the 3D model exceeds
 1000km. On the other hand, no explosion is obtained  for the 1D model.
The right panel of Fig.~\ref{f3} shows a comparison of the average shock radii as a 
function of the postbounce time. In the 2D model, the shock expands rather 
continuously after bounce. These trends in the 1D and 2D models are
 qualitatively consistent with those found in \citen{Buras06a}\footnote{
 The reason why the shock of our 2D model expands on average much faster than theirs 
might be our neglect of general relativity, inelastic neutrino-electron scattering and cooling 
by heavy-lepton neutrinos. All of these important missing 
 ingredients in our 3D simulations could give a more optimistic condition for explosions.
 Apparently these ingredients should be appropriately implemented, which we hope 
 to be tractable in the next-generation 3D simulations.}.

 Comparing the shock evolutions between the 2D (green line in 
 the right panel of Fig.~\ref{f3}) and 3D (red line) models, we find that 
the shock expands much faster in 2D.
 The pink line labeled by "3D low" is the result of the low resolution 3D simulation, 
in which the azimuthal grid number is reduced to half the number for the standard model.
 Note that the 3D computational grid consists of 300 logarithmically spaced
radial zones and 64 polar ($\theta$) and 32 azimuthal ($\phi$) uniform mesh points
to cover the entire sphere with a radius of 5000km. Compared with the standard 3D model 
(red line), the shock expansion is less energetic for the low resolution model
 (later than $\sim 150$ms). These results indicate that a successful explosion is most easily 
 obtained in 2D and hampered by low resolutions. At first glance, this may be at odds with
the results obtained in the parametric 3D explosion models (e.g., \citen{Nordhaus10}), 
in which the authors claimed that explosions would be easier in 3D than in 2D. The reason 
for the discrepancy is summarized shortly.

\begin{figure}[htbp]
    \centering
    \includegraphics[width=.45\linewidth]{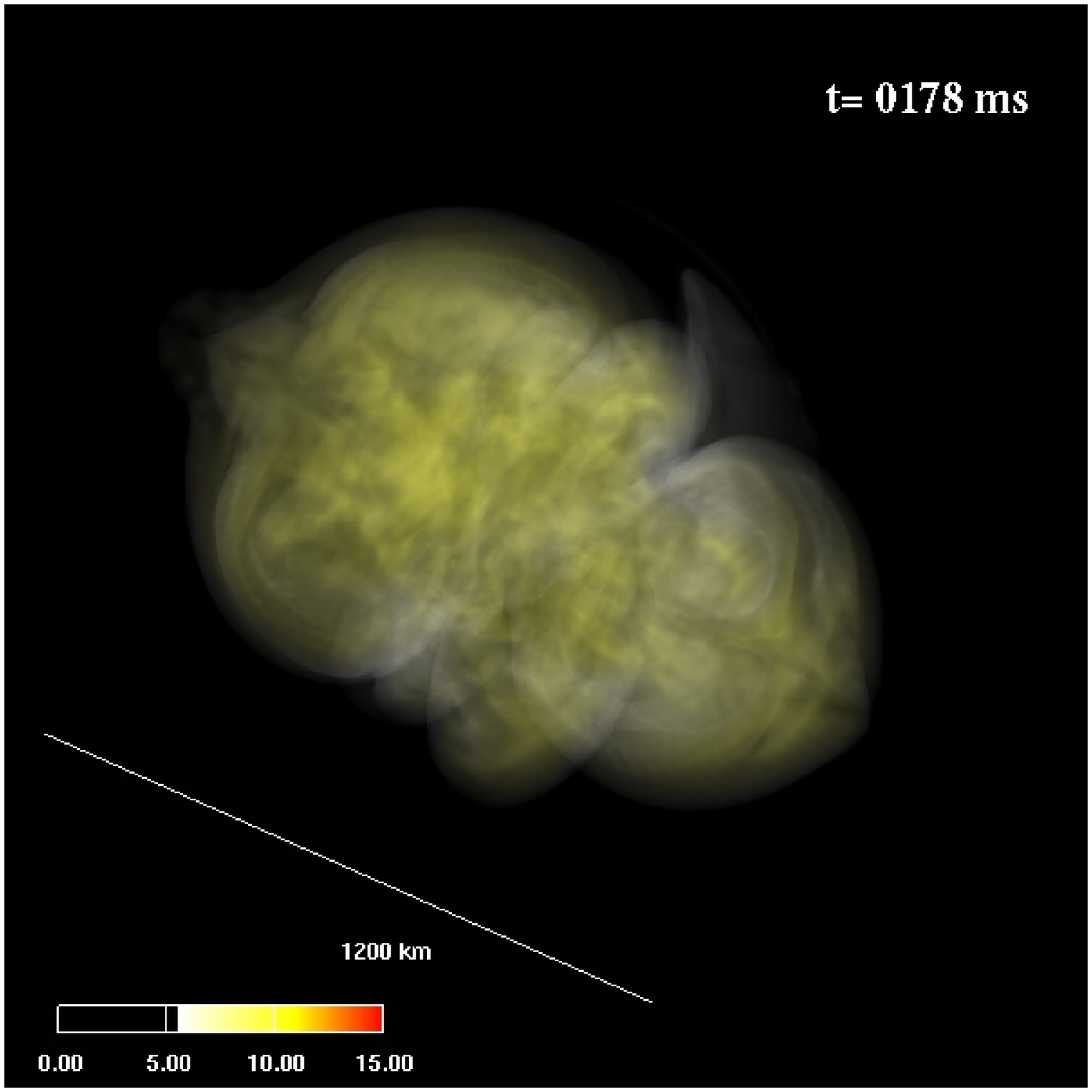}
    \includegraphics[width=.45\linewidth]{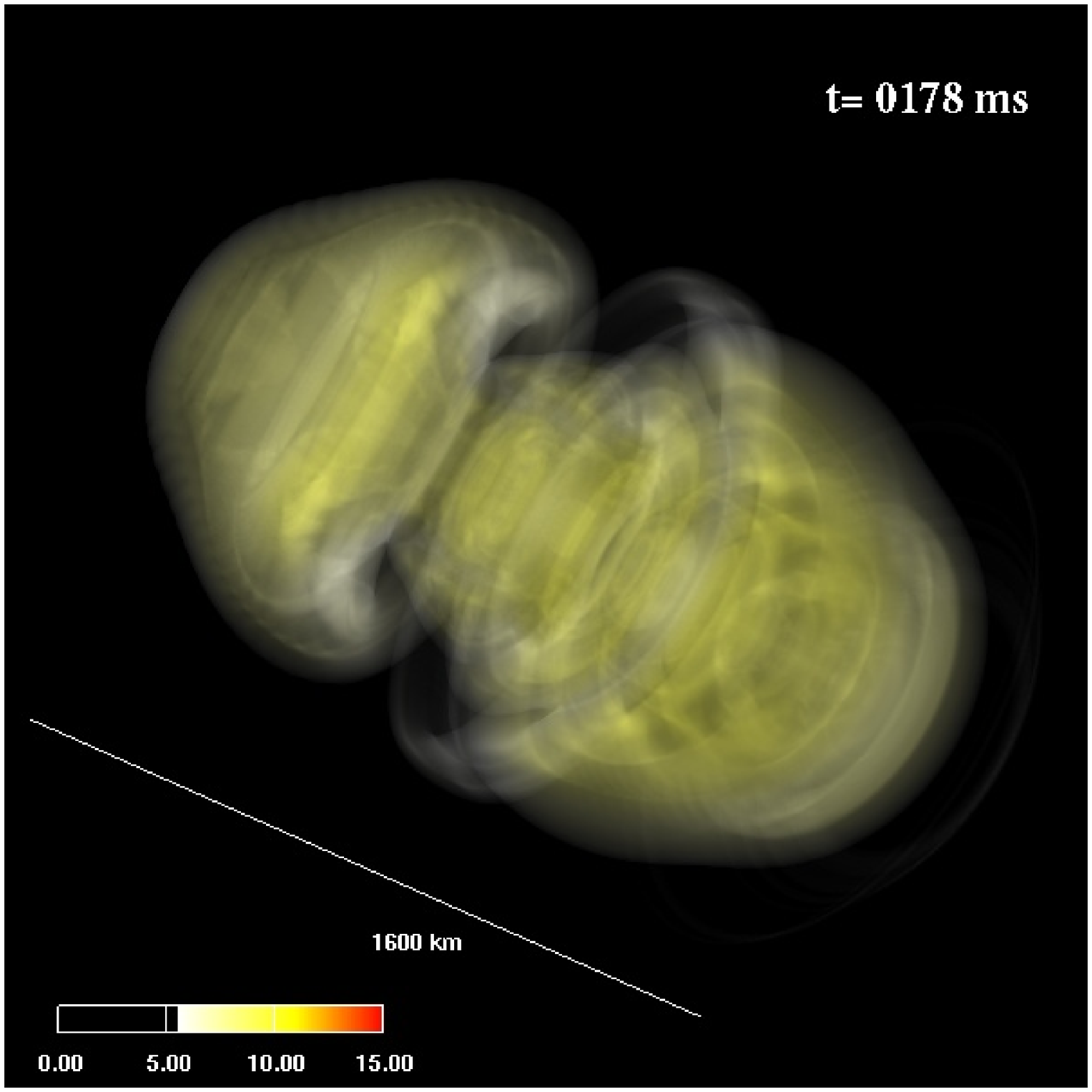}
 \caption{Blast morphologies at $t=178$ms after bounce in our 3D (left) and 2D (right) models presented as the volume rendering of entropy.  
The progenitor is a $11.2 M_{\odot}$ star~\cite{woos02}. The polar axis is tilted by about $\pi/4$ in both panels 
(figures taken from \citen{Takiwaki11}, reproduced by permissions
 of the AAS).}
\label{f5_added}
\end{figure}

\begin{figure}[htbp]
    \centering
    \includegraphics[width=.39\linewidth]{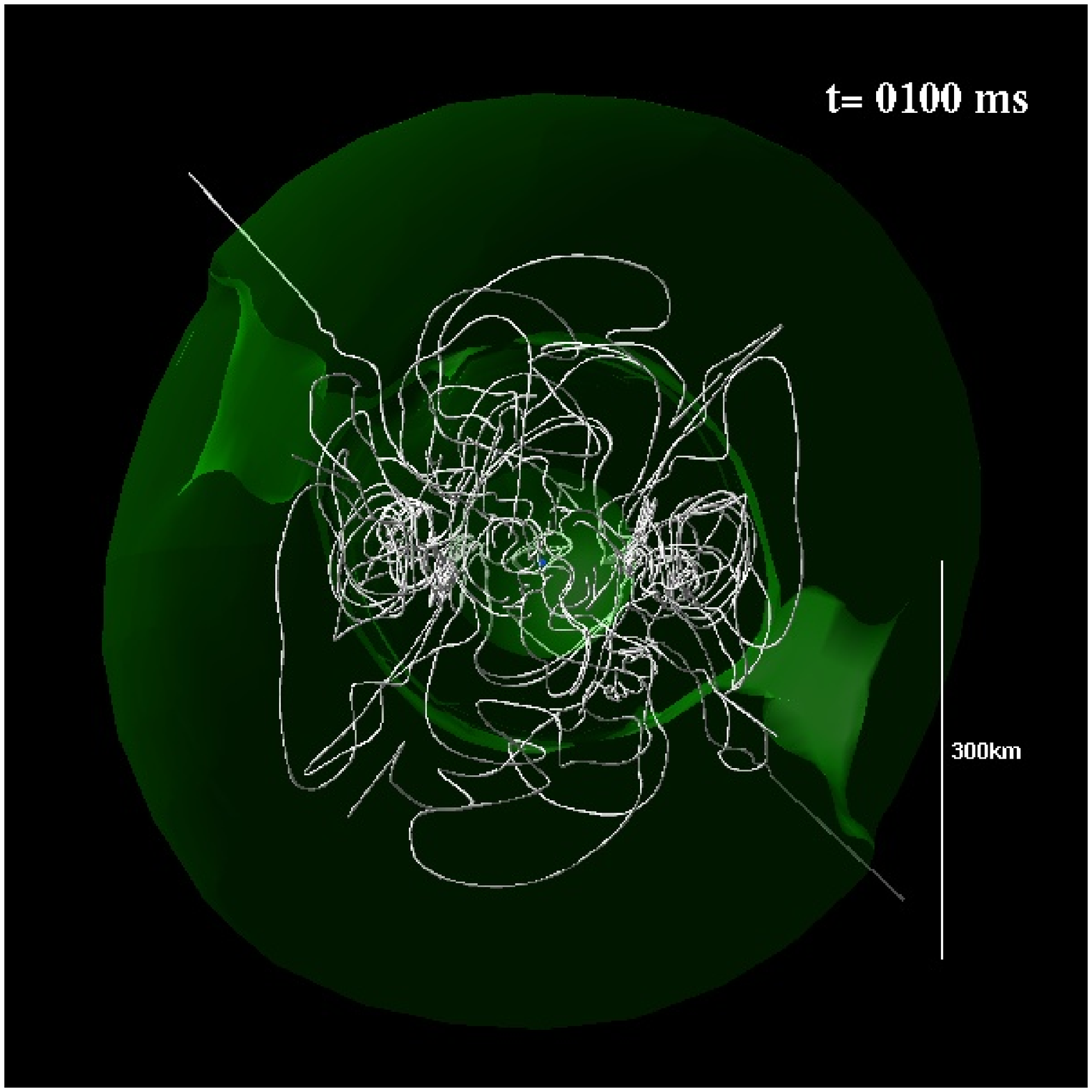}
    \includegraphics[width=.49\linewidth]{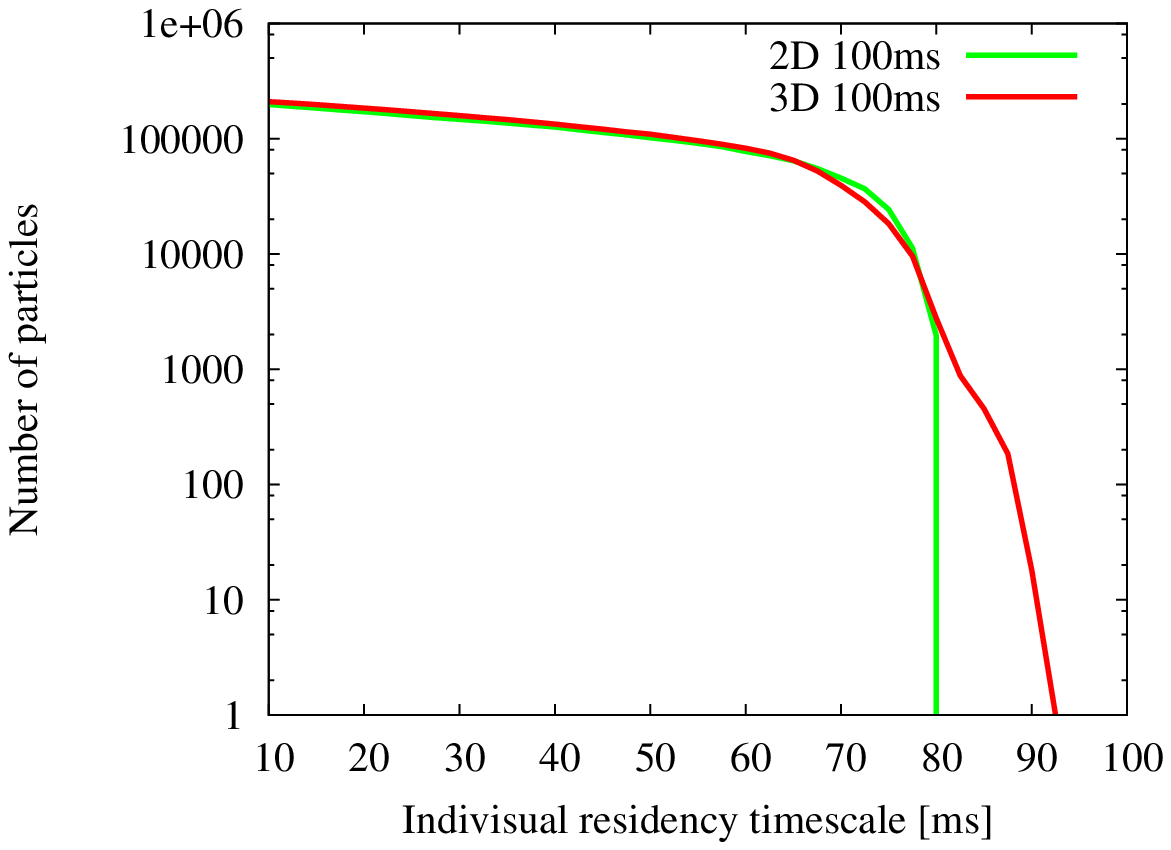}
 \caption{Left panel: streamlines of selected tracer particles 
advecting through the shock to the PNS in the 3D model. The observer is located on the polar axis. Several surfaces of constant 
entropy are also displayed. An outer greenish one marks the shock surface whereas the PNS is presented as a central sphere. The 
linear scale is given at the bottom right corner. Right panel: the number of tracer-particles traversing the gain region as a function of 
their individual residency time for the 2D and 3D models given at $t=100$ms after bounce. The maximum residency time in the 3D model 
is longer than that in the corresponding 2D model (figures taken from \citen{Takiwaki11}, reproduced by permissions
 of the AAS). }
\label{to_ref1}
\end{figure}

\begin{figure}[htbp]
    \centering
    \includegraphics[width=.49\linewidth]{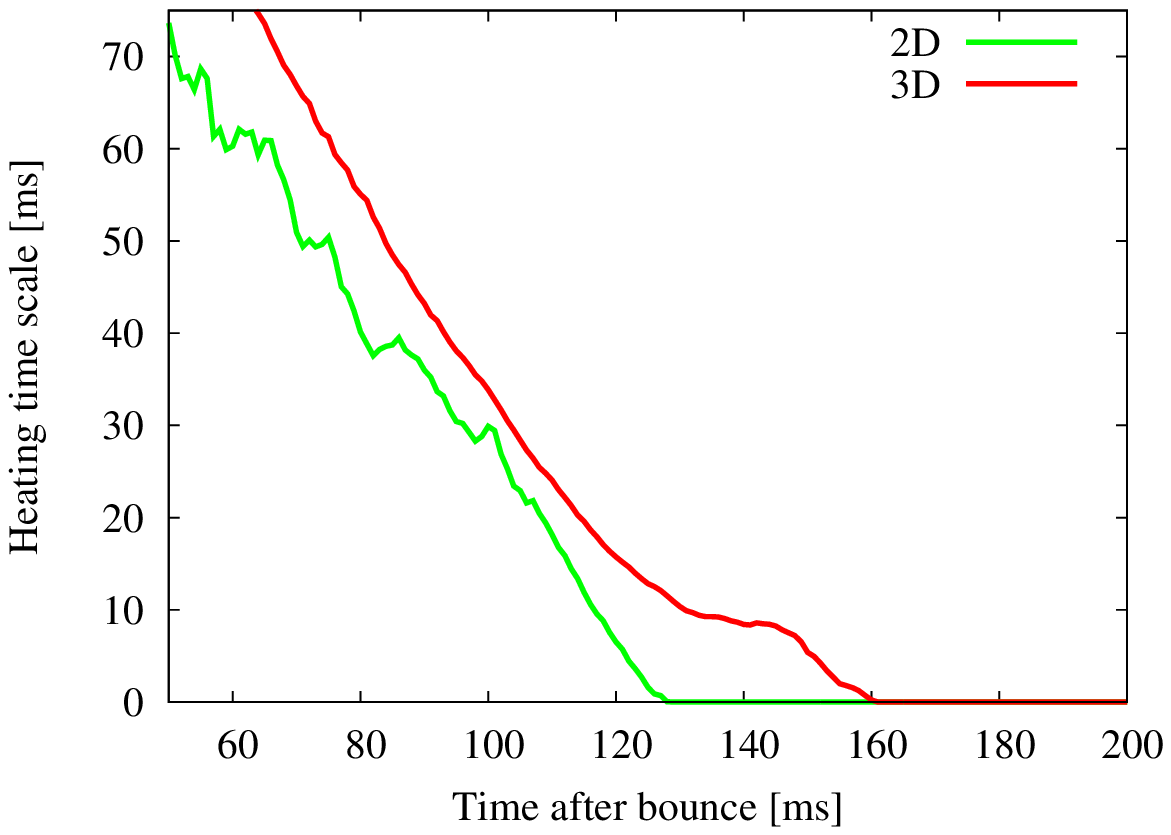}
    \includegraphics[width=.49\linewidth]{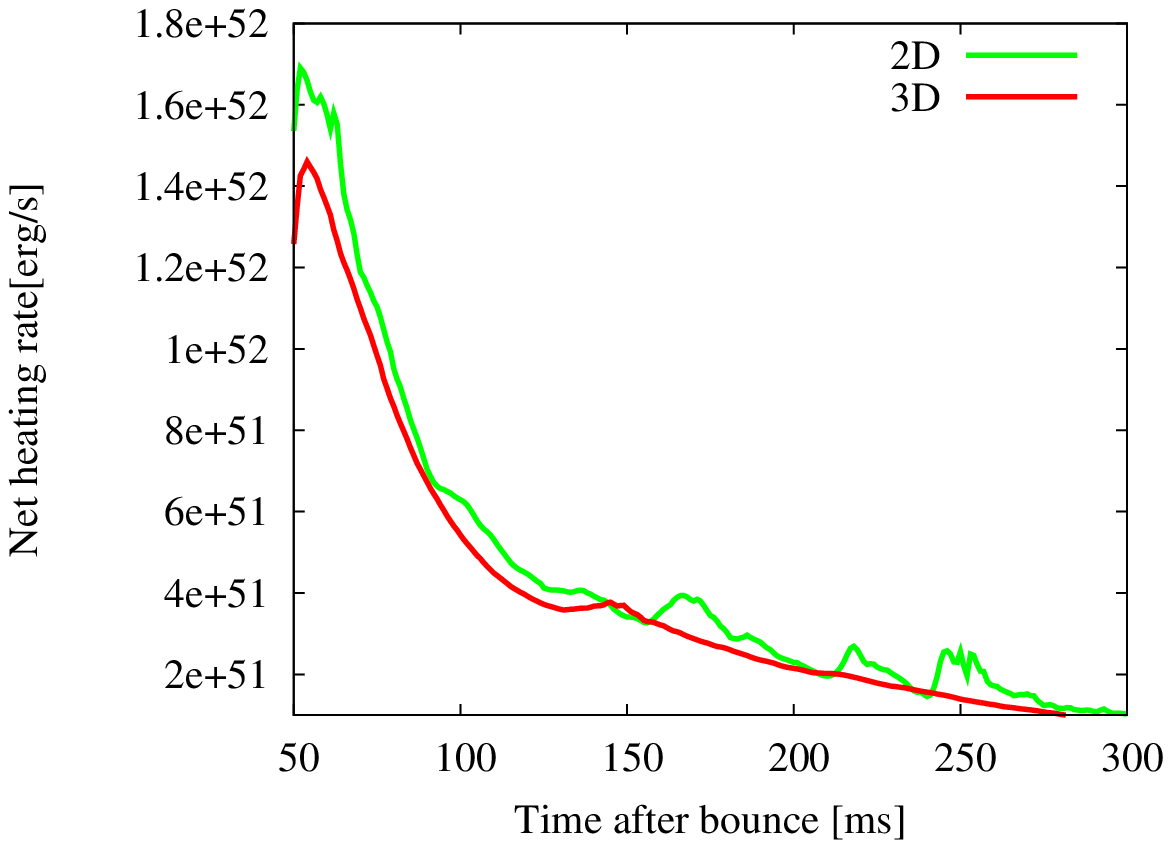}\\
 \caption{Time evolutions of neutrino-heating time scale (left) 
and volume-integrated net neutrino-heating rate (right) in the 2D (green line) 
and 3D (red line) models (figures taken from Ref. \cite{Takiwaki11}, reproduced by permissions of the AAS). }
\label{f11}
\end{figure}

 Fig.~\ref{f5_added} compares the blast morphologies in the 3D (left panel) and 2D (right) models.
 In the former, non-axisymmetric structures are clearly seen.
 Performing a tracer-particle analysis, we find that the maximum residency time is longer in 3D 
than in 2D owing to the non-axisymmetric motions (see Fig.~\ref{to_ref1}). This 
is one of advantageous aspects of 3D models to obtain the neutrino-driven explosions.
Another merit in 3D is that convective matter motions below the gain radius 
is much more violent than in 2D, which enhances the neutrino luminosity in 3D (see \citen{Takiwaki11} 
for more details). The negative point, on the other hand, is lower energies of emitted neutrinos 
owing to the enhanced cooling. The competition of these effects eventually leads to 
a longer neutrino-heating time scale in our 3D models with an outcome of a smaller 
 net-heating rate compared with the corresponding 2D model (Fig.~\ref{f11}). 
 Note here that the IDSA scheme, with which the feedback from the mass accretion to the neutrino luminosity
 is automatically and self-consistently incorporated unlike the light-bulb approximation that assumes a constant luminosity, 
is quite efficient and a good choice for the first-generation 3D simulations.

Although it is encouraging that the shock expansion becomes more energetic with better resolution (recall that the explosions
obtained so far are all under-energetic~\cite{Kotake11} and the present model is no exception),
this implies that a systematic and time consuming convergence test is required to draw a robust conclusion (e.g. \citen{Hanke11}). 
More advanced treatments of neutrino transport as well as of gravity will be also needed, which will probably be a subject 
done on forthcoming petaflops-class supercomputers. 

\clearpage
\subsection{3D fully general relativistic simulations with an approximate
 neutrino transport}\label{3d_2}

In addition to the 3D effects mentioned in the previous section, 
impacts of GR on the neutrino-heating mechanism stand out 
among the biggest open questions in the supernova theory. It should be remembered 
that researchers in the pioneering era of supernova simulations tackled GR 
from very early on~\cite{May66}, using a newly derived formulation~\cite{Misner64}. 
One year after Colgate \& White published their seminal paper~\cite{Colgate66} , Schwartz~\cite{Schwarz67} 
reported the first fully GR simulation of stellar collapse 
to study the supernova mechanism, implementing a gray neutrino transport in the 
1D GR hydrodynamics code\footnote{Cited from his paper, 
"{\it In this calculation, the neutrino luminosity of the core is found to be $10^{54}$ erg/s, or 1/2 a solar rest mass
 per second !! .... This is the mechanism which the supernova explodes".} The 
 neutrino luminosity rarely becomes so high in modern simulations, but it is 
 surprising that the potential impact of GR on the neutrino-heating mechanism
 was already indicated in the very first GR simulation.}.
 Using the GR Boltzmann equations derived by Lindquist~\cite{Lindquist66},
 Wilson~\cite{Wilson71} developed a 1D GR-radiation-hydrodynamics code 
including more realistic (at that time) descriptions
 of the collisional term than the one adopted by Schwartz~\cite{Schwarz67}. 
1D GR hydrodynamical simulations with the so-called leakage scheme for neutrino cooling
were also performed to explore hydrodynamical properties up to the prompt-shock 
stagnation.~\cite{VanRiper79,VanRiper81,VanRiper82}. These pioneering studies, albeit 
 with a much more simplified neutrino physics than today, did provide a 
 bottom-line of our current understanding of the supernova mechanism
(see Bruenn et al.~\cite{Bruenn01} for a complete list of references for the early GR studies).
 In the middle of the 1980s, Bruenn~\cite{Bruenn85} developed a code that coupled 1D GR hydrodynamics to 
the MGFLD transport with a relativistic correction of order $(v/c)$ and included the so-called standard
 set of neutrino interactions. By the late 1990s, the ultimate 1D simulations, in which the GR Boltzmann transport
 is coupled to 1D GR hydrodynamics, became 
feasible~\cite{Mezzacappa93a,Mezzacappa93b,Yamada97,Yamada99,Bruenn01,Liebendorfer01,Sumiyoshi05}. 

Bruenn et al.~\cite{Bruenn01} demonstrated clearly that the average neutrino energies of all neutrino flavors 
 are higher in GR than in Newtonian gravity during the shock-heating phase. 
 They also pointed out that the redshift and gravitational time dilation, the agents to counteract, are
rather minor. Employing the best weak interaction rates available at present,
 Lentz et al.~\cite{Lentz11} reported very recently the update of Bruenn et al.~\cite{Bruenn01}, in which
they showed that the neglect of the observer corrections in the transport equation 
 particularly does harm to neutrino-driven explosions. 
  In these full-fledged 1D simulations, a commonly observed disadvantageous aspect of GR is 
that the residency time of material in the gain region is shorter owing to stronger gravitational pull. All these effects
taken into account, GR is negative in the neutrino-heating mechanism in 1D. In fact, switching from Newtonian to 
GR hydrodynamics, we find that the maximum shock radius becomes $\sim 20$\% smaller in the postbounce phase (e.g., \citen{Lentz11}). 

 In the most advanced multi-D simulations with spectral neutrino transport mentioned earlier, 
 GR effects are addressed at best by a modified gravitational potential that is adopted from the 1D  
 post-Newtonian correction~\cite{Buras06a,Buras06b,Marek09,bruenn}.
 A possible drawback of this prescription is that the total-energy conservation is compromised owing to the 
term added artificially to the Poisson equation for self-gravity. Since the supernova engine is powered by the gravitational energy, 
we have to avoid any potential inaccuracies in energy conservation. On the other hand, there have been a number of fully general 
relativistic simulations of massive star collapse thus far both in 2D (e.g., \citen{Shibata05a})
and in 3D (e.g., \citen{Shibata05b,Ott2007}, and references therein). The so-called conformal-flatness approximation (CFC) has been
also employed~\cite{Dimmelmeier02,Isa12}. In these computations the treatment of neutrino transport was overly compromised, e.g., with the transfer
being entirely replaced by a prescribed $Y_e$ formula~\cite{Matthias05} or the so-called leakage scheme being 
adopted.~\cite{Sekiguchi10,ott12}\footnote{Very recently, M\"uller et al.~\cite{Bernhard12} reported explosions for $11.2$ and $15 M_{\odot}$ 
stars based on their 2D GR simulations in CFC with detailed neutrino transport similar to \citen{Buras06a} being implemented~\cite{Bernhard10}.}

 In this section, we present results from our first generation multi-D hydrodynamical 
simulations in full GR that incorporate an approximate neutrino transport~\cite{kuroda12}. 
 The code is a marriage of an adaptive-mesh-refinement (AMR),
 conservative 3D GR MHD code developed by Kuroda and Umeda~\cite{Kuroda10}, 
and the approximate neutrino transport code that we newly developed in this work. 
Our GR code is based on the Baumgarte-Shapiro-Shibata-Nakamura (BSSN) 
formalism~\cite{Shibata95,Baumgarte99}. Hydrodynamics can be 
 solved either in full GR or in special relativity (SR), a feature that allows us to 
 investigate purely GR effects on the supernova dynamics. 
  Using the so-called M1 closure scheme with an analytic variable Eddington factor, 
 we solve the radiation energy and momentum. This part of the code is partially based on the Thorne's momentum 
formalism~\cite{Thorne81}, which was recently extended by Shibata et al.~\cite{Shibata11} so that it should be  
more suitable for neutrino transport. To simplify the source terms of the transport equations, on the other hand,
a multi-flavour neutrino leakage scheme is also employed partially.
The new code is designed so that it could evolve the Einstein equation and GR radiation-hydrodynamical equations
self-consistently, satisfying the Hamiltonian and momentum constraints. The AMR technique implemented in the 3D 
 code enables us to follow the dynamics from the onset of gravitational core-collapse of a 15 $M_{\odot}$ star through bounce 
up to $\sim 100$ms after bounce in this study. We compute four models with different combinations of SR/GR and
1D/3D, which we label as 1D-SR, 1D-GR, 3D-SR and 3D-GR, respectively. Limited to the early postbounce phase ($t\lesssim 100$ms), 
we discuss exploratory results in the following sections to illuminate GR effects in the multi-D neutrino-heating mechanism. 

\begin{figure}[htbp]
\begin{center}
\includegraphics[width=150mm]{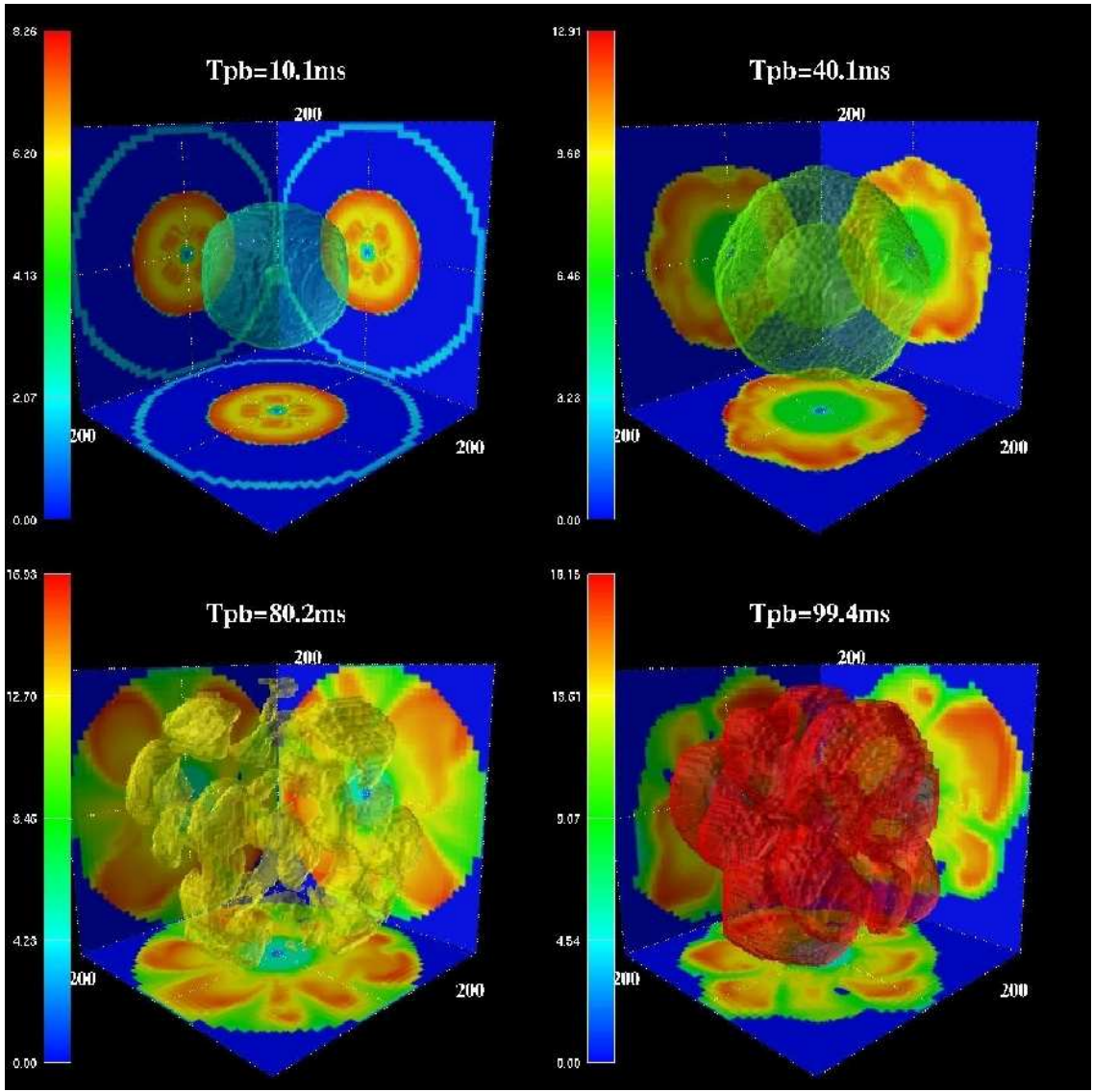}
\end{center}
\caption{3D snapshots of entropy per baryon at four different times 
(top left; $t = 10$ms, top right; $t =40$ms, bottom left; $t =80$ms,
 and bottom right; $t =100$ms) for model 3D-GR. 
The contours in the $x=0$, $y=0$, and $z=0$ planes are projected on the back right, 
back bottom and back left sidewalls, respectively, to visualize the 3D structures. In each plot, 
an arbitrarily chosen iso-entropy surface is displayed. The linear scale is indicated along each axis 
in unit of km (figures taken from Ref.\citen{kuroda12},
 reproduced by permissions of the AAS).}
\label{f7}
\end{figure}

\begin{figure}[htbp]
\begin{center}
\includegraphics[width=80mm,angle=-90.]{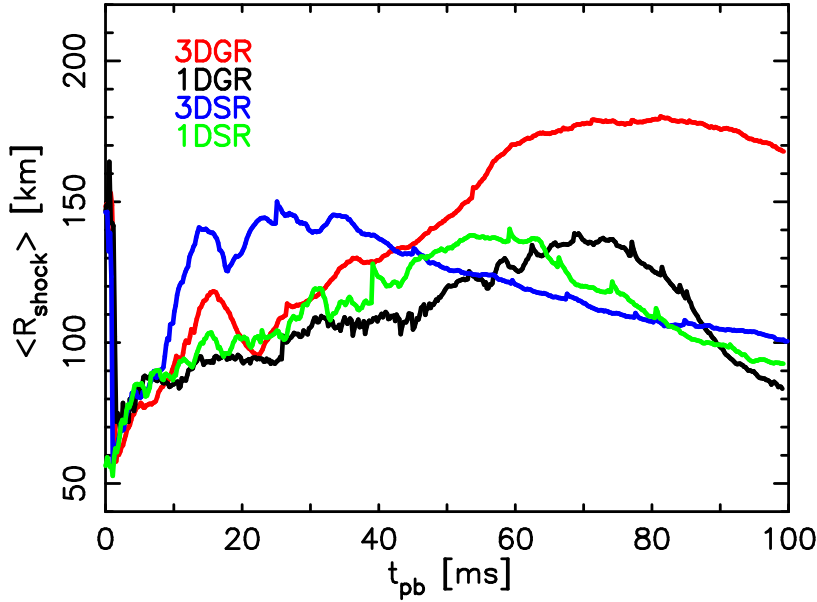}
\end{center}
  \caption{Evolutions of average shock radii as a function of time after bounce
for models 1D-SR (green line), 3D-SR(blue line), 1D-GR (black line), and 3D-GR (red line), 
respectively. The maximum shock radii and the time scale of shock recession for model
1D-GR is similar to those obtained in the previous 1D simulations for the same progenitor model
that incorporated neutrino transport more in detail (e.g., \citen{Lentz11,Bernhard10}).}
\label{shock}
\end{figure}

\subsubsection{Hydrodynamical features in full GR 3D simulations}

Four snapshots in Fig.~\ref{f7} are helpful to characterize 
 the postbounce features in our 3D-GR model\footnote{
The 3D computational domain is a cube of $10000^3$ km$^3$ volume fit in the Cartesian coordinates.
 The maximum refinement level in AMR is 5 at the beginning and then incremented as 
the collapse proceeds. The criterion for incrementation is renewed when the central density exceeds 
$10^{12,13,13.5}$ g/cm$^{3}$, yielding an effective resolution of $\Delta x\sim 600$m at bounce 
(see \citen{Kuroda10} for more details).}.
 The top left panel shows the distribution of entropy per baryon at $t \approx 10$ms, when the bounce shock stalls 
at a radius of $\sim 90$km (shown as a central blueish sphere). 
 Comparing the top left with top right panel in Fig.~\ref{f7}, we see that 
the shock (a greenish sphere in the top right panel) becomes bigger.
This implies that the bounce shock turns into a so-called ``passive'' shock, which expands outward gradually with all 
matter advecting inward after passing through the shock (e.g., \citen{Buras06a}). 
 At this stage, there forms a gain region, in which neutrino-heating dominates over local cooling.
 The neutrino-driven convection gradually develops from this point on. The sidewalls in the top right panel also demonstrate
 the growth of the postshock convection. The entropy behind the standing shock becomes higher with time owing to 
the neutrino-heating, which can be inferred from yellowish bubbles in the bottom left panel.
 These high entropy bubbles ($s \gtrsim 10{\rm k_B}$) rise and sink behind the standing shock. The shock deformation 
is dominated by unipolar and bipolar modes, which may be
 a characteristic feature of the SASI.
The neutrino-heated region becomes larger with time in a non-axisymmetric way, which is 
evident in the bubbly structures 
that are shown as reddish regions in the bottom right panel.

\begin{figure}[htbp]
\begin{center}
\includegraphics[width=140mm]{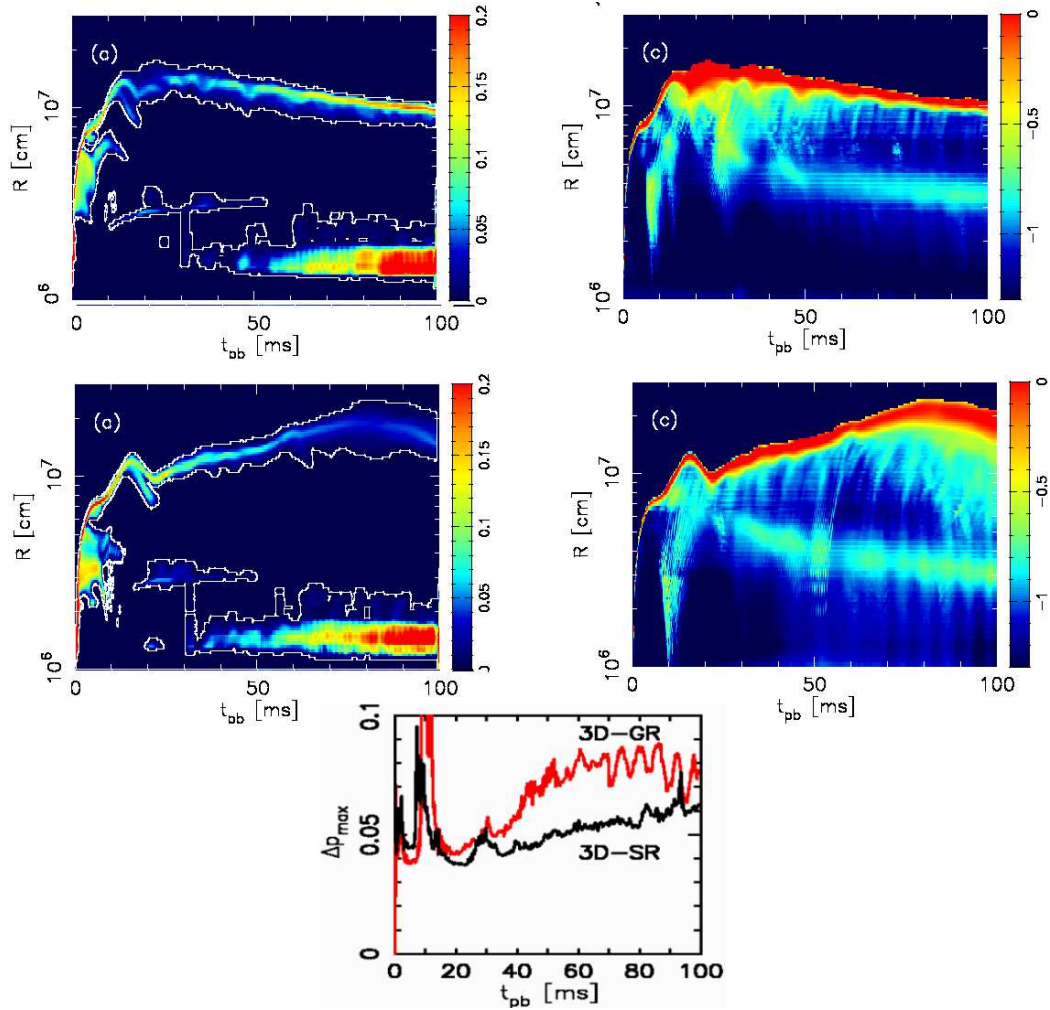}
\end{center}
\caption{Postbounce evolution of angle-averaged Brunt-V$\ddot{\rm a}$is$\ddot{\rm a}$l$\ddot{\rm a}$ frequencies 
($\omega_{\rm BV}$ in ms$^{-1}$) (left panels) and pressure dispersions $\Delta p$ (right panels) for models 3D-SR (top row)
 and 3D-GR (middle row), respectively, and the maximum pressure 
 dispersion $\Delta p_{\rm max}$ (bottom row). Only convectively 
unstable regions (i.e. $\omega_{\rm BV}>0$) are shown and the white lines represent the boundaries of convective regions with $\omega_{\rm BV}=0$
in the left panels. In searching the maximum value of pressure 
dispersions for the bottom panel, we restrict the region to $20\le r \le 50$km, i.e., 
the vicinity of the coupling radius (figures taken from \cite{kuroda12},
 reproduced by permissions of the AAS).}
\label{pic:F7}
\end{figure}

In these simulations up to 100~ms after bounce, the largest shock radius is recorded in
model 3D-GR (red line in Fig.~\ref{shock}). The other models have already seen the shock recession by this time.
Before we focus on the reason for it in the next section, let us compare the activities of convection and SASI between 
the computations in SR and GR. Figure \ref{pic:F7} displays the angle-averaged Brunt-V$\ddot{\rm a}$is$\ddot{\rm a}$l$\ddot{\rm a}$ (BV) 
frequency $\omega_{\rm BV}$~\cite{Buras06a} (left panels) and 
pressure dispersion $\Delta p$\footnote{This is defined as $
\Delta p\equiv \frac{\sqrt{\langle p^2 \rangle -\langle p 
\rangle^2}}{\langle p \rangle}$,
 where  $\langle A \rangle $ represents the angle average of quantity $A$.} in a logarithmic scale 
(right panels)
for models 3D-SR (top two panels) and 3D-GR (middle two panels), respectively. 

Irrespective of SR or GR,  there are typically three convectively unstable regions 
in the postbounce phase: (1) the greenish region behind the shock\footnote{Note that the shock is indicated by 
 a white thin line that rises quickly after bounce and declines after the passive shock stalls at a radius of $r \sim 150$km
.}  at $t \lesssim 20$ms that corresponds to the so-called prompt convection,  (2) a narrow horizontal strip behind the 
shock that corresponds to the convection sometimes referred to as Bethe convection, (3) a thick horizontal strip above the 
 PNS at a radius of $r \sim 10 - 20$km that emerges at $t \gtrsim 60$ms.
Comparison between the two panels in the left column for models 3D-SR and 3D-GR shows that the unshocked core 
(the central part of PNS surrounded by the convective region) is more compact in the GR model  at $t \gtrsim 50$ms.
The PNS convection develops only very weakly before $t \sim 60$ms. This is common to both SR and GR cases and due to the stabilizing 
effect of a positive entropy gradient prevailing outside the PNS surface ($r \sim 10$km).
The PNS convection becomes vigorous gradually with time afterward as the negative lepton gradient develops in the nascent PNS. 

Next we pay attention to the right two panels of Fig.~\ref{pic:F7} to infer the activities of SASI. 
In these panels, we may recognize two horizontal strips: one is colored in red and shows strong pressure perturbations behind the shock; 
the other is colored in green and roughly corresponds to the bottom of cooling layer that recedes from $r \sim 80$km to $r \sim 30$km 
gradually in time from $t \sim 30$ms to $t \sim 100$ms. The accreting flows that advect from the standing shock on to PNS receive
abrupt decelerations near the bottom of the cooling layer. Strong pressure perturbations are produced there as mentioned above and  
propagate outward subsequently until  they hit the shock.
The up-going stripes in the figure seem to indicate these outward propagation of 
pressure waves. The features just mentioned may be reminiscent of the 
so-called advectic-acoustic cycle (e.g., \citen{Foglizzo00,Foglizzo02,Scheck08} and references therein) and are common to the SR and 
GR models.

The bottom panel of Fig.~\ref{pic:F7} compares the maximum pressure 
 dispersions that the 
 advecting vortices produce in the vicinity of the deceleration region.\footnote{Scheck
 et al.~\cite{Scheck08} referred to this region as the "coupling radius", at which the coupling of 
vortices and acoustic waves takes place.}  As is evident in the figure, the maximum pressure dispersion is generally
 larger in the GR model (red line) than the SR counterpart (black line) in the early postbounce phase ($t\lesssim 100$ms) we study here. 
This is presumably because stronger gravitation pull in GR makes the coupling radius smaller, leading to the production of 
 more energetic acoustic waves. Although it is not straightforward to say something very 
 solid only from this figure, what we observed so far in our 3D-GR model, i.e., the generation of stronger acoustic waves and 
larger shock radii in GR, suggests at least that 3D is not unfavorable for the neutrino-driven explosion. We now move on to 
more detailed discussions on potential impacts of 3D and GR on the neutrino-heating mechanism. 

\subsubsection{3D versus GR: impacts on the neutrino-heating mechanism}
\label{4_3}

Recalling that the neutrino heating rate can be expressed as $Q^{+}_{\nu} \propto L_{\nu}\langle \epsilon_{\nu}^2 \rangle$~\cite{Janka01}, 
we first analyze the neutrino luminosities ($L_{\nu}$) and mean energies ($\langle\varepsilon_\nu\rangle$). We then compare the 
dwell time with the neutrino-heating time in the gain region and discuss which one, 3D-SR or 3D-GR, is more likely to satisfy 
the criterion for shock revival.

\begin{figure}[htpb]
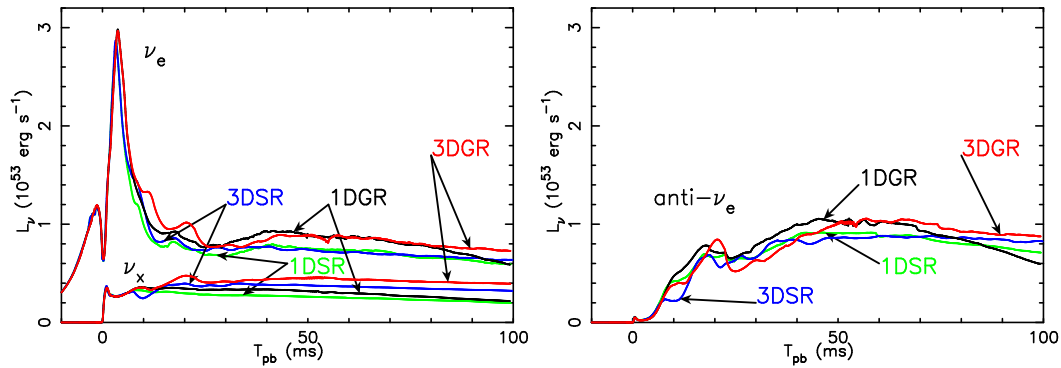

\begin{center}
\includegraphics[width=48mm,angle=-90.]{Lnu24.eps}
\includegraphics[width=48mm,angle=-90.]{Lnu3.eps}
\end{center}
  \caption{Luminosities of all neutrino flavors for all the models as a function of time. $\nu_e$ and $\nu_x$ are displayed in the left panel
 and ${\bar{\nu}}_e$ is presented in the right panel (figures taken from \cite{kuroda12}, reproduced by permissions of the AAS).}
\label{lnu}
\end{figure}

Figure \ref{lnu} shows for all the computed models the evolutions of neutrino luminosities of all the species: $\nu_e$, $\nu_x$ (left panel) and
 ${\bar{\nu}}_e$ (right panel). The spike in the $\nu_e$ luminosity is the so-called neutronization burst that occurs when the shock passes 
through the neutrino sphere for $\nu_e$. The peak $\nu_e$ luminosities for the GR models are $L_{\nu_e}\sim3\times10^{53}$erg s$^{-1}$, 
slightly larger than those for the SR models, ($L_{\nu_e}\sim2.9\times10^{53}$erg s$^{-1}$) but are rather insensitive to dimensionality.
This trend is qualitatively similar to what was found in \citen{Bruenn01}. On the other hand, recent studies with weak interactions 
being treated in more detail in approximate Boltzmann transport have demonstrated that the peak $\nu_e$ 
luminosity is $\sim 10$\% smaller in GR than in Newtonian gravity (e.g., \citen{Lentz11,Bernhard10}). This may carry an important 
message that the Boltzmann transport should be implemented in the full GR simulations to obtain a $\sim 10$\% accuracy, which is not 
small at all when speaking about the neutrino-heating mechanism. 
 
After the neutronization burst ($t \gtrsim 10$ms), the $\nu_e$ luminosities increase with time in the GR models whereas
they are almost constant in the SR models in the early postbounce phase up to $t\sim100$ms (green and blue lines in Fig.~\ref{lnu}). 
The $\bar{\nu}_e$ luminosities are highest in model~3D-GR after $t \sim 50$ms (red line in the right panel of Fig.~\ref{lnu}).
This is also the case for the $\nu_x$ luminosities (left panel). Although the luminosities change with time, the luminosities of
different neutrino flavors satisfy the following orders in general: 
\begin{itemize}
\item[]{$\nu_e$: 3D-GR $>$ 1DGR, 3D-SR $\sim$ 1D-SR,}
\item[]{$\bar{\nu}_e$: 3D-GR $>$ 1DGR, 3D-SR $>$ 1D-SR,}
\item[]{${\nu}_x$: 3D-GR $>$ 1DGR, 3D-SR $>$ 1D-SR.}
\end{itemize}
In short, both 3D and GR raise the neutrino luminosity in the early postbounce phase. More specifically,  
the maximal boost by GR, $\sim 50 \%$, is obtained for $\nu_x$ in 3D as is found in the left panel of Fig.~\ref{lnu}
whereas the maximum gain by 3D is less than $\sim 20$\%, which is obtained for $\bar{\nu}_e$ in the comparison between   
models 3D-GR and 1D-GR. These results indicate that GR holds a comparatively more important key to the neutrino luminosity.

\begin{figure}[htbp]
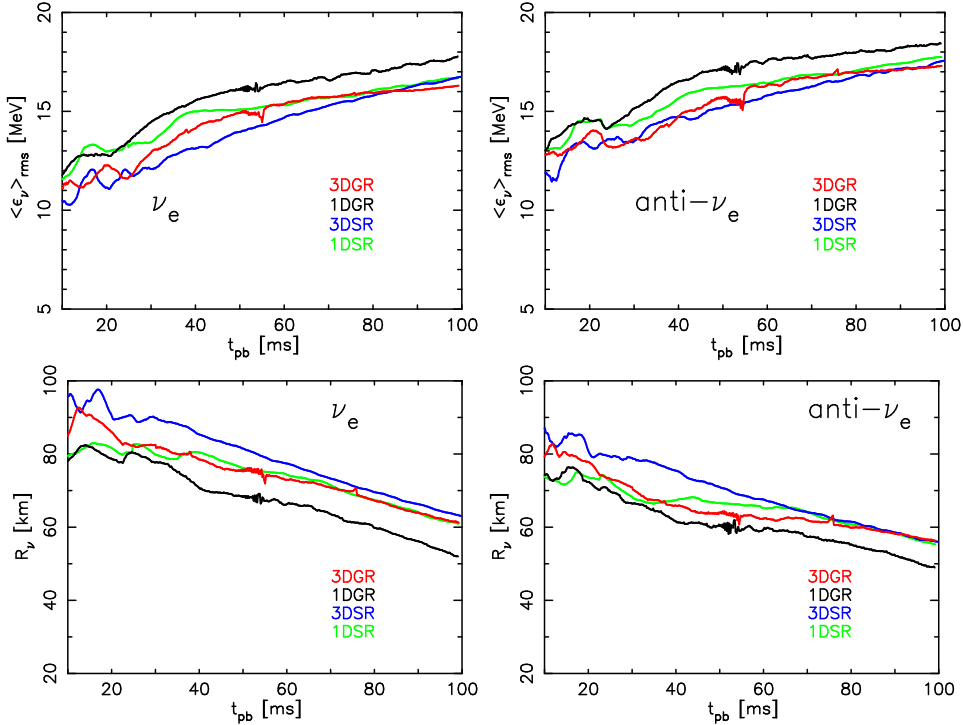

\begin{center}
\includegraphics[width=48mm,angle=-90.]{F3_E_nuE2.eps}
\includegraphics[width=48mm,angle=-90.]{F3_E_nuE3.eps}
\end{center}
\begin{center}
\includegraphics[width=48mm,angle=-90.]{F3_R_nuE2.eps}
\includegraphics[width=48mm,angle=-90.]{F3_R_nuE3.eps}
\end{center}
\caption{Evolutions of the angle averaged RMS energies 
(upper panels) and the radii of neutrino spheres (lower panels) 
for $\nu_e$ (left panels) and $\bar{\nu}_e$ (right panels). The colors of lines are the same as those in Fig.~ \ref{lnu} (figures taken from \cite{kuroda12}, reproduced by permissions of the AAS).}
\label{pic:F3}
\end{figure}

The top two panels in Fig.~\ref{pic:F3} present the angle averaged RMS neutrino energies for 
$\nu_e$ (left panel) and $\bar{\nu}_e$ (right panel) after the neutronization burst ($t
 \gtrsim 10$ms). We obtain the highest energies in model~1D-GR (black line) and the second highest in 
model~1D-SR. Then comes model~3D-GR followed by model~3D-SR. In accord with the previous 1D 
results~\cite{Lentz11,Bernhard10,matthias04,Bruenn01}, our 3D models (albeit limited to the early postbounce phase) 
support the expectation that we will obtain higher neutrino energies when switching from SR to GR.
The deeper gravitational well in GR is the reason for the higher neutrino energies. In fact, PNS becomes more compact and,
as a consequence, hotter in GR, which then leads to smaller and hotter neutrino spheres. This is evident when one compares
the radii of neutrino spheres between the GR and SR models in the bottom panels of Fig.~\ref{pic:F3}. 
Smaller neutrino energies in the 3D models compared with the corresponding 1D counter parts (top panels) 
are due to larger neutrino spheres in the former (bottom panels). In fact, the shock reaches larger radii in 3D, assisted by the 
convection and SASI (e.g., Fig.~\ref{shock}), which also helps shift the positions of neutrino spheres outwards.
The enlarged neutrino spheres in multi-D models are qualitatively consistent with the 2D post-Newtonian results by 
\citen{Buras06a}, which incorporated the advanced neutrino transport.

\begin{figure}[htpb]
\begin{center}
\includegraphics[width=65mm,angle=-90.]{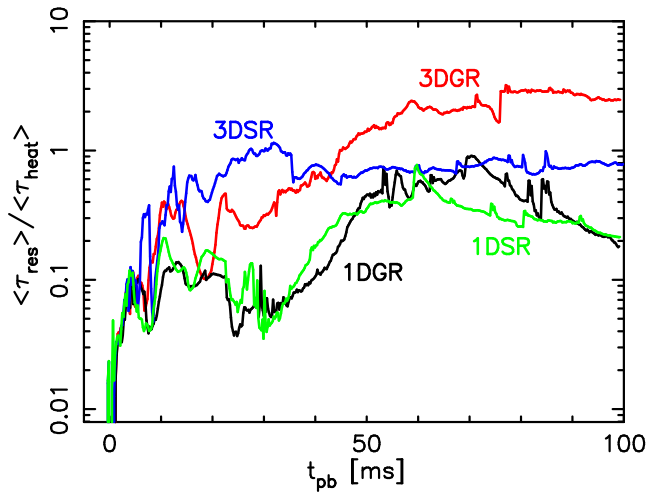}
\end{center}
  \caption{The ratio of the residency timescale to the heating timescale 
 for the set of our models as functions of post-bounce time (see text for the definition of the timescales). This figure is taken from \cite{kuroda12}, reproduced by permissions of the AAS.}
\label{pic:Res2Heat}
\end{figure}

Not all GR effects are good for the neutrino-heating mechanism. In fact, stronger gravitational pull in GR
tends to shorten the residency times of accreting matter in the gain region. In the following we discuss 
whether the net GR effect, after all these effects being taken into account, is positive or not in the multi-D context.
Figure~\ref{pic:Res2Heat} shows the ratio of the residency time scale to the neutrino-heating time scale for all the 
 computed models. Although the computed time $\sim 100$ms is way too short for the stalled shock to be revived, 
our results clearly suggest that the shock revival is most likely to occur in model 3D-GR (red line). Models~3D-SR, 
1D-SR and 1D-GR follow it in this order. Thanks to larger degrees of freedom, the residency time scale is much longer 
in the 3D models than in the 1D models. In addition, the increases in neutrino luminosity and RMS energy via the GR 
effects (Fig.~\ref{pic:F3}) raise the time scale ratio by a factor of $\lesssim$ 2 in model 3D-GR (red line) from 
the SR counter part (blue line). Our results hence suggest that the combination of 3D and GR will provide the most 
favorable condition for the neutrino-driven explosion. 

It is expected from Fig.~\ref{pic:Res2Heat} that the shock revival will never occur in the 1D models, which have 
already shown the sign of a rapid shock recession by the end of the simulations. On the other hand, the time scale ratio
remains high in the 3D models for the last $\sim 30$ms before the simulations are terminated. For the 15 $M_{\odot}$ 
progenitor employed in this study, it is expected that the neutrino-driven explosions take place at 
$t \sim 200$ms at the earliest~\cite{bruenn} and that it might be delayed to $t \gtrsim 600$ms~\cite{Marek09} as
already mentioned. The parametric explosion models showed that the earlier the shock revival occurs, the stronger the explosion
becomes~\cite{Nordhaus10,yamamoto12}). The shock revival times obtained in the previous 2D simulations~\cite{Marek09,bruenn,Suwa10,suwa12} 
could have been shorter if the combination of GR and 3D had been included. We anticipate that this can be a possible 
remedy to turn the relatively under-powered 2D explosions into more powerful ones. It is worth pointing out that 
 the combination of GR and 3D\footnote{It should be mentioned that MHD effects also remain to be studied
(e.g., \citen{kota04b,taki04,taki09,burr07,fogli_B,Kuroda10,martin11,taki_kota} and see also \citen{kota06} 
for collective references).} should affect not only the supernova dynamics,
 but also the observational multi-messenger signatures (e.g., Ref. \citen{kotake12} 
for a recent review), such as gravitational-waves 
(e.g., \citen{ewald11,kotake09a,kotake09b,kotake11gw,ottprl}), neutrino emission (e.g., \citen{icecube,marek09b,lund}),
 and nucleosynthetic yields (e.g., \citen{fujimoto,friedel}).
To give reliable predictions to these important observables, the multi-energy and multi-angle neutrino transport should be 
incorporated in full GR simulations together with more detailed weak interactions. This work is only a very first step 
on the long and winding road.


\section{Progresses in Boltzmann neutrino transport}\label{KS-Boltzmann}

\subsection{Numerical simulations with Boltzmann equations: overview}\label{KS-review}

We briefly overview here the recent progresses in the numerical treatment of 
neutrino transfer with exact Boltzmann equations in the CCSN simulations.~\cite{suz94,Sumi12}.  
Although 3D computations of hydrodynamics are now made practicable thanks to large computing resources
available these days, the neutrino transport in three spatial dimensions is still a great challenge.  
It is required to solve the time evolution of neutrino distributions in the six dimensional phase space 
with three components of neutrino momentum (an energy and two angles) in addition to three spacial 
dimensions. Even in spherical (axial) symmetry, three (five) dimensional computations are needed.  
Although various approximations have been proposed so far, 
solving exact Boltzmann equations is highly 
recommended so that we could remove uncertainties in the neutrino transfer, the key ingredient in the
neutrino-heating mechanism. Simplifications such as just dropping 
energy or angle dependence are not reliable in principle, since neutrino interactions are strongly energy-dependent and angular distributions 
are essentially important to accurately estimate neutrino-heating rates in the semi-transparent region.  

Under spherical symmetry, the direct solution of the Boltzmann equations for neutrino transfer is now 
possible~\cite{Rampp00,mez01,Rampp02,thom03} even in GR~\cite{Matthias01,matthias04,Sumiyoshi05,oconnor} 
with current computing resources. In fact, with these first-principle-based codes,  
the influences of EOS and various neutrino reactions on the supernova dynamics have been examined in detail over the 
years~\cite{lan03a,hix03,Sumiyoshi05,Buras06a}. As mentioned already, 
it has been consistently demonstrated that no explosion is obtained under spherical 
symmetry. It is well established through the improvements 
in the numerical treatment of neutrino transport during these years, however, that the accurate computation of 
neutrino transfer is indispensable to determine the luminosity and energy spectrum of 
neutrinos emitted from the PNS and the heating rates behind the stalled shock, which will in turn affect the 
shock revival~\cite{Janka96,Yamada99}. The Boltzmann neutrino transport 
is indispensable also for a reliable theoretical prediction of neutrinos signals
\cite{thom03,Sumiyoshi05,Sumiyoshi07,fischer09,fischer10}, which 
should be compared with future observations by terrestrial neutrino detectors~\cite{Totani98,and05,nak10a,keehn12}.  

With an assumption of axisymmetry, very elaborate, state-of-the-art, approximations have been developed 
in the last couple of years. The flux limited diffusion (FLD) method~\cite{liv04,wal05,Burrows06} and 
the "ray-by-ray" extension of the 1D Boltzmann transport scheme~\cite{Buras06a,Marek09} 
have been extensively employed to investigate multi-D effects on the explosion mechanism.  
Each approximation has its pros and cons: in the flux-limited diffusion approximation, for example,
the transport in the semi-transparent region is not very reliable; in the "ray-by-ray" approximation, on the other hand, 
the neutrino transfer equations are solved along each radial ray independently and, as a consequence, although the
computations are highly efficient, the forward-peaked distributions of neutrinos in the transparent region tend to be 
overestimated. Some of more recent works combine approximate 1D transport schemes such as FLD or
IDSA with the ray-by-ray technique~\cite{bruenn,idsa,Suwa10}.  
These 2D simulations have demonstrated the critical role of hydrodynamical instabilities such as convections and SASI
in the neutrino-heating mechanism. It should be also mentioned that the exact Boltzmann equations were also  
solved in 2D by \citen{ott_multi,bra11} with the discrete-ordinate method, albeit for a limited number of models.

Spatially 3D simulations of core-collapse are still in its infancy.  
In most of the earliest 3D simulations~\cite{Blondin03,Ohnishi06,blo07b,Iwakami08} 
the neutrino transfer were just neglected or simplified considerably and the authors
paid attention to novel features of 3D hydrodynamics, in particular instabilities
such as convections and SASI. In the so-called light bulb approximation, for example, 
the neutrino luminosity and energy spectrum are not solved but prescribed parametrically 
to seek favorable conditions for shock revival~\cite{Nordhaus10,rantsiou}.  
More recently, by combining the ray-by-ray approximation with FLD~\cite{bruenn} 
or IDSA~\cite{idsa,Suwa10,Takiwaki11}, 3D hydrodynamical simulations with spectral 
neutrino transport were performed~\cite{Takiwaki11}. These 3D computations heralded a new stage of 
supernova simulations. As already mentioned in section \ref{3d_1},  these models did not have sufficient resolution yet and further improvements are needed.
In this section we will make an attempt to go beyond such approximate 
neutrino transport schemes and solve the exact Boltzmann equations in 3
spatial dimensions, i.e., in 6D phase space. For the moment, relativity is 
neglected in the multi-D Boltzmann transport. 

But before going to the multi-D transport, we will first summarize briefly our
1D GR radiation-hydrodynamical core-collapse simulations with a Boltzmann solver to 
demonstrate what insights can be obtained into microphysics with these simulations. 
We then report our recent progresses in the coding of multi-D Boltzmann solver, 
presenting the results of some test calculations. 

\subsection{1D GR neutrino-radiation hydrodynamics with a Boltzmann solver}\label{KS-1DGR}

The first-principle simulations by solving the GR hydrodynamical equations 
and exact Boltzmann equations for neutrinos under spherical symmetry enable us to avoid 
uncertainties in numerics and investigate physics, in particular, the influences of microphysics 
on the dynamics in a quantitative manner (see also more recent works~\cite{sage09,Lentz11,Hempel12}).  
This is indeed a good example to see the close connections between the latest knowledge on 
nuclear/particle physics in laboratory and the understanding of astrophysical phenomena.  
We discuss here how EOS at high densities impacts the postbounce dynamics and what
information on EOS can be extracted from them in return. 
In the following, we summarize our 1D results for 15M$_{\odot}$ and 40M$_{\odot}$ 
stars, paying particular attention to 
the light curve and energy spectrum of neutrinos obtained in the long-term 
 postbounce evolution~\cite{Sumiyoshi05,sum06,Sumiyoshi07,sum08,sum09,nak10b,nak11}.

\subsubsection{Competing effects of nuclear EOS on the core dynamics}\label{KS-1DSN}

In the case of the 15M$_{\odot}$ star, we computed the evolution up to $\sim 1$ s after bounce~\cite{Sumiyoshi05} to find out 
the fate of the stalled shock and see the thermal evolution of PNS (see also the long term evolutions by \citen{fischer10}).  
We adopt two EOS's as in the previous sections, i.e., Shen's EOS~\cite{Shen98,Shen98b,Shen11} and Lattimer\& Swesty's EOS with an
incompressibility of $180$MeV~\cite{latt91}. We found that neither EOS produced explosions and that the shock radii are rather 
similar in the two cases  despite the different features of the EOS's (Shen's EOS is harder than Lattimer \& Swesty's).  
It turns out that a number of effects are counteracting each other.

\begin{figure}[htbp]
    \centering
    \includegraphics[width=.49\linewidth]{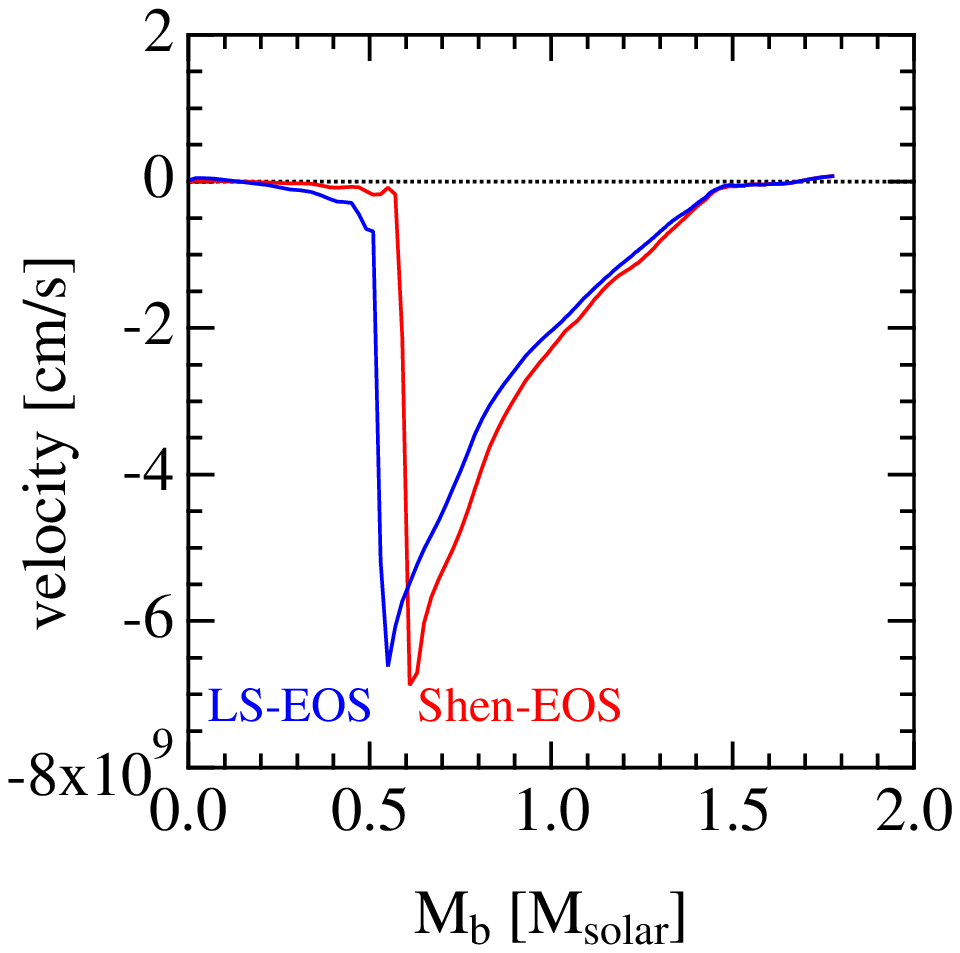}
    \includegraphics[width=.49\linewidth]{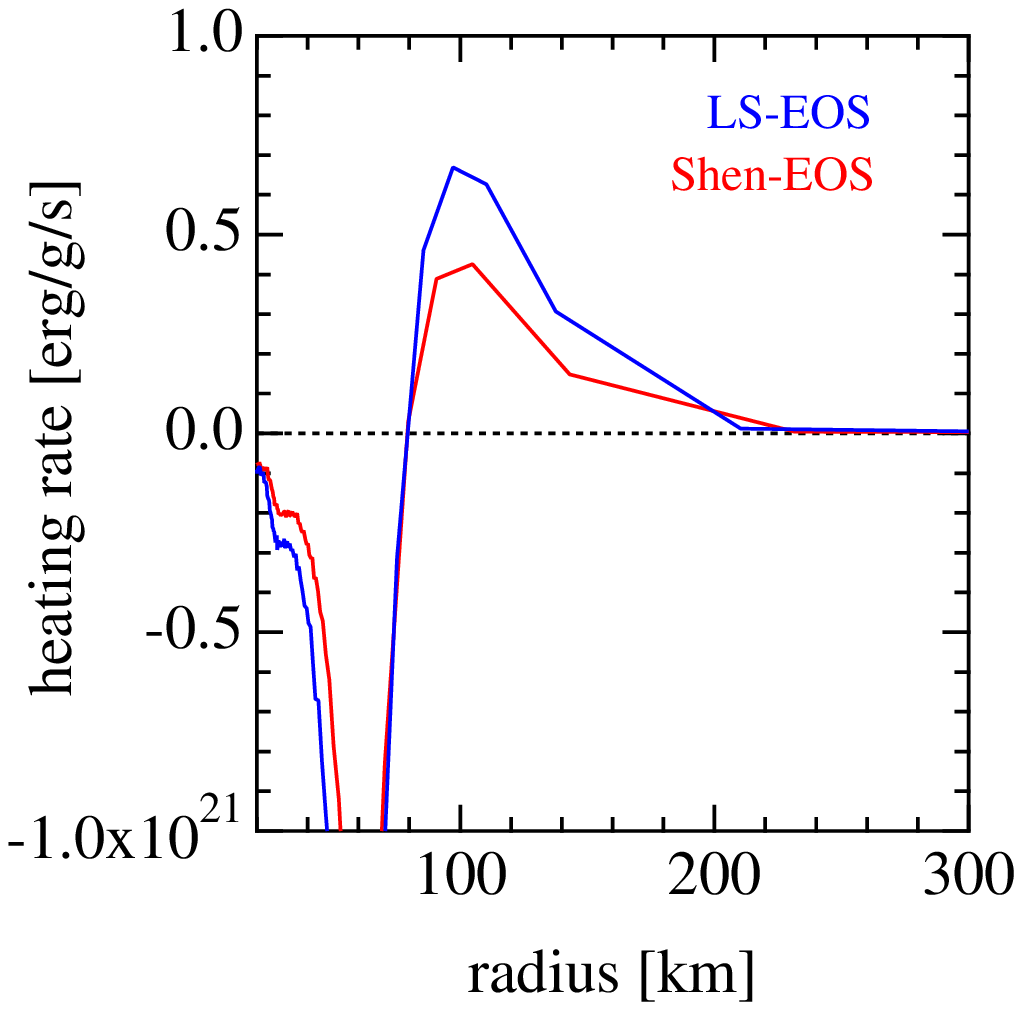}
 \caption{The velocities at core bounce (left) and heating rates at $t=150$ ms after bounce (right) in the
collapse of 15M$_{\odot}$ star. The red and blue lines show the results obtained with Shen's and Lattimer \& Swesty's EOS's, 
respectively.}
\label{fig:15M}
\end{figure}

On one hand, the larger symmetry energy of Shen's EOS leads to a smaller abundance of free protons, 
which then reduces electron captures~\cite{swe94} and make the inner core at bounce more massive, 
as can be seen in Fig.~\ref{fig:15M}. Note that a larger inner core is favorable for explosion. 
The difference in core mass amounts to $\sim 0.1$M$_{\odot}$, which can sap the shock energy of $\sim 10^{51}$erg 
by dissociation of heavy nuclei. On the other hand, stiffer Shen's EOS produces core bounce at a slightly lower density 
and, as a consequence, gives a lower core temperature than Lattimer \& Swesty's.  
This then reduces the neutrino luminosity and leads to lower heating rates behind the 
shock as demonstrated in Fig.~\ref{fig:15M}. This way the advantage earned during the collapsing phase  
is almost canceled out in the post-bounce phase. Note that the electron captures on nuclei 
may dominate over those on free protons~\cite{lan03a}. In order to address this issue appropriately, 
we need to implement a multi-nuclei EOS at sub-nuclear densities~\cite{Hempel11,Furu11,Botvina,Blin11}. Note that 
the standard EOS's (including Shen and Lattimer \& Swesty)
 employ the so-called single nucleus approximation, 
in which an ensemble of heavy nuclei is represented by a single, supposedly most abundant nucleus. Note again that the most abundant nucleus need not
have the largest electron capture rates. In fact, one has to take an ensemble average of the electron capture rate 
multiplied by the abundance of individual nucleus. It is also stressed that microphysics such as EOS and 
weak interaction rates is important also in multi-D simulations, since they set the initial shock energy and 
the emission and absorption of neutrinos just in the same way we have seen above.  




\subsubsection{Neutrino signals from failed supernovae: extraction of information on EOS}\label{KS-1DBH}

The long-term simulations of core-collapse of the 40M$_{\odot}$ star provides us with an opportunity to study 
black formations and neutrino signals from them. In fact, since the size of Fe core is much larger compared with that for
15M$_{\odot}$ star, it is expected that there is no chance of explosion and that continued mass accretion from outer envelopes
will eventually cause the second collapse to black hole.~\cite{fryer99,Maeda03,nom05}  
Indeed the mass of PNS increases rapidly and it reaches the critical mass in $\sim 1$ s and triggers the dynamical 
collapse to the black hole~\cite{matthias04,sum06,fischer09}.  
During the thermal evolution of the central object, the neutrinos 
are copiously emitted by electron and positron captures as well as thermal productions.  
The neutrino emission is terminated soon after the event horizon is formed and the emission region is swallowed into it. 
Hence the duration of the neutrino emission is essentially determined 
by the time of the second collapse of the accreting PNS into the black hole.  

\begin{figure}[htbp]
    \centering
    \includegraphics[width=.50\linewidth]{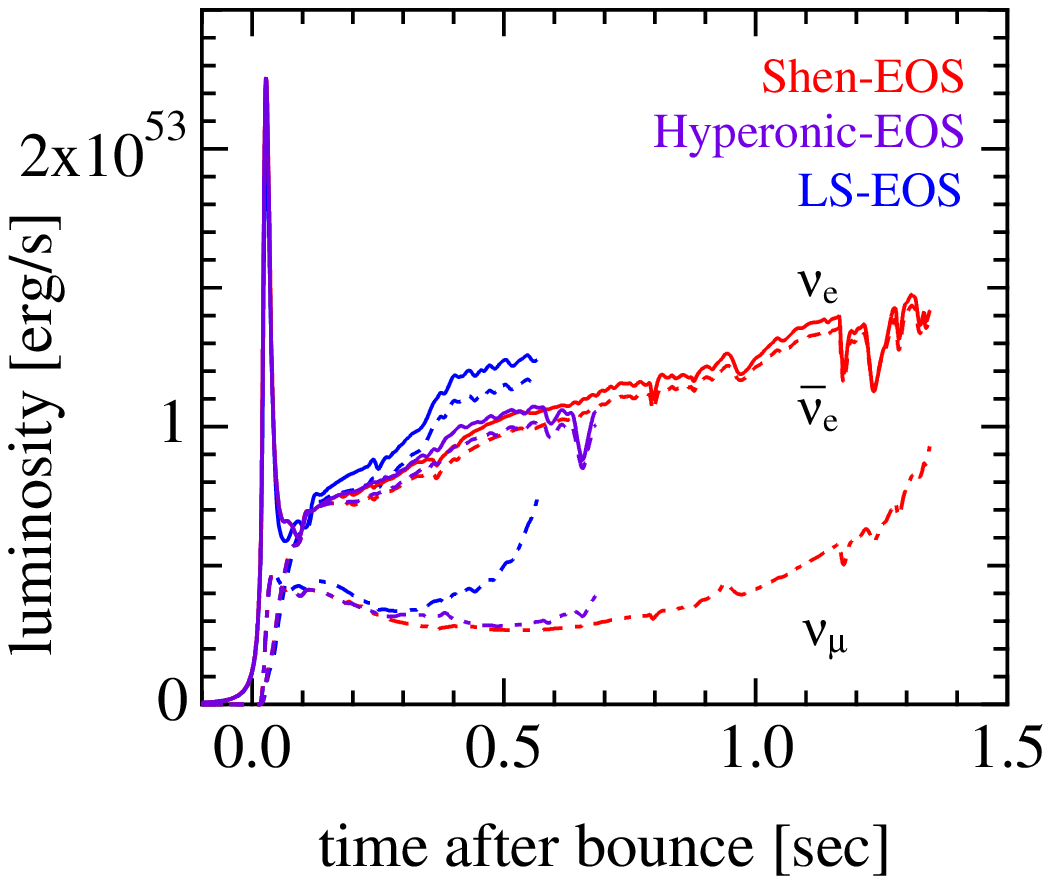}
    \includegraphics[width=.46\linewidth]{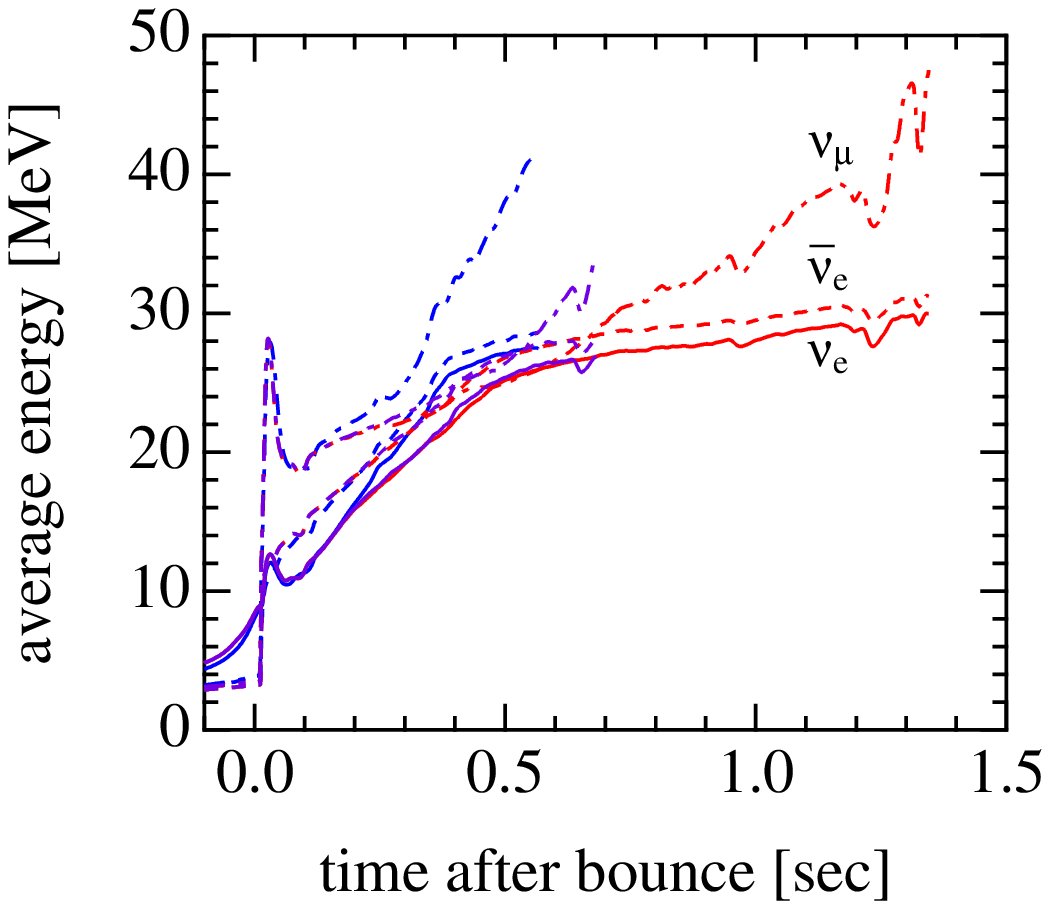}
 \caption{Time profiles of the neutrino emission in the collapse 
 of the $40$M$_{\odot}$ star. The neutrino luminosities (left) and average energies 
(right) of three neutrino species are shown for three EOS's: Shen's EOS, Lattimer \& Swesty's EOS and Hyperonic EOS.  }
\label{fig:40M}
\end{figure}

In order to specify
 common and different characters of hydrodynamics and neutrino emission in 
the black-hole forming collapse, we investigated other progenitor models in the mass range of 
$40-50$M$_{\odot}$.~\cite{sum06,Sumiyoshi07,sum08} In addition to the standard EOS's, we also 
employed hyperonic EOS~\cite{sum09,nak11} as well as quark EOS.~\cite{nak10b} 
In Fig.~\ref{fig:40M} we show the comparison of the three EOS's: Shen's EOS, Lattimer \& Swesty's EOS 
(with an incompressibility of $180$MeV) 
and hyperonic EOS. It is evident that the energies and luminosities of neutrinos 
rise rapidly due to the increase of temperatures inside the slowly contracting PNS 
and the persistent mass accretion. The duration of the neutrino emission is only 0.6-1.3 s.   
These features are different from those for the ordinary neutrino emission in supernovae, which last $\sim 20$ s 
with gradually decreasing energies and luminosities. Hence, it is possible to distinguish the black hole formations 
from the neutron star formations by the neutrino signals. In fact, taking properly into account the detector 
properties as well as neutrino oscillations, we estimate that Super-Kamiokande will record $\sim$10$^{4}$ events 
for a galactic event~\cite{nak10a,keehn12}, which are comparable to those for ordinary supernovae~\cite{nak08b}.  

As is evident in Fig.~\ref{fig:40M}, the features of neutrino emission in the 
black-hole forming collapse  
are sensitive to EOS at supernuclear densities. A softer EOS ends up with shorter neutrino emission, since the 
critical mass for the second collapse to black hole is lower.~\cite{sum06,Sumiyoshi07}. 
The reason for the softening of EOS may be due to interactions between 
baryons or to the emergence of new degrees of freedom such as hyperons~\cite{sum09,nak11}
or quarks~\cite{nak10b}. We further analyzed the neutrino signals obtained with the rather soft 
nucleonic EOS (Lattimer \& Swesty'S EOS with an incompressibility of $220$MeV) and with the Hyperonic EOS~\cite{nak10a} by a statistical method.  
We demonstrated that although the durations of neutrino emission are similar to each other, they are still distinguishable
by Super-Kamiokande if they occur in the Galaxy. These differences of signals in turn can be used as a useful probe into 
the EOS at very high densities, which may be accessible to the next-generation
 terrestrial experiments~\cite{nak10a,keehn12}. 
It should be mentioned that the features of dynamics and neutrino emission depend also on the profile of progenitors mainly 
through the accretion rate~\cite{sum08,fischer09}. Hence, it is important to perform more systematic simulations and 
construct the templates of signals that can be compared with future observations.~\cite{and05,abe11,kistler11,keehn12}  

\subsubsection{Parallel computing in 1D simulations}
It is profitable to comment on the aspect of supercomputing in the 1D studies. In our numerical simulations described above, 
the numerical inversion of block tridiagonal matrices that appear in the discretization and linearization of basic equations 
is the major computational load. We implemented the block cyclic reduction~\cite{Sumiyoshi98} as an efficient parallelled 
algorithm of matrix inversion that replaces the conventional, serial algorithm of the Feautrier method~\cite{Mihalas99}. 
The numerical technique developed for 1D spherically symmetric neutrino transfer can be easily extended to multi-D by 
the use of the approximate ray-by-ray implementation of Boltzmann solvers.  

\subsection{Multi-D neutrino transport: numerical solutions of 6D Boltzmann equations}\label{KS-3D}

Thanks to recent expansions of supercomputing resources, it has now become feasible to numerically solve the Boltzmann
equations for neutrino transfer in three spacial dimensions. We have indeed developed a Boltzmann solver with 
multi-energy and multi-angle groups that is meant for multi-D simulations~\cite{Sumi12}.  
The code is based on the so-called discrete-ordinate (S$_{n}$) method in six dimensions
 (see Ref. \citen{abdi} for an alternative approach by Monte Carlo scheme).
  
We describe the neutrino distribution in the space coordinate 
with radial $N_r$-, polar $N_{\theta}$-, and azimuthal $N_{\phi}$-grid points 
and in the neutrino momentum space with energy $N_{\varepsilon}$-grid points 
and angle $N_{\theta_{\nu}}$- and $N_{\phi_{\nu}}$-grid points.  
A fully implicit differencing is adopted for time advancement.  
We choose the inertial frame to write down the Boltzmann equations, in which the advection 
terms have the simplest expressions. Then we need to consider the Lorentz transformations to evaluate the collision terms, 
which become simplest in the comoving frame. For the moment, however, neglecting all the corrections of the order of $v/c$ or higher,
we do not distinguish these two frames\footnote{We have an idea to rigorously treat the Doppler effects and angular aberrations 
in the collisional integrations and will publish it elsewhere}.   

The basic set of neutrino reactions\cite{Bruenn85,Sumiyoshi05} including pair processes but not inelastic scatterings 
is implemented in the collision terms. Three species of neutrinos ($\nu_e$, $\bar{\nu}_e$, $\nu_{\mu/\tau}$) are treated. 
The standard EOS tables are employed to obtain the thermodynamical quantities and composition of matter, which are 
necessary to evaluate the collision terms. In the following test computations we adopt Shen's or Lattimer \& Swesty's EOS. 

\begin{figure}[htbp]
    \centering
    \includegraphics[width=.49\linewidth]{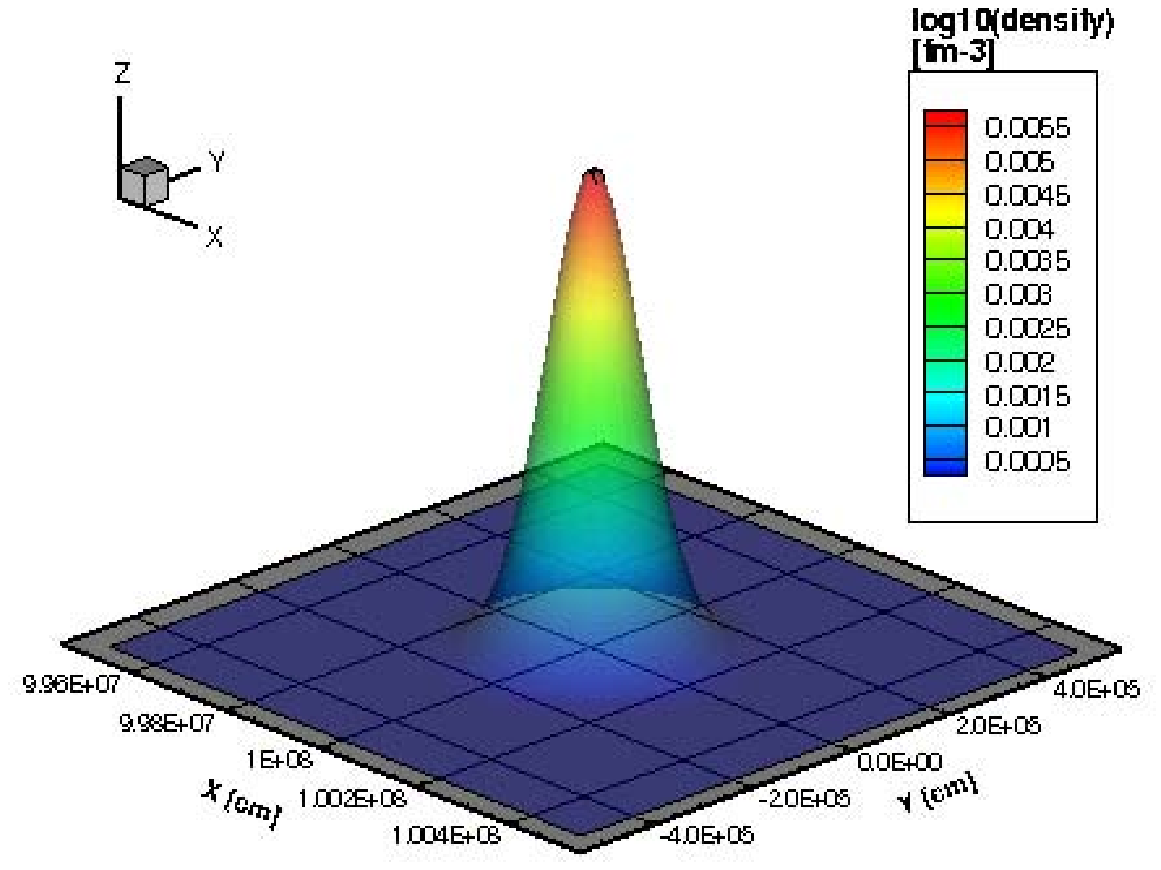}
    \includegraphics[width=.49\linewidth]{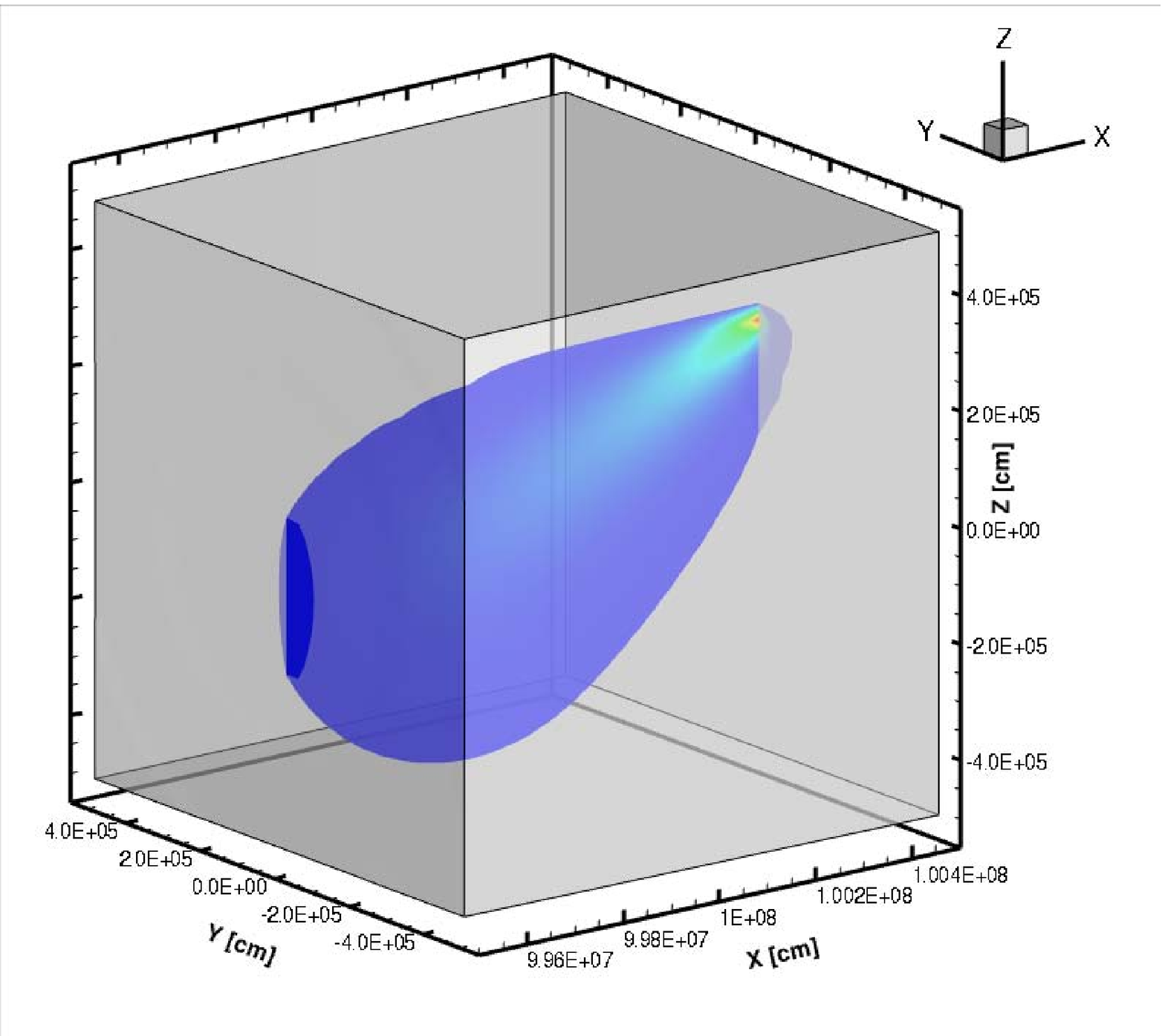}
 \caption{Examples from the test suit of the time evolution of neutrino distributions 
 by the 3D Boltzmann solver.  
 Snapshot of neutrino density from the time evolution of 
 the diffusion of 2D Gaussian packet by surface plot (left) 
 and
 the neutrino beam injected from the point source in 3D box (right).  
 The color expresses the neutrino density.   
 The number of grid points is 
$N_r \times N_{\theta} = 100 \times 96$ (left) and 
$N_r \times N_{\theta} \times N_{\phi} = 50 \times 48 \times 48$ (left) with 
$N_{\theta_{\nu}} \times N_{\phi_{\nu}} \times N_{\varepsilon} = 12 \times 12 \times 4$.  
}
\label{fig:gauss-beam}
\end{figure}

\subsubsection{Some basic tests}\label{KS-3Dmodel}

A suit of numerical tests have been done to validate the newly developed Boltzmann solver. 
We have first computed multi-D transport in uniform matter both in the diffusion and free streaming regimes.
In the left panel of Fig.~\ref{fig:gauss-beam}, the results of 2D/3D diffusion of Gaussian packets are displayed. We have
found a satisfactory agreement with the exact solutions. In the right panel of the same figure, on the hand, we have shown
the free propagation of neutrinos in a designated direction. Substantial numerical diffusion are apparent in this case. 
Note that this test is too demanding and no such situation becomes important in supernova simulations. 
The intermediate regime, the most important one, have been examined by utilizing the formal solutions 
for more realistic matter distributions as in the following. 

We prepare artificially deformed spheroidal/ellipsoidal cores based on the post-bounce profile obtained
in the 1D GR neutrino-radiation hydrodynamical core-collapse simulations of the 15M$_{\odot}$ star~\cite{Sumiyoshi05}. With the background profiles of density, proton fraction and temperature being fixed, 
we follow the time evolution from a certain initial condition for a sufficiently long period ($\sim$10 ms)
to obtain the steady neutrino distributions. In this computation we adopt the Shen's EOS table for the evaluation of 
the collision terms. The neutrino densities and fluxes are evaluated by appropriate integrations of the neutrino 
distribution functions over the momentum space. Various angle- and energy moments of the neutrino distribution
functions including the flux factor and Eddington tensors are also examined (see \citen{Sumi12} for detailed analyses).  
Moreover, we have also obtained from the collision terms the detailed information on neutrino reactions 
such as mean free paths, deleptonization rates and cooling/heating rates.  

\begin{figure}[htbp]
    \centering
    \includegraphics[width=.90\linewidth]{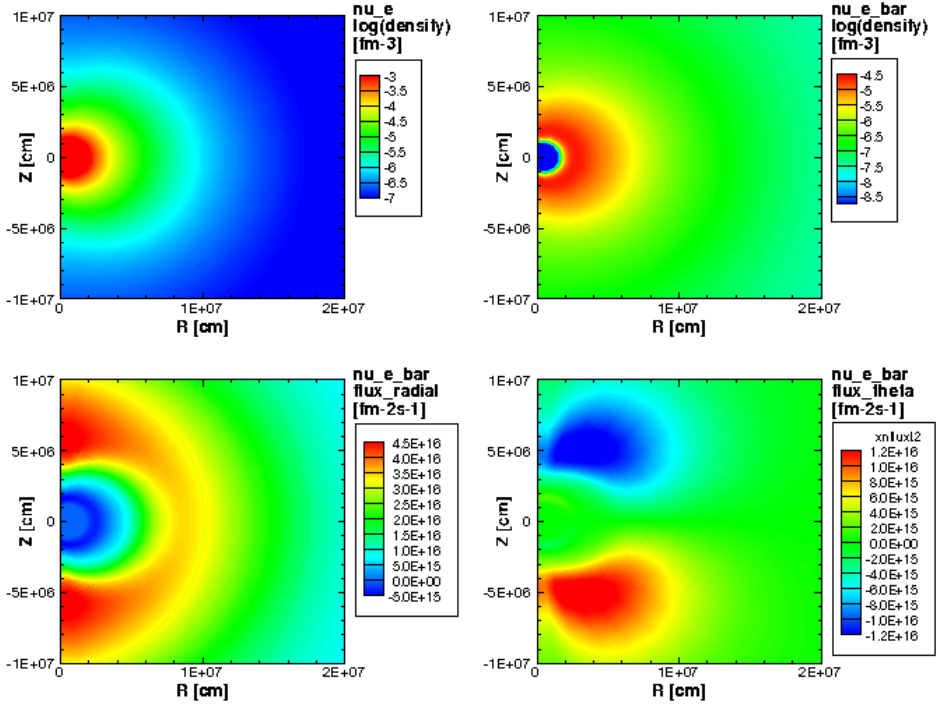}
 \caption{Color maps of neutrino densities and fluxes in the half of meridian slice for an axisymmetrically 
deformed spheroidal core. The top panels display the densities of electron-type neutrinos (left panel) and
anti-neutrinos (right panel) whereas the bottom left and right panels present the radial and polar components of 
flux vectors of electron-type anti-neutrinos, respectively.  
The number of grid points is 
$N_r \times N_{\theta} \times N_{\phi} = 200 \times 18 \times 9$ with 
$N_{\theta_{\nu}} \times N_{\phi_{\nu}} \times N_{\varepsilon} = 6 \times 12 \times 14$.  
}
\label{fig:2d-deform}
\end{figure}

In Fig.~\ref{fig:2d-deform}, we show the results for the axially symmetric, spheroidal
core. An oblate PNS sits at the center and the shock is standing around $140-200$km in this model.  
The electron-type neutrinos (top left panel) and anti-neutrinos (top right panel) are abundant 
at and off center, respectively, reflecting the degeneracy of electrons.  
It is as expected intuitively for the oblate shape of the core that the radial flux is enhanced near 
the polar axis. The polar fluxes ($\theta$-component of the flux vectors) are substantial at  
intermediate polar angles, since the neutrino fluxes are inclined towards the polar axis. 

\begin{figure}[htbp]
    \centering
    \includegraphics[width=0.9\linewidth]{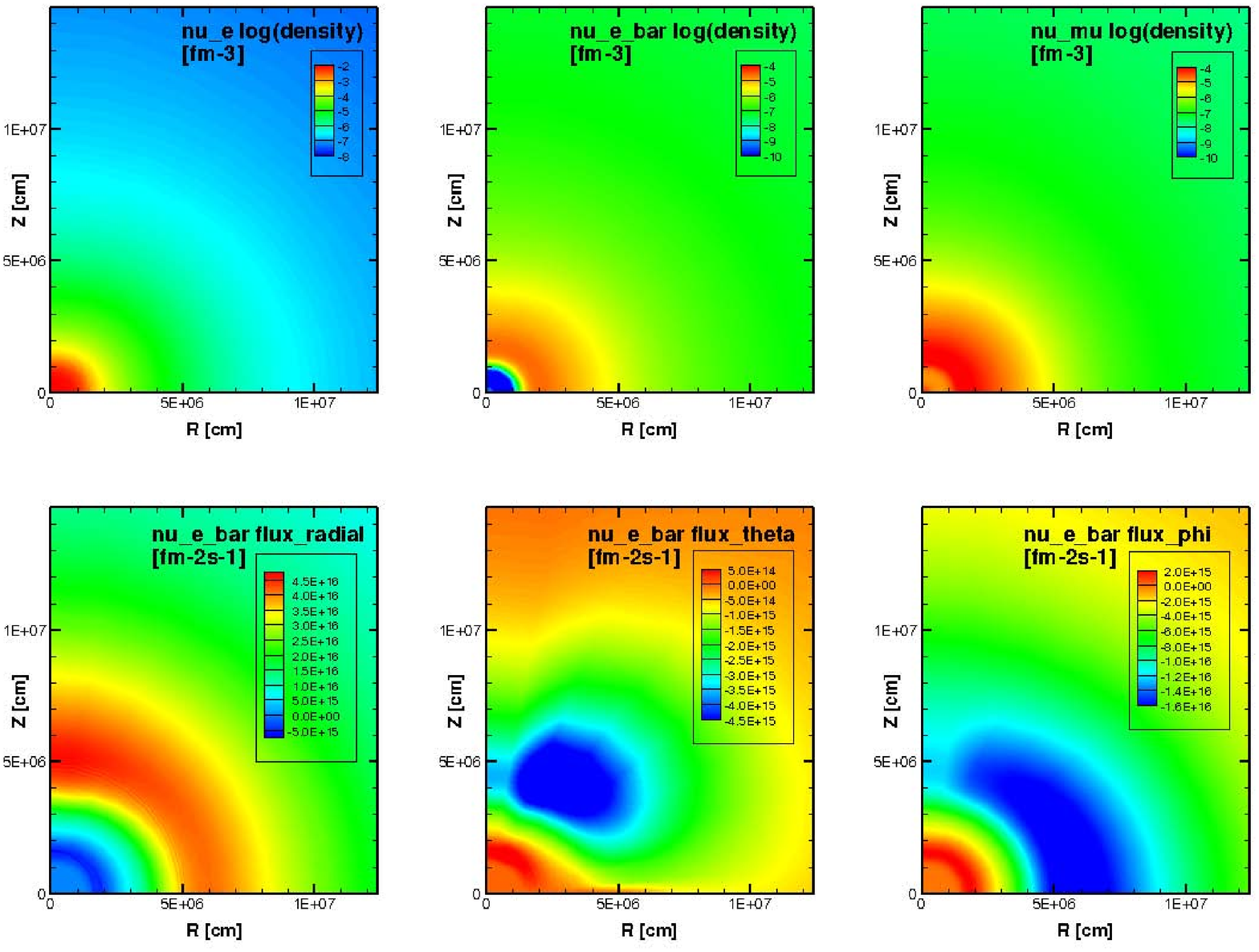}
 \caption{Color maps of neutrino densities and fluxes in the quadrant of meridian slice for a non-axisymmetrically 
deformed ellipsoidal core.  
 The densities of electron-type neutrinos (left panel), anti-neutrinos (middle panel) 
 and mu-type neutrinos (right panel) are shown on the top rows whereas the radial (left panel), polar (middle panel) 
and azimuthal (right panel) components of the flux vectors of electron-type anti-neutrinos are shown, respectively, 
on the bottom row.  
The number of grid points is 
$N_r \times N_{\theta} \times N_{\phi} = 200 \times 9 \times 9$ with 
$N_{\theta_{\nu}} \times N_{\phi_{\nu}} \times N_{\varepsilon} = 6 \times 12 \times 14$.  
}
\label{fig:3d-deform}
\end{figure}

In Fig.~\ref{fig:3d-deform}, we show the 3D case, in which the core is deformed to
a non-axisymmetric, ellipsoidal shape by adding an azimuthal ($\phi$) dependence to the 
spheroidal configuration employed above. Displayed in the figure are densities and fluxes of 
three species of neutrinos in the first octant of a meridian slice with an azimuthal angle of $\phi=0.44$ radian. 
The asymmetry between the pole and the equator in this slice is slightly smaller than that in 
Fig.~\ref{fig:2d-deform}. The density of electron-type neutrinos is high at center 
whereas electron-type anti-neutrinos and mu-type neutrinos are abundant off center, where 
the temperatures are higher than at center. This feature is similar to what we saw in the previous 
axisymmetric case. The fluxes of electron-type anti-neutrinos reflect the 3D deformation, though 
(lower panels of the figure). The radial flux around the polar axis is larger than that near 
the equatorial plane although the asymmetry is less remarkable compared with the 2D case.  
The polar fluxes are again non-negligible and largest at intermediate polar angles.  
Owing to the azimuthal dependence of deformation, the azimuthal fluxes are non-vanishing and 
substantial indeed in the broad region around $r=50-100$km. 
These non-radial fluxes are important to accurately describe the global behavior of neutrino transfer 
in 3D and may affect the neutrino heating rates and, as a consequence, explosions.  
We remark that these non-radial fluxes can be automatically and properly treated by the 
3D Boltzmann solver unlike the ray-by-ray approximation, in which the neutrino fluxes are 
assumed to be radial. It is also pointed out that the non-radial fluxes are non-negligible up to $\sim100$km
and it is dubious that FLD can give the flux vectors correctly there (see also 
Ref. \citen{ott_multi}). 

\subsubsection{Several demonstrations for more realistic backgrounds}\label{KS-3Dtaki}

\begin{figure}[htbp]
    \centering
    \includegraphics[width=0.49\linewidth]{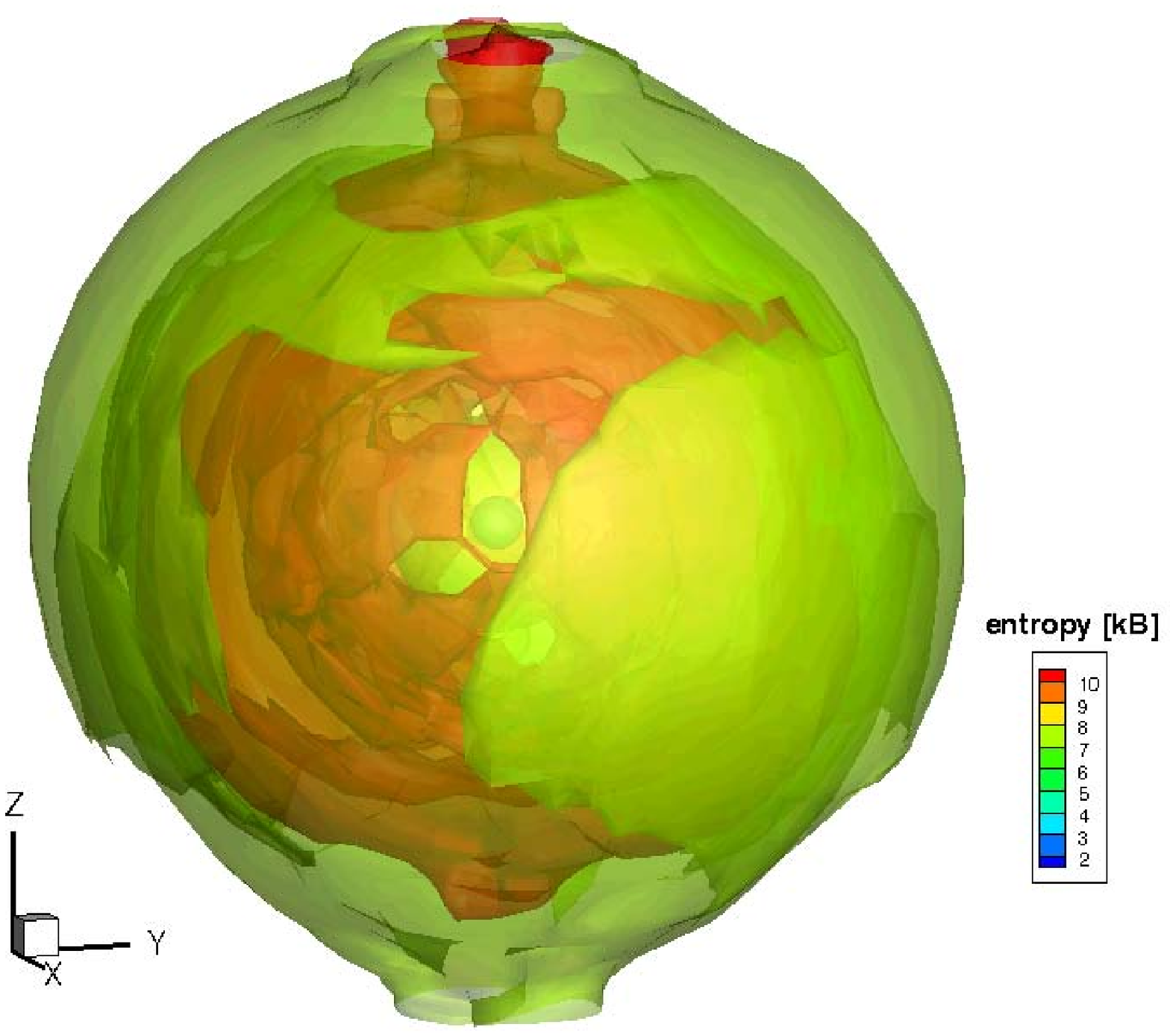}
    \includegraphics[width=0.49\linewidth]{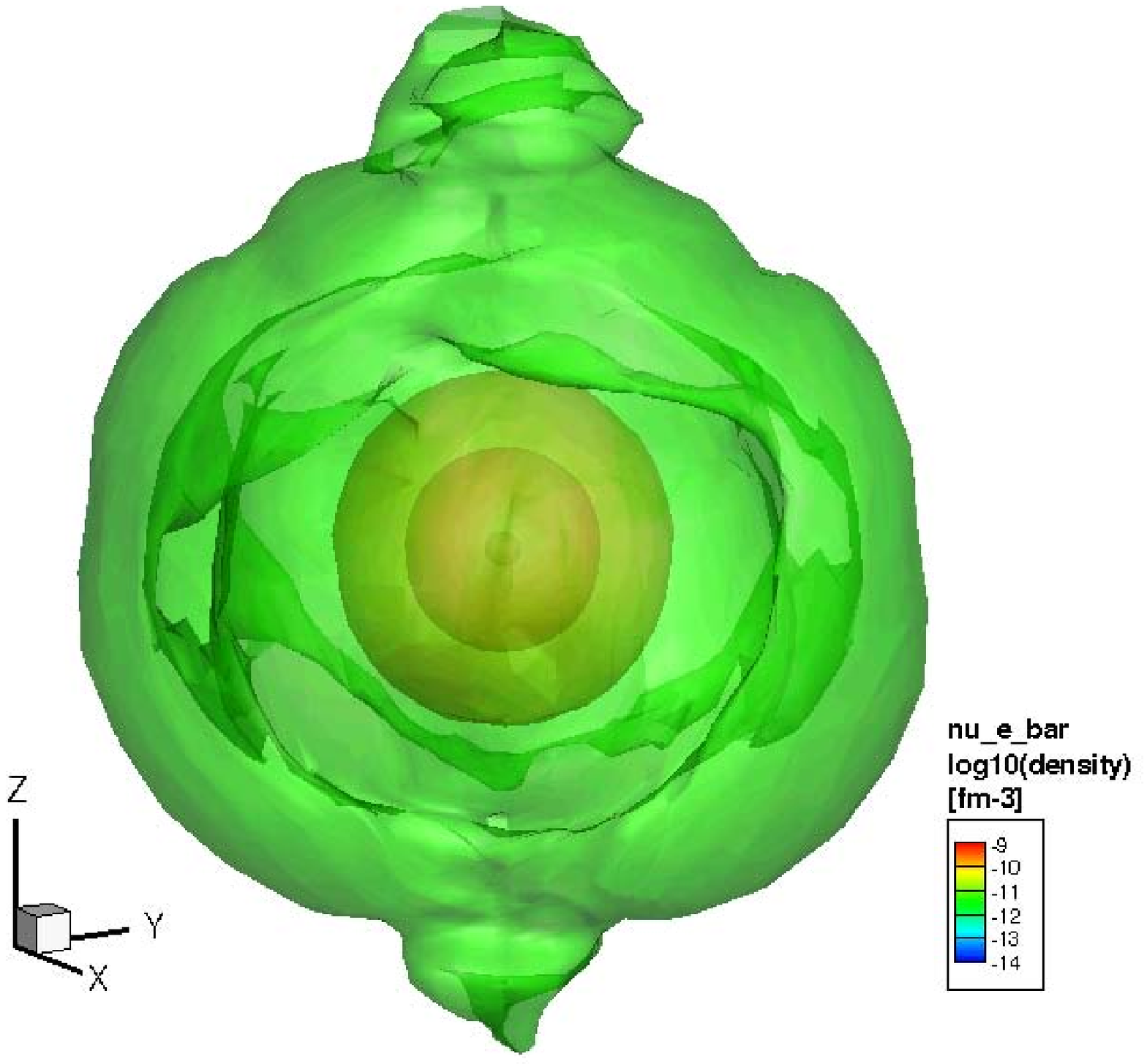}
 \caption{Iso-entropy surfaces at $100$ ms after bounce in the collapse of the 11.2M$_{\odot}$ star (left panel)  
 and surfaces of a constant density of electron-type anti-neutrinos in the evolution by the Boltzmann 
equations (right).  
The number of grid points is 
$N_r \times N_{\theta} \times N_{\phi} = 128 \times 16 \times 32$ with 
$N_{\theta_{\nu}} \times N_{\phi_{\nu}} \times N_{\varepsilon} = 6 \times 12 \times 14$.  
}
\label{fig:3d-takiwaki}
\end{figure}

We proceed to some more test computations done under more realistic settings. 
We employ the snapshot at $100$ ms after bounce that is obtained in the simulations of the $11.2$M$_{\odot}$ 
star~\cite{Takiwaki11} discussed in \S \ref{3d}. In the left panel of Fig.~\ref{fig:3d-takiwaki}, 
we show the 3D entropy distribution in the supernova core. It is clear that the matter distribution is deformed 
both globally and locally due to the hydrodynamical instabilities below the shock (shown
 as a greenish sphere), 
which is located around 200$-$300 km. Adopting Lattimer \& Swesty's EOS for this test and fixing the background, 
we follow the time evolutions of neutrino distribution functions in the 6D phase space from an almost vanishing 
population until time-independent solutions are obtained.
The Boltzmann solver treats the building up of equilibrium distributions in the optically thick region,  
neutrino cooling and heating in the intermediate region and outward free streaming in the optically thin 
outer layers simultaneously. A snapshot of some surfaces with a constant density of anti-neutrinos 
is shown in the right panel of Fig.~\ref{fig:3d-takiwaki}. It is seen that the neutrino distribution is 
rather spherical near the center and becomes asymmetric globally and locally in the outer layers, tracing 
the matter distributions (compare with the left panel).
  
\begin{figure}[htbp]
    \centering
    \includegraphics[width=.47\linewidth]{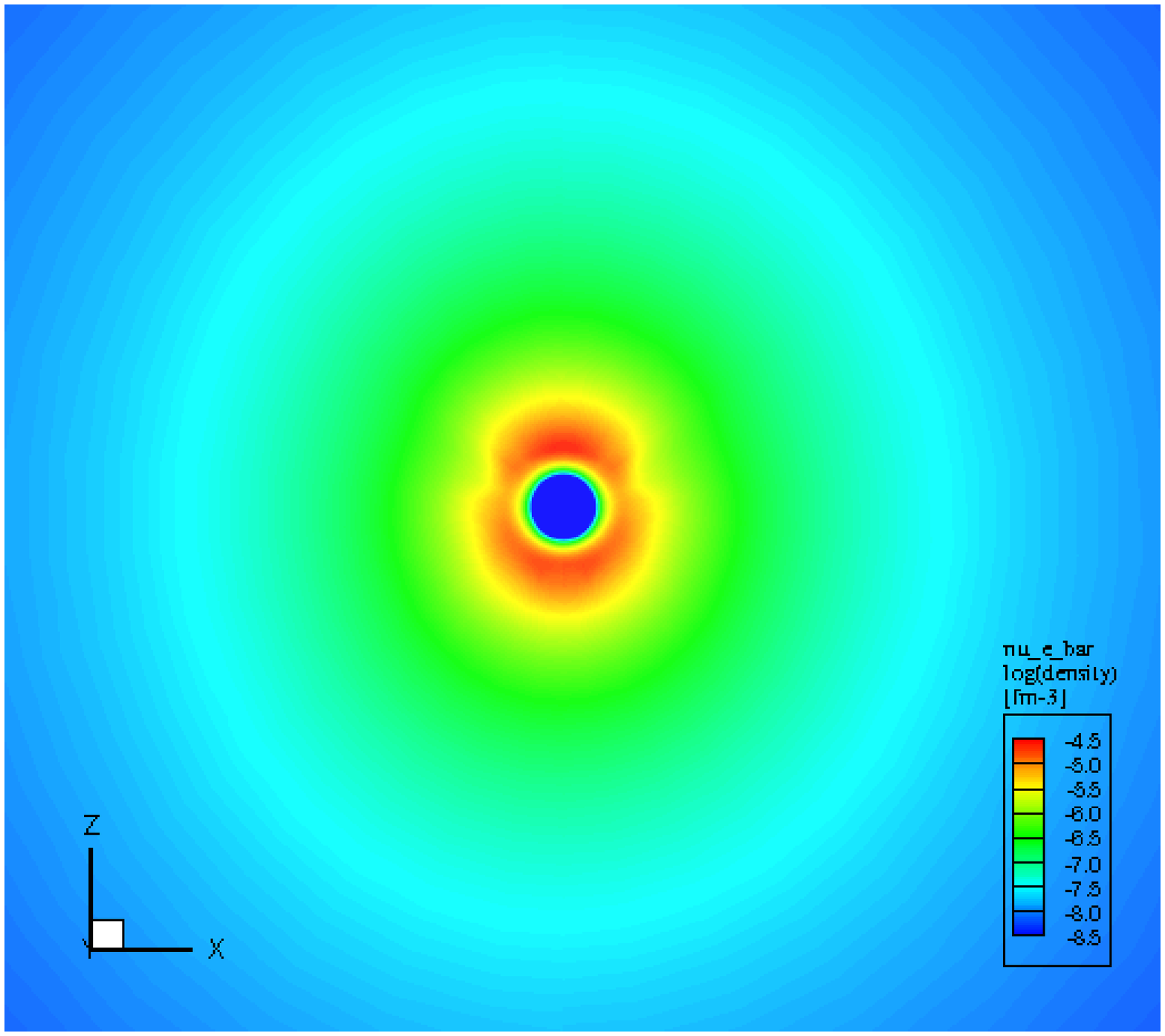}
    \includegraphics[width=.47\linewidth]{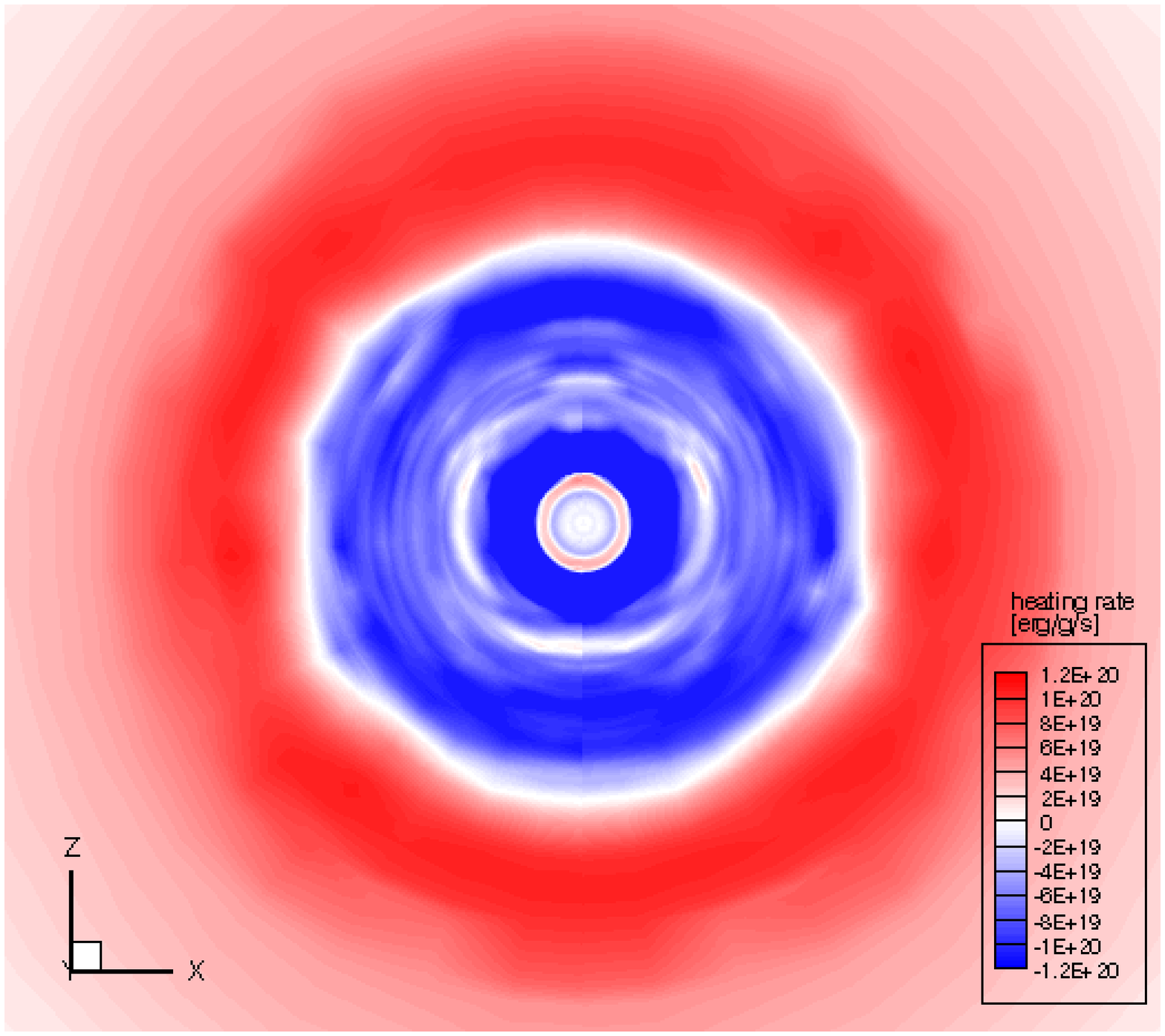}
 \caption{The density of electron-type anti-neutrinos (left) and net heating rate (right) at $100$ ms after bounce 
on a slightly off-center slice. The background matter distribution is obtained from the 3D simulation of $11.2$M$_{\odot}$
progenitor and fixed in the computations of neutrino transport.  }
\label{fig:3D-takiwaki2}
\end{figure}

Fig.~\ref{fig:3D-takiwaki2} shows the computed density of electron-type anti-neutrinos (left panel) and net heating 
rates (right panel) on a slightly off center slice, respectively. The electron-type anti-neutrinos are abundant again in the off center 
region, where the temperature is high and all flavors of neutrinos are produced thermally.  
The iso-density surfaces are prolate, reflecting the matter distributions.   
The gain radius is located around $r\sim100$ km and the global asymmetry of heating/cooling regions is moderate. 
The heating/cooling rates are calculated directly from the integrations of the collision terms.  
 For implementing the radiation module in a hydrodynamic code, these quantities describing the changes in energy and compositions of 
matter enter in the right-hand-side of the hydrodynamic equations (with negative sign).
 This is what we are currently undertaking (Nagakura et al. in preparation 
~\cite{nagakura}). And then, the next important task is 
   to implement the velocity dependent terms in the transport 
equations, which is indispensable for the accurate treatment of the relativistic effects.
 After that, we plan to study hydrodynamical instabilities in the 
supernova core,
 then move on to perform a full-scale simulation starting 
 from the onset of gravitational collapse, through core bounce to shock-stall,
 until shock revival and explosion in a consistent manner.
In doing so, highly competitive supercomputing resources in Japan, in particular 
the ``K computer''\footnote{it is named after the Japanese word of "Kei", meaning 
 10 quadrillion (10 petaflops). Note that the given name of the corresponding
 author of this paper has nothing to do with the supercomputer (except for 
 the fact that the person is going to use it for SN simulations).}, the fastest 
one in the world as of November 2011, will be helpful.   

\subsubsection{Neutrino transport in highly non-spherical environments: collapsars}\label{KS-3Dseki}

As an additional demonstration of the capability of our new 3D code, 
we take the collapsar model for gamma ray bursts\cite{macfadyen99,Woosley2011}, 
which are more energetic and asymmetric than CCSNe.  
Gravitational collapse of very rapidly rotating, massive cores results in 
the formation of black hole with a surrounding disk, which is presumably 
responsible for jet formations. To pin down the physical processes to form the 
relativistic jets in the collapsar model is a long-standing issue. 
Annihilations of neutrino pairs emitted from the disks are one of the plausible 
mechanisms~ (e.g., \cite{paz90,mezree,Harikae2010,Zalamea2011} and references therein).
 Mass ejections from the disk and/or jet through neutrino interactions are attracting
 broad attention in the studies of nucleosynthesis of heavy elements 
\cite{fuji1,ono1,ono2,basel}. 
Neutrino transport in the collapsar model is hence an indispensable ingredient 
for the investigations both of jet formations and nucleosynthesis.
Since the black hole and disk system is highly non-spherical, numerical approaches 
are almost mandatory. Although extensive studies mainly by 
utilizing a ray-tracing technique have been reported so far to this end 
\cite{ruffert,ruff98,birkl,harikae,kotake12_2}, the spectral treatment of neutrino transfer has been 
a major undertaking.
 As a very first step toward better description of neutrino transport 
in collapsar, we employ our 3D code here to obtain time-independent 
neutrino distributions (see also Ref. \citen{dess09})
 for a matter configuration extracted from a 
hydrodynamical simulation of collapsar.

\begin{figure}[htbp]
    \centering
    \includegraphics[width=0.457\linewidth]{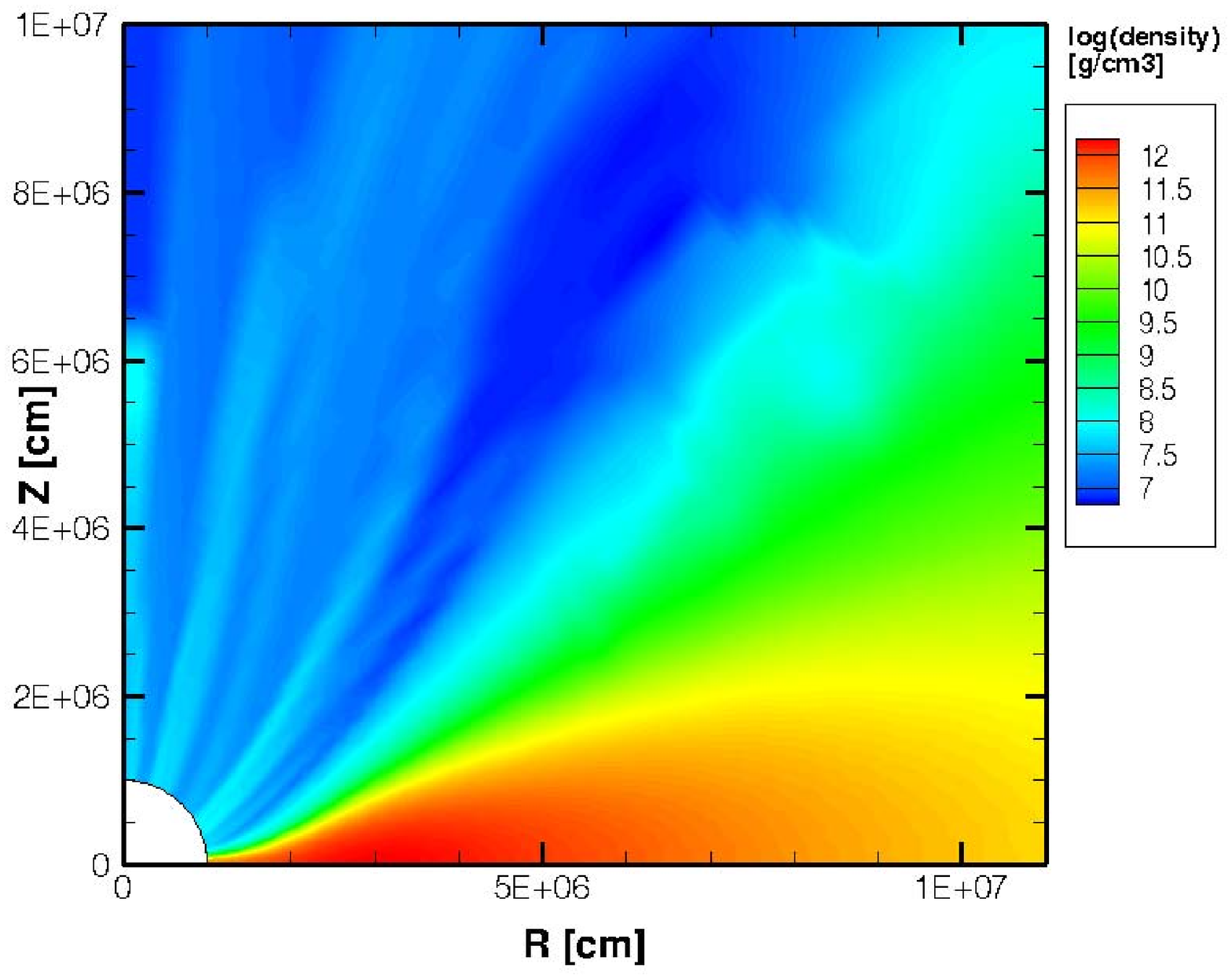}
    \includegraphics[width=0.5\linewidth]{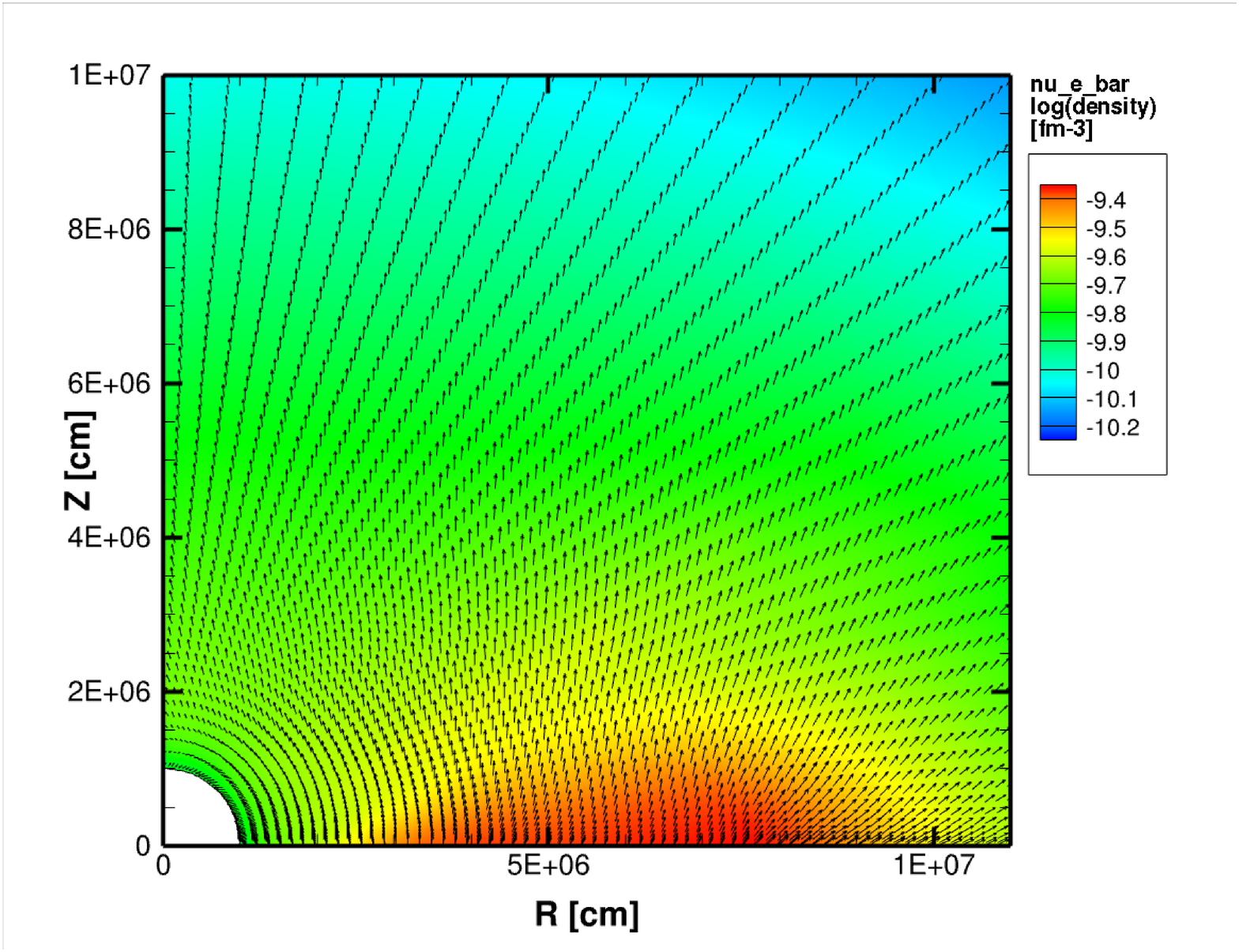}
 \caption{The density profile in the quadrant of a meridian section used for the computation of neutrino transport with
the Boltzmann solver (left) and the resultant steady distributions of electron-type anti-neutrinos (right). 
The neutrino densities are shown by a color map in log scale and fluxes are presented by arrows in the right panel.
The number of grid points is 
$N_r \times N_{\theta} \times N_{\phi} = 100 \times 45 \times 3$ with 
$N_{\theta_{\nu}} \times N_{\phi_{\nu}} \times N_{\varepsilon} = 6 \times 12 \times 14$.  
}
\label{fig:3d-sekiguchi}
\end{figure}

Distributions of density, temperature and electron fraction are provided by 2D GR simulations 
of gravitational collapse of a $100$M$_{\odot}$ star with a rapid rotation~\cite{Shibata11,Sekiguchi11}.  
The density profile after black hole formation is shown in the left panel of Fig.~\ref{fig:3d-sekiguchi}.  
The black hole sits at the center, which is removed in the neutrino calculation and shown as a white circle in the figure, and a dense disk
surrounds it. Not shown explicitly in the figure, there are accretions and outflows outside the disk.  
Fixing the matter distribution, we compute the time evolutions of neutrino distribution functions until the steady 
state is reached. We adopt the Shen's EOS table for this simulation.
  
The density of electron-type anti-neutrinos is shown in the right panel of Fig.~\ref{fig:3d-sekiguchi}.  
It is found that the electron-type anti-neutrinos are abundant in the outer part of the disk, where the temperatures 
are high, whereas the density of electron-type neutrinos (not shown in the figure) is high near the the black hole.    
The neutrino fluxes (shown as arrows in the figure) reflect the geometry of the system, very roughly agreeing with 
the local gradient of neutrino density, and completely non-radial. This is a situation that the ray-by-ray approach 
is certainly inappropriate. Detailed information such as the location of neutrino spheres and energy spectra of all
species of neutrinos is essential to investigate the dynamics of outflows and nucleosynthesis inside them.  
We will address these issues in the near future with the new Boltzmann solver, which will be combined with a 3D GR hydrodynamics code described in section \ref{3d_2}. 

\subsubsection{Progresses in the new algorithms for large-matrix inversion}

It is worth mentioning that the exact 3D Boltzmann solver demonstrated above 
is a product of our collaboration with computational scientists, who know how to make 
best use of computing resources. 
They are specialists indeed in the mathematical modeling, algorithms and parallel computing. In this section, we briefly 
describe our recent progresses in this aspect of the development of the 3D Boltzmann solver.  

\begin{figure}[htbp]
    \centering
    \includegraphics[width=0.4\linewidth]{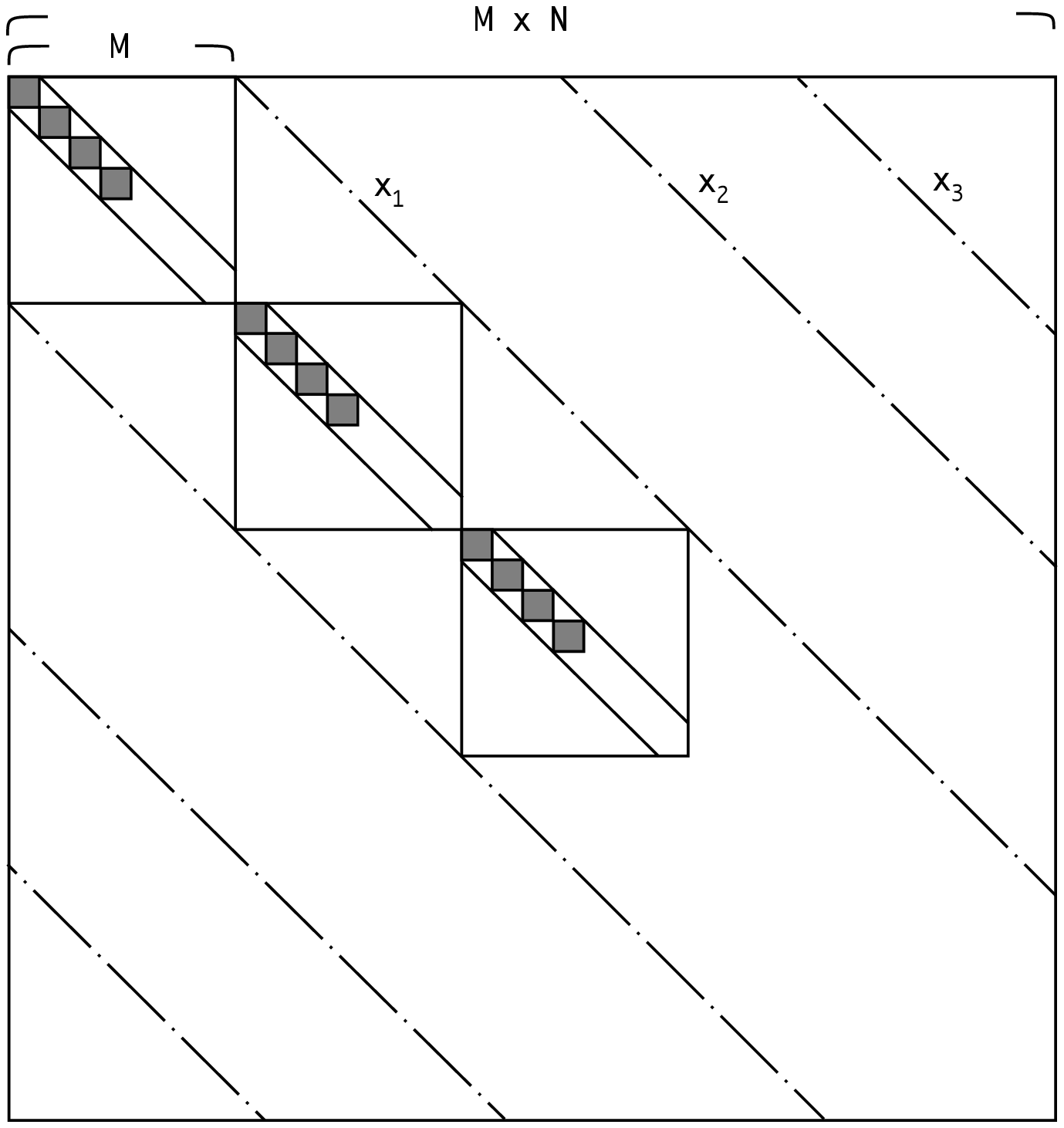}
    \includegraphics[width=0.5\linewidth]{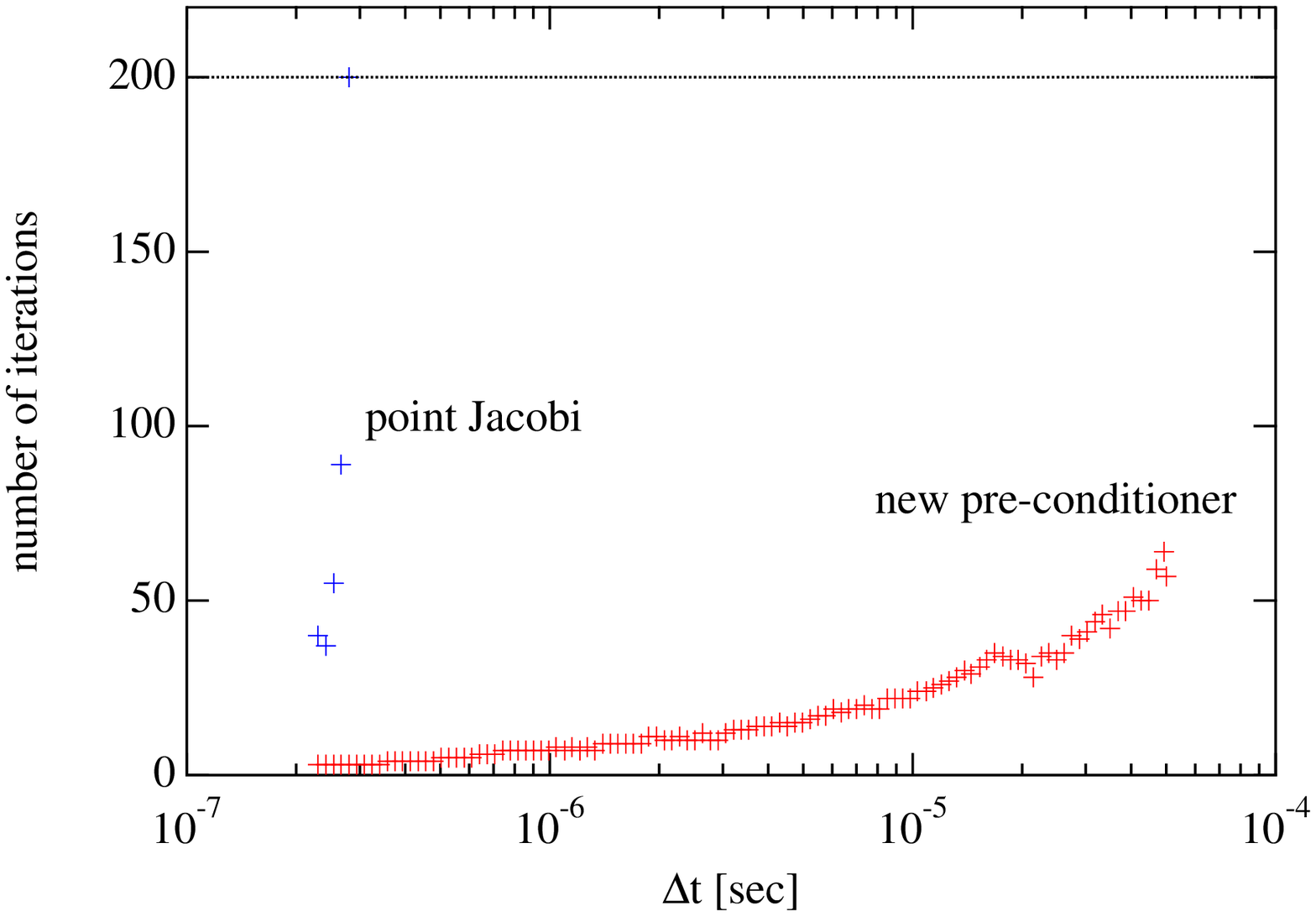}
 \caption{Left: pattern of the sparse matrix appearing in the linear system obtained for the implicit discretization of 
the Boltzmann equations.  
N and M denote the numbers of spatial grids (N$_{r}$ N$_{\theta}$ N$_{\phi}$)
and neutrino grids
(N$_{\theta_{\nu}}$N$_{\phi_{\nu}}$N$_{\varepsilon}$), respectively.
For the studies on the current supercomputers without energy couplings,
the size of diagonal black matrices (gray) is
N$_{\theta_{\nu}}$N$_{\phi_{\nu}}$.  
 Right: number of iterations as a function of the time step for different pre-conditioners, i.e., the point 
Jacobi method (blue crosses) and newly developed method (red crosses).  
The number of grid points for the numerical experiment is 
$N_r \times N_{\theta} \times N_{\phi} = 200 \times 9 \times 9$ with 
$N_{\theta_{\nu}} \times N_{\phi_{\nu}} \times N_{\varepsilon} = 6 \times 12 \times 14$.  
}
\label{fig:matrix}
\end{figure}

As mentioned earlier, our scheme is based on the finite differencing of the Boltzmann equations ($S_n$ method), which is implicit in 
time. The resulting equations are a linear system with a large sparse matrix. The main computational load
comes from the inversion of this matrix, which has to be done at every time step. In the left panel of Fig.~\ref{fig:matrix}, 
we show the positions of non-zero elements in the matrix we obtain from the discretization of the original equations.  
It is found that the matrix consists of dense sub-matrices along the diagonal line (shown as gray boxes in the figure) as well as 
non-zero elements on six off-diagonal lines (labeled as $x_1$, $x_2$ and $x_3$ in the figure). The sub-matrices originate 
from the collision terms that express local emission and absorption
of neutrinos as well as coupling by neutrino scattering moving in different directions, whereas the off-diagonal lines
of non-zero elements corresponds to the spatial advection terms. The total number of grid points in the 6D phase space amounts 
to $\sim 10^{9}$ in a typical simulation. The size of dense sub-matrix is $\sim 100$ in the current studies 
and will be larger for higher angular resolutions and/or full energy-couplings.   

For the inversion of matrices of this size, iterative methods~\cite{saad} are the first choice.  
We employ as a standard option the Bi-CGSTAB method, utilizing a program in the Templates library~\cite{barrett}, 
together with the point-Jacobi method as a pre-conditioner.  
We obtain convergence typically within 20 iterations with a residual error of $10^{-8}$. This is of course a function
of the time step, $\Delta t$. As we increase it, convergence becomes slower because the diagonal elements become less
dominant. As a matter of fact, sometimes no convergence is obtained even after 200 iterations in the simulations for realistic matter 
distributions extracted from dynamical models in \S \ref{KS-3Dtaki}. We certainly need to find a better way to improve convergence.  

Recently we have found a new method to optimize the pre-conditioning. 
Taking out a set of the matrices from our simulations of 3D neutrino transfer, 
which presents the slow-convergence problem in 
the standard approach with the point-Jacobi method, their properties have been analyzed in detail to propose a 
parameter-optimized damped Jacobi-type pre-conditioner, details of which will be published elsewhere~\cite{imakura}.
The convergence efficiency is compared between the two pre-conditioners for the same matrices extracted from the 
3D Boltzmann simulations. In the right panel of Fig.~\ref{fig:matrix}, we show the numbers of iteration as a function of 
the time step, $\Delta t$, for a representative case. As mentioned 
already, the convergence becomes very slow for 
$\Delta t \gtrsim 10^{-7}$s and no convergence is obtained for $\Delta t > 3 \times 10^{-7}$s even after 200 iterations
  when the point-Jacobi method is employed.  
On the contrary, with the new pre-conditioning method, the convergence is improved drastically. The time steps can be 
increased by a factor of 100 up to $\gtrsim 10^{-5}$s. This is favorable particularly for long-term computations.  
It is true that the computational cost of the new method is higher but it is just by a factor of $\sim 10$ compared with the
standard method. Our efforts have hence paid off and we have achieved a speed-up by a factor of $\sim 10$.  
We are currently applying the new method to various cases to see if such a good performance is retained or not. 
We will also continue to seek for even better methods, since we expect that the matrix will be larger in 
the productive runs of neutrino-radiation hydrodynamical simulations in 3D.  

\section{Beyond the ``K computer''}

Rapid growth of the supercomputing capability in Japan  these years
enables us to perform large-scale simulations such as those presented above.  
3D supernova simulations with sufficient resolutions definitely require the K computer and more even beyond Exa-flops 
scale platforms. We remark also that allocations of sufficiently long cpu-time on such facilities are also indispensable for 
long-term computations such as those of delayed neutrino-driven explosions.   

As reported in \S\ref{3d} and \S\ref{KS-3Dtaki}, 
the first 3D simulations of core-collapse supernovae with spectral neutrino transport
by the ray-by-ray IDSA were performed on the currently available supercomputers. It was demonstrated that the numerical grid 
deployed in the computations was not fine enough to draw a solid conclusion on the 3D explosion mechanism.  
Scaled-up simulations are scheduled on the K computer in Kobe, Japan. 
The 3D neutrino transfer with the Boltzmann solver requires even larger memory and speed specifications.  
Summarized in Table \ref{table:spec} are the target sizes of numerical grids we deploy for the 3D Boltzmann transport 
as well as the required memory sizes and expected operation numbers 
 per time-step on both the current and future platforms. 
Note that we have to follow more than 10$^{5}$ time-steps for a productive run.  

\begin{table}[htdp]
\caption{Target grid sizes and required computational resources for the current and future supercomputers.  
The currently available (typical) supercomputers, the K computer, Exa-scale supercomputers and beyond are
listed. The numbers of radial, polar and azimuthal grid points in the space (N$_{r}$, N$_{\theta}$, N$_{\phi}$) and 
those of angles and energy grid points in the momentum space (N$_{\theta_{\nu}}$, N$_{\phi_{\nu}}$, N$_{\varepsilon}$) are 
given together with the required memory sizes for the storage of the distribution functions of 3 species of neutrinos 
(4th column denoted as $f_{\nu}$'s) and the matrix in the system of linear equations (5th column denoted as Matrix) 
as well as the floating-point operations (6th column denoted as Operations). The last row refers to the case, in which 
energy couplings by inelastic scatterings are fully taken into account.
}
\label{table:spec}
\begin{center}
\begin{tabular}{cccrrr} \hline \hline
Platforms & Space (N$_{r}$N$_{\theta}$N$_{\phi}$) & Neutrino (N$_{\theta_{\nu}}$N$_{\phi_{\nu}}$N$_{\varepsilon}$)  & $f_{\nu}$'s & Matrix & Operations \\ \hline

Current            & $256 \times  32 \times  64$ & $ 8 \times 12  \times 14$  &  $2 \times 10^{1}$ GB   &    2                TB  &   $6 \times 10^{12}$ \\
K computer         & $512 \times  64 \times 128$ & $12 \times 24  \times 20$  &  $6 \times 10^{2}$ GB   &   $2 \times 10^{2}$ TB  &   $2 \times 10^{15}$ \\
Exa-scale          & $512 \times 128 \times 256$ & $24 \times 24  \times 24$  &   6                TB   &    3                PB  &   $8 \times 10^{16}$ \\
E$_{\nu}$-coupling & $512 \times 128 \times 256$ & $24 \times 24  \times 24$  &   6                TB   &   $8 \times 10^{1}$ PB  &   $4 \times 10^{19}$ \\

\hline \hline

\end{tabular}
\end{center}
\label{default}
\end{table}%

We assume in the table that the computational load mainly comes from the inversion of the sub-matrices that account for 
the local emission and absorption as well as scattering of neutrinos; the operation numbers for the currently adopted 
inversion scheme, i.e., the Bi-CGSTAB method with the preconditioner discussed above, are proportional to 
$N_{\varepsilon} N_{angle}^{3}$, in which $N_{\varepsilon}$ and $N_{angle}$ are the number of energy and angle grid points 
for neutrino transport, respectively; the latter gives the size of the dense sub-matrices on the diagonal line, since 
the scatterings couple different angular grid points. It is found that on the currently available supercomputers, 
we can afford only moderate resolutions for the 3D transfer, limited by the necessary memory size, which amounts to 
2TB for the storage of the sparse matrices in the linear system (plus 20GB for the distribution functions of three species
of neutrinos).  The number of floating-point operations per time-step is estimated to be $6\times10^{12}$ for a single species of neutrinos.  

With the K computer, which is now in operation, we can deploy twice finer spatial, energy and angle grids;  
the memory size is $2\times10^{2}$TB for the storage of matrices and $6\times10^{2}$GB for the neutrino distribution
functions; the floating-point operations per time-step becomes 2$\times$10$^{15}$, which certainly requires 10Pflops-class 
supercomputers. Even with the K computer, though, it will be hard to follow the evolution of supernova cores  over 1 s in 3D 
and such long-term simulations may be limited to axisymmetric 2D models. Nevertheless, the K computer makes it possible for 
us to accomplish systematic, high-resolution 2D Boltzmann simulations, which are still important on their own right.  

To perform long-term 3D simulations by the 3D Boltzmann solver, on the other hand, we need supercomputers of Exaflops scale, which
are expected to come as a next-generation platform. As shown in the table, the 3D neutrino radiation hydrodynamics 
simulations with a sufficient resolution will be feasible only if a memory of 3 PB is available and the computational speed is 
fast enough to handle $8 \times 10^{16}$ operations per time-step. So far we have ignored inelastic scatterings, which would
couple different energy grid points and increase the size of the dense sub-matrices by a factor of $N_{\varepsilon}$. Different
energies are also coupled by Doppler effect and gravitational redshift, which are neglected in the current version of our Boltzmann solver. 
If these effects are taken into account and the resultant enlarged sub-matrices are to be inverted in the same way, the required 
memory and operation numbers are gigantic as given in the last row of the table, since they are proportional to
$N_{\varepsilon}^2$ and $N_{\varepsilon}^3$, respectively. Last but not least, the implementation of GR in the Boltzmann solver, the ultimate goal of our project, should
be addressed at an appropriate point during this scale-up.   

It is now obvious to readers that the supernova research is a subject of supercomputing science that keeps step with the 
advancement of hard and softwares for supercomputing. Hopefully, the next generation supercomputers will provide us with the
opportunity to finally reach the goal. We hope also that our quest for the supernova mechanism will in turn contribute to 
pushing the limit of computational science in the decade to come.  

\section*{Acknowledgements}
We are grateful to S. Furusawa, K. Nakazato, K. Kiuchi, N. Ohnishi and H. Suzuki 
for fruitful collaborations and profitable discussions on supernova simulations.  
KK, YS, and TT are grateful to K. Sato for continuing 
encouragements and also wish to thank their collaborators, W. Nakano-Iwakami,
M. Liebend\"orfer, M. Hashimoto, S. Harikae, N. Yasutake, N. Nishimura, K. Nakamura, and K. Shaku. 
KS thanks Y. Sekiguchi for providing data from his 2D/3D simulations of core-collapse.  
We express our gratitude to C. Ott, A. Mezzacappa, H. -Th. Janka, E. M\"uller and W. Hillebrandt 
for valuable discussions on the current state-of-the-art simulations of supernovae.  
KS acknowledges the collaboration with H. Matsufuru, A. Imakura, T. Sakurai and S. Hashimoto and their advice on parallel computing.  

The numerical computations in this work were performed in part on the supercomputers 
at the center for Computational Astrophysics, CfCA, the National Astronomical Observatory of Japan,  
Research Center for Nuclear Physics (RCNP) in Osaka University, 
The University of Tokyo, Yukawa Institute for Theoretical Physics (YITP) in Kyoto University, 
Japan Atomic Energy Agency (JAEA) and High Energy Accelerator Research Organization (KEK).  

This work is partially supported by the Grant-in-Aid for Scientific Research on Innovative Areas (Nos. 20105004, 20105005) 
and the Grant-in-Aid for the Scientific Research (Nos. 19104006, 20740150, 21540281, 22540296, 23540323, and 23340069) 
from the Ministry of Education, Culture, Sports, Science and Technology (MEXT) in Japan.  

The numerical study on core-collapse supernovae using the supercomputer 
facilities is supported by the HPCI Strategic Program of MEXT, Japan.

%


\begin{thebibliography}{99}

\bibitem{Colgate66} S. A.~Colgate and R. H.~White, \AJ{143,1966,626}.
\bibitem{kota06} K.~{Kotake}, K.~{Sato} and K.~{Takahashi}, Rep.~Prog.~Phys.~\textbf{69} (2006), 971, arXiv:astro-ph/0509456.
\bibitem{Kotake11} K.~Kotake, accepted to Comptes Rendus Physique (2012); arXiv:1110.5107.
\bibitem{smartt09} {Smartt}, S.~J. Annual Review of Astronomy and Astrophysics {47,2009,63}
\bibitem{utrobin11}
{Utrobin}, V.~P. \& {Chugai}, N.~N. \aap~\textbf{532} (2011), A100

\bibitem{Sato-and-Suzuki} K.~Sato and H.~Suzuki, \PRL{58,1987,2722}.
\bibitem{raffelt12} G.~Raffelt, Proceedings ISAPP School "Neutrino Physics and Astrophysics", 26 July-5 August 2011, 
Villa Monastero, Varenna, Italy; arXiv:1201.1637.
\bibitem{jank07} H.~-Th.~Janka, K.~{Langanke}, A.~{Marek}, G.~{Mart{\'{\i}}nez-Pinedo} and B.~M\"{u}ller, \JL{\physrep,442,2007,38}.
\bibitem{Wilson85} J. R.~Wilson, Numerical Astrophyics (1985), p422.
\bibitem{Bethe85} H. A.~Bethe and J. R.~Wilson, \AJ{295,1985,14}.
\bibitem{Rampp00} M.~Rampp and H.-Th.~Janka, \AJ{539,2000,L33}.
\bibitem{Liebendorfer01} M.~Liebend$\ddot{\rm o}$rfer, A.~Mezzacappa, F.-K.~Thielemann, O. E.~Messer, W. R.~Hix and S. W.~Bruenn, \PRD{63,2000,103004}.
\bibitem{thom03} T.~A.~{Thompson}, A.~{Burrows} and P.~A.~{Pinto}, \AJ{592,2003,434}.
\bibitem{Sumiyoshi05} K.~Sumiyoshi, S.~Yamada, H.~Suzuki, H.~Shen, S.~Chiba and H.~Toki, \AJ{629,2005,922}.
\bibitem{horowitz02} C. J.~Horowitz, \PRD{65,2002,043001}. 
\bibitem{buras03} R.~Buras, M.~Rampp, H.-Th.~Janka and K.~Kifonidis, \PRL{90,2003,241101}.
\bibitem{burrows06npa} A.~Burrows, S.~Reddy and T. A.~Thompson, \NPA{777,2006,356}. 
\bibitem{Kitaura06} F. S.~Kitaura, H.-Th.~Janka and W.~Hillebrandt, \aap~\textbf{450} (2006), 345.
\bibitem{wang01} L.~{Wang}, D.~A.~{Howell}, P.~{H{\"o}flich} and J.~C.~{Wheeler}, \AJ{550,2001,1030}.
\bibitem{Maeda08} K.~Maeda et al., Science~\textbf{319} (2008), 1220.
\bibitem{Tanaka09} M.~Tanaka, K. S.~Kawabata, K.~Maeda, M.~Iye, T.~Hattori, E.~Pian, K.~Nomoto, P. A.~Mazzali and N.~Tominaga, \AJ{699,2009,1119}.
\bibitem{Herant92} M.~Herant, W.~Benz and S.~Colgate, \AJ{395,1992,642}.
\bibitem{shimizu93} T.~Shimizu, S.~Yamada and K.~Sato, Publ. Astron. Soc. J.~\textbf{45} (1993), L53.
\bibitem{Burrows95} A.~Burrows, J.~Hayes,  \& B. A.~Fryxell, \AJ{450,1995,830}.
\bibitem{Janka96} H.-Th.~Janka and E.~M\"{u}ller, \aap~\textbf{306} (1996), 167.
\bibitem{fryer04a} C.~L.~Fryer, \AJ{601,2004,L175}.
\bibitem{Blondin03} J. M.~Blondin and A.~Mezzacappa and C.~DeMarino, \AJ{584,2003,971}.
\bibitem{scheck06} L.~{Scheck}, K.~{Kifonidis}, H.-T.~{Janka} and E.~{M{\"u}ller}, \aap~\textbf{457} (2006), 963.
\bibitem{Ohnishi06} N.~Ohnishi, K.~Kotake and S.~Yamada, \AJ{641,2006,1018}, arXiv:astro-ph/0509765.
\bibitem{ohnishi07} N.~Ohnishi, K.~Kotake and S.~Yamada, \AJ{667,2007,375}, arXiv:astro-ph/0606187.
\bibitem{Foglizzo06} T.~Foglizzo, L.~Scheck and H.-Th.~Janka, \AJ{652,2006,1436}.
\bibitem{Iwakami08} W.~Iwakami, K.~Kotake, N.~Ohnishi, S.~Yamada and K.~Sawada, \AJ{678,2008,1207}, arXiv:0710.2191.
\bibitem{iwakami2}  W.~Iwakami, K.~Kotake, N.~Ohnishi, S.~Yamada and K.~Sawada, \AJ{700,2009,232}, arXiv:0811.0651.
\bibitem{Murphy08} J. W.~Murphy and A.~Burrows, \AJ{688,2008,1159}.
\bibitem{rodrigo09_2} R.~{Fern{\'a}ndez} and C.~{Thompson}, \AJ{703,2009,1464}.
\bibitem{Buras06a} R.~Buras, M.~Rampp, H.-Th.~Janka and K.~Kifonidis, \aap~\textbf{447} (2006), 1049.
\bibitem{woos02} S.~E.~{Woosley}, A.~{Heger} and T.~A.~{Weaver}, Rev. Mod. Phys.~\textbf{74} (2002), 1015.
\bibitem{WW95} S. E.~Woosley and T. A.~Weaver, \apjs~\textbf{101} (1995), 181.
\bibitem{Marek09} A.~Marek and H.-Th.~Janka, \AJ{694,2009,664}.
\bibitem{bruenn} S.~W.~{Bruenn}, A.~{Mezzacappa}, W.~R.~{Hix}, J.~M.~{Blondin}, P.~{Marronetti}, O.~E.~B.~{Messer}, C.~J.~{Dirk} and S.~{Yoshida}, AIP Conference Proceedings~\textbf{180} (2009), pp. 1-5; arXiv:1002.4914.
\bibitem{idsa} M.~{Liebend{\"o}rfer}, S.~C.~{Whitehouse} and T.~{Fischer}, \AJ{698,2009,1174}.
\bibitem{Suwa10} Y.~Suwa, K.~Kotake, T.~Takiwaki, S. C.~Whitehouse, M.~Liebend\"{o}rfer and K.~Sato, Publ. Astron. Soc. J.~\textbf{62} (2010), L49,
 arXiv:0912.1157.

\bibitem{Kotake03} Kotake, K., Yamada, S., \& Sato, K.\ 2003, \apj, 595, 304 

\bibitem{yamamoto12} Y.~Yamamoto and S.~Yamada, "Formations of Compact Objects: from the cradle to the grave", March 7-9 (2012), Waseda Univ., http://www.heap.phys.waseda.ac.jp/cnf1203/program.html.
\bibitem{latt91} J.~M.~{Lattimer} and F.~D.~{Swesty}, \NPA{535,1991,331}.
\bibitem{shlo06} S.~{Shlomo}, V.~M.~{Kolomietz} and G.~{Col{\`o}}, European Physical Journal A~\textbf{30} (2006), 23.
\bibitem{demo10} P.~B.~{Demorest}, T.~{Pennucci}, S.~M.~{Ransom}, M.~S.~E.~{Roberts} and J.~W.~T.~{Hessels}, Nature \textbf{467} (2010), 1081.
\bibitem{oconnor} E.~{O'Connor} and C.~D.~{Ott}, \AJ{730,2011,70}.
\bibitem{kiuc08} K.~{Kiuchi} and K.~{Kotake}, \mnras~\textbf{385} (2008), 1327, arXiv:0708.3597.
\bibitem{wolff} W.~Hillebrandt, K.~Nomoto and R. G.~Wolff, \aap~\textbf{133} (1984), 175.
\bibitem{Janka-pr} H.-Th.~Janka, private communications. 
\bibitem{suwa12} Y.~Suwa, "Formations of Compact Objects: from the cradle to the grave", March 7-9 (2012), Waseda Univ., http://www.heap.phys.waseda.ac.jp/cnf1203/program.html.
\bibitem{steiner} A.~W.~Steiner, J.~M.~Lattimer and E.~F.~Brown, \AJ{722,2010,33}. 
\bibitem{lattimer12} Lattimer, J.~M., \& Lim, Y.\ 2012, arXiv:1203.4286 
\bibitem{blo07} J.~M.~{Blondin} and A.~{Mezzacappa}, Nature~\textbf{445} (2007), 58.
\bibitem{Nordhaus10} J.~Nordhaus, A.~Burrows, A.~Almgren and J.~Bell, \AJ{720,2010,694}.
\bibitem{Hanke11} F.~Hanke, A.~Marek, B.~M\"{u}ller and H.-Th.~Janka, submitted to \apj~(2011); arXiv:1108.4355.
\bibitem{Burrows06} A.~Burrows, E.~Livne, L.~Dessart, C. D.~Ott and J.~Murphy, \AJ{640,2006,878}.
\bibitem{weinberg} N. N.~Weinberg and E.~Quataert, \mnras~\textbf{387} (2008), L64.
\bibitem{yoshida} S.~Yoshida, N.~Ohnishi and S.~Yamada, \AJ{665,2007,1268}.
\bibitem{leblanc}
{LeBlanc}, J.~M. \& {Wilson}, J.~R. 1970, Astrophys. J., 161, 541
{LeBlanc}, J.~M. \& {Wilson}, J.~R. 1970, Astrophys. J., 161, 541
\bibitem{yamasawa} S.~Yamada and H.~Sawai, \AJ{608,2004,907}.
\bibitem{kota04b} Kotake, K., Sawai, H., 
Yamada, S., \& Sato, K.\ 2004, \apj, 608, 391 

\bibitem{taki04} T.~{Takiwaki}, K.~{Kotake}, S.~{Nagataki} and K.~{Sato}, \AJ{616,2004,1086}, arXiv:astro-ph/0408388.
\bibitem{kota05} K.~Kotake, S.~Yamada and K.~Sato, \AJ{618,2005,474},
arXiv:astro-ph/0409244.
\bibitem{kotake_prd} K.~Kotake, S.~Yamada, K.~Sato, K.~Sumiyoshi, H.~Ono and H.~Suzuki, \PRD{69,2004,124004}, arXiv:astro-ph/0401563.
\bibitem{suwa07a} Suwa, Y., Takiwaki, T., 
Kotake, K., \& Sato, K.\ 2007, \apjl, 665, L43 
\bibitem{suwa07} Suwa, Y., Takiwaki, T., 
Kotake, K., \& Sato, K.\ 2007, \pasj, 59, 771
\bibitem{taki09} T.~{Takiwaki}, K.~{Kotake} and K.~{Sato}, \AJ{691,2009,1360}, arXiv:0712.1949.
\bibitem{burr07} A.~{Burrows}, L.~{Dessart}, E.~{Livne}, C.~D.~{Ott} and J.~{Murphy}, \AJ{664,2007,416}.
\bibitem{fogli_B} J.~Guilet, T.~Foglizzo and S.~Fromang, \AJ{729,2011,71}.
\bibitem{martin11} M.~{Obergaulinger} and H.-Th.~{Janka}, submitted to \aap~(2011); arXiv:1101.1198.
\bibitem{taki_kota} T.~Takiwaki and K.~Kotake, \AJ{743,2011,30},
arXiv:1004.2896.
\bibitem{kotake12} Kotake, K., Takiwaki, 
T., Suwa, Y., et al.\ 2012, arXiv:1204.2330 
\bibitem{endeve12} Endeve, E., Cardall, 
C.~Y., Budiardja, R.~D., et al.\ 2012, arXiv:1203.3108 
\bibitem{aki} S.~Akiyama, J. C.~Wheeler, D. L.~Meier and I.~Lichtenstadt, \AJ{584,2003,954}.
\bibitem{yoon} S.~-C.~Yoon and N.~Langer, \JL{\aap,443,2005,643}.
\bibitem{woos06} S.~E.~Woosley and A.~Heger, \AJ{637,2006,914}.
\bibitem{shapiro} Z. B.~Etienne, Y. T.~Liu and S. L.~Shapiro, \PRD{74,2006,044030}.
\bibitem{martin_mri}
{Obergaulinger}, M., {Cerd{\'a}-Dur{\'a}n}, P., {M{\"u}ller}, E., \& {Aloy},
  M.~A. 2009, \aap, 498, 241
\bibitem{takahara88} M.~{Takahara} and K.~{Sato}, \PTP{80,1988,861}.
\bibitem{sage09} I.~{Sagert}, T.~{Fischer}, M.~{Hempel}, G.~{Pagliara}, J.~{Schaffner-Bielich}, A.~{Mezzacappa}, F.~{Thielemann} and M.~{Liebend{\"o}rfer}, \PRL{102,2009,081101}.
\bibitem{fischer} T.~{Fischer}, I.~{Sagert}, G.~{Pagliara}, M.~{Hempel}, J.~{Schaffner-Bielich}, T.~{Rauscher}, F.-K.~{Thielemann}, R.~{K{\"a}ppeli}, G.~{Mart{\'{\i}}nez-Pinedo} and M.~{Liebend{\"o}rfer}, \apjs~\textbf{194} (2011), 39. 
\bibitem{thomp05} T.~A.~{Thompson}, E.~{Quataert} and A.~{Burrows}, \AJ{620,2005,861}.
\bibitem{masa11} Y.~{Masada}, T.~{Takiwaki} and K.~{Kotake}, submitted to \apj~(2011).
\bibitem{suzu08} T.~K.~{Suzuki}, K.~{Sumiyoshi} and S.~{Yamada}, \AJ{678,2008,1200}.
%
\bibitem{Shen98}  H.~Shen, H.~Toki, K.~Oyamatsu and K.~Sumiyoshi, \NPA{637,1998,435}.
\bibitem{Shen98b} H.~Shen, H.~Toki, K.~Oyamatsu and K.~Sumiyoshi, \PTP{100,1998,1013}.
\bibitem{Shen11} H.~Shen, H.~Toki, K.~Oyamatsu and K.~Sumiyoshi, \apjs~\textbf{197} (2011), 20.
\bibitem{Hempel11} M.~Hempel and J.~Schaffner-Bielich, \NPA{837,2010,210}.
\bibitem{Furu11} S.~Furusawa, S.~Yamada, K.~Sumiyoshi and H.~Suzuki, \AJ{738,2011,178}.
\bibitem{GShen11} G.~Shen, C.~J.~Horowitz and S.~Teige, \PRC{83,2011,035802}.
\bibitem{ish08} C.~Ishizuka, A.~Ohnishi, K.~Tsubakihara, K.~Sumiyoshi and S.~Yamada, \JP{G35,2008,085201}.
\bibitem{nak10b} K.~{Nakazato}, K.~{Sumiyoshi} and S.~{Yamada}, \AJ{721,2010,1284}.
\bibitem{nak10a} K.~{Nakazato}, K.~{Sumiyoshi}, H.~{Suzuki} and S.~{Yamada}, \PRD{81,2010,083009}.
\bibitem{sum06} K.~Sumiyoshi, S.~Yamada, H.~Suzuki and S.~Chiba, \PRL{97,2006,091101}.
\bibitem{fischer09} T.~Fischer, S. C.~Whitehouse, A.~Mezzacappa, F.-K.~Thielemann and M.~Liebend{\"o}rfer, \aap~\textbf{499} (2009), 1.
\bibitem{Furu2}  S.~Furusawa, S.~Yamada, K.~Sumiyoshi and H.~Suzuki, (2012) in preparation.
\bibitem{Botvina} A. S.~Botvina and I. N.~Mishustin, \NPA{843,2010,98}.
\bibitem{woos_blom}
{Woosley}, S.~E. \& {Bloom}, J.~S. Annual Review of Astronomy and Astrophysics {44,2006,507}
\bibitem{modjaz11}
{Modjaz}, M. 2011, Astronomische Nachrichten, 332, 434
\bibitem{jankamueller96} H.~{Janka} and E.~{M\"uller}, \aap~\textbf{306} (1996), 167.
\bibitem{annop} A.~{Wongwathanarat}, N.~J.~{Hammer} and E.~{M{\"u}ller}, \aap~\textbf{514} (2010), A48.
\bibitem{rantsiou} E.~{Rantsiou}, A.~{Burrows}, J.~{Nordhaus} and A.~{Almgren}, \AJ{732,2011,57}.
\bibitem{fern} R.~{Fern{\'a}ndez}, \AJ{725,2010,1563}.
\bibitem{endeve} E.~{Endeve}, C.~Y.~{Cardall}, R.~D.~{Budiardja} and A.~{Mezzacappa}, \AJ{713,2010,1219}.
\bibitem{kotake09a} K.~Kotake, W.~Iwakami, N.~Ohnishi and S.~Yamada, \AJ{697,2009,L133}, arXiv:0904.4300.
\bibitem{ewald11} E.~{M\"{u}ller}, H.-Th.~{Janka} and A.~{Wongwathanarat}, \aap~\textbf{537} (2012), 63.
\bibitem{kneller} H.~{Duan} and J.~P.~{Kneller}, \JP{G36,2009,113201}.
\bibitem{Takiwaki11} T.~Takiwaki, K.~Kotake and Y.~Suwa, accepted to ApJ 2011; arXiv:1108:3989.
\bibitem{goshy} A.~{Burrows} and J.~{Goshy}, \AJ{416,1993,L75}.
\bibitem{Janka01} H.-Th.~Janka, \aap~\textbf{368} (2001), 527.
\bibitem{May66} M.~M.~May and R.~H.~White, \PRL{141,1966,1232}.
\bibitem{Misner64} C.~W.~Misner and D.~H.~Sharp, \PR{136,1964,571}.
\bibitem{Schwarz67} R.~A.~Schwartz, Annals of Physics~\textbf{43} (1967), 42.
\bibitem{Lindquist66} R.~W.~Lindquist, Annals of Physics~\textbf{37} (1966), 487. 
\bibitem{Wilson71} J.~R.~Wilson, \AJ{163,1971,209}.
\bibitem{VanRiper79} K.~A.~van Riper, \AJ{232,1979,558}.
\bibitem{VanRiper81} K.~A.~van Riper and J.~M.~Lattimer, \AJ{249,1981,270}. 
\bibitem{VanRiper82} K.~A.~van Riper, \AJ{257,1982,793}.
\bibitem{Bruenn01} S.~W.~Bruenn, K. R.~De Nisco and A.~Mezzacappa, \AJ{560,2001,326}.
\bibitem{Bruenn85} S.~W.~Bruenn, \apjs~\textbf{58} (1985), 771.
\bibitem{Mezzacappa93a} A.~Mezzacappa and S. W.~Bruenn, \AJ{405,1993,669}.
\bibitem{Mezzacappa93b} A.~Mezzacappa and S. W.~Bruenn, \AJ{410,1993,740}.
\bibitem{Yamada97} S.~Yamada, \AJ{475,1997,720}.
\bibitem{Yamada99} S.~Yamada, H.-T.~Janka and H.~Suzuki, \aap~\textbf{344} (1999), 533.
\bibitem{Lentz11} E.~J.~Lentz, A.~Mezzacappa, O.~E.~Bronson Messer, M.~Liebend{\"o}rfer, W. R.~Hix and S.~Bruenn, \apj~in press (2012); arXiv:1112.3595.
\bibitem{Buras06b} R.~Buras, H.-Th.~Janka, M.~Rampp and K.~Kifonidis, \aap~{457} (2006), 281.
\bibitem{Shibata05a} M.~Shibata and Y.~Sekiguchi, \PRD{71,2005,024014}.
\bibitem{Shibata05b} M.~Shibata and Y.~Sekiguchi, \PRD{72,2005,044014}.
\bibitem{Ott2007} C.~D.~Ott, H.~Dimmelmeier, A.~Marek, H.-T.~Janka, I.~Hawke, B.~Zink and E.~Schnetter, \PRL{98,2007,261101}.
\bibitem{Dimmelmeier02} H.~Dimmelmeier, J. A.~Font and E.~M\"{u}ller, \aap~\textbf{388} (2002), 917.
\bibitem{Isa12}
Cordero-Carri{\'o}n, I., Cerd{\'a}-Dur{\'a}n, P., 
\& Ib{\'a}{\~n}ez, J.~M, \PRD{85,2012,044023}
\bibitem{Matthias05} M.~Liebend{\"o}rfer, \AJ{633,2005,1042}.
\bibitem{Sekiguchi10} Y.~Sekiguchi, \PTP{124,2010,331}.
\bibitem{ott12} Ott, C.~D., Abdikamalov, 
E., O'Connor, E., et al.\ 2012, arXiv:1204.0512 
\bibitem{Bernhard12} B.~M\"{u}ller, H.-T.~Janka and A.~Marek, submitted to \apj~(2012); arXiv:1202.0815.
\bibitem{Bernhard10} B.~M\"{u}ller, H.-T.~Janka and H.~Dimmelmeier, \apjs~\textbf{189} (2010), 104.
\bibitem{kuroda12} T.~Kuroda, K.~Kotake and T.~Takiwaki, accepted to \apj~(2012); arXiv:1202.2487.
\bibitem{Kuroda10} T.~Kuroda and H.~Umeda, \apjs~\textbf{191} (2010), 439.
\bibitem{Shibata95} M.~Shibata and T.~Nakamura, \PRD{52,1995,5428}.
\bibitem{Baumgarte99} T. W.~Baumgarte and S. L.~Shapiro, \PRD{59,1999,024007}.
\bibitem{Thorne81} K.~S.~Thorne, \mnras~\textbf{194} (1981), 439.
\bibitem{Shibata11} M.~Shibata, K.~Kiuchi, Y.~Sekiguchi and Y.~Suwa, \PTP{125,2011,1255}.
\bibitem{Foglizzo00} T.~Foglizzo and M.~Tagger, \aap~\textbf{363} (2000), 174.
\bibitem{Foglizzo02} T.~Foglizzo, \aap~\textbf{392} (2000), 353.
\bibitem{Scheck08} L.~Scheck, H.-Th.~Janka, T.~Foglizzo and K.~Kifonidis, \aap~\textbf{477} (2008), 931.
\bibitem{matthias04} M.~Liebend{\"o}rfer, O.~E.~B.~Messer, A.~Mezzacappa, S.~W.~Bruenn, C.~Y.~Cardall and F.-K.~Thielemann, \apjs~\textbf{150} (2004), 263.
\bibitem{kotake09b} K.~Kotake, W.~Iwakami, N.~Ohnishi and S.~Yamada, \AJ{704,2009,951}, arXiv:0909.3622.
\bibitem{kotake11gw} K.~Kotake, W.~Iwakami-Nakano and N.~Ohnishi, \AJ{736,2011,124}, arXiv:1106.0544.
\bibitem{ottprl} C.~D.~Ott, C.~Reisswig, E.~Schnetter, E.~O'Connor, U.~Sperhake, F.~L{\"o}ffler, P.~Diener, E.~Abdikamalov, I.~Hawke, A.~Burrows, \PRL{106,2011,161103}.
\bibitem{icecube} R.~Abbasi, Y.~Abdou, T.~Abu-Zayyad et al., \aap~\textbf{535} (2011), A109.
\bibitem{marek09b} A.~Marek, H.-Th.~Janka and E.~M\"uller, \aap~\textbf{496} (2009), 475.
\bibitem{lund} T.~Lund, A.~Marek, C.~Lunardini, H.-T.~Janka and G.~Raffelt, \PRD{82,2010,063007}.
\bibitem{fujimoto} S.~Fujimoto, K.~Kotake, M.~Hashimoto, M.~Ono and N.~Ohnishi, \AJ{738,2011,61}, arXiv:1106.2606.
\bibitem{friedel} F.-K.~{Thielemann}, A.~{Arcones}, R.~{K{\"a}ppeli}, M.~{Liebend{\"o}rfer}, T.~{Rauscher}, C.~{Winteler}, C.~{Fr{\"o}hlich}, I.~{Dillmann}, T.~{Fischer}, G.~{Martinez-Pinedo}, K.~{Langanke}, K.~{Farouqi}, K.-L.~{Kratz}, I.~{Panov} and I.~K.~{Korneev}, Prog. Part. Nucl. Phys.~\textbf{66} (2011), 346.
\bibitem{suz94} H.~Suzuki, in Physics and Astrophysics of Neutrinos (1994), (eds. M. Fukugita and A. Suzuki, Springer-Verlag, Tokyo) p.763.
\bibitem{Sumi12} K.~Sumiyoshi and S.~Yamada, \apjs~\textbf{199} (2012), 17.
\bibitem{mez01} A.~Mezzacappa, M.~Liebend\"orfer, O. E. B.~Messer, W. R.~Hix, F.-K.~Thielemann, S. W.~Bruenn, \PRL{86,2001,1935}.
\bibitem{Rampp02} M.~Rampp and H.-Th.~Janka, \aap~\textbf{396} (2002), 361.
\bibitem{Matthias01} M.~Liebend{\"o}rfer, A.~Mezzacappa and F.-K.~Thielemann, \PRD{63,2001,103004}.
\bibitem{lan03a} K.~Langanke, G.~Mart\'inez-Pinedo, J. M.~Sampaio, D. J.~Dean, W. R.~Hix, O. E. B.~Messer, A.~Mezzacappa, M.Liebend{\"o}rfer, H.-Th.~Janka and M.~Rampp, \PRL{90,2003,241102}.
\bibitem{hix03} W. R.~Hix, O. E. B.~Messer, A.~Mezzacappa, M.~Liebend{\"o}rfer, J. M.~Sampaio, K.~Langanke, D. J.~Dean and G.~Mart\'inez-Pinedo, \PRL{91,2003,201102}.
\bibitem{Sumiyoshi07} K.~Sumiyoshi, S.~Yamada and H.~Suzuki, \AJ{667,2007,382}.
\bibitem{fischer10} T.~Fischer, S. C.~Whitehouse, A.~Mezzacappa, F.-K.~Thielemann and M.~Liebend{\"o}rfer, \aap~\textbf{517} (2010), A80.
\bibitem{Totani98} T.~Totani, K.~Sato, H. E.~Dalhed and J.R.~Wilson, \AJ{496,1998,216}.
\bibitem{and05} S.~Ando, J. F.~Beacom and H.~Yuksel, \PRL{95,2005,171101}.
\bibitem{keehn12} J. G.~Keehn and C.~Lunardini, \PRD{85,2012,043011}.
\bibitem{wal05} R.~{Walder}, A.~{Burrows}, C.~D.~{Ott}, E.~{Livne}, I.~{Lichtenstadt} and M.~{Jarrah}, \AJ{626,2005,317}.
\bibitem{liv04} E.~{Livne}, A.~{Burrows}, R.~{Walder}, I.~{Lichtenstadt} and T.~A.~{Thompson}, \AJ{609,2004,277}.
\bibitem{ott} check below 
\bibitem{ott_multi} C.~D.~{Ott}, A.~{Burrows}, L.~{Dessart} and E.~{Livne}, \AJ{685,2008,1069}.
\bibitem{bra11} T.~D.~{Brandt}, A.~{Burrows}, C.~D.~{Ott} and E.~{Livne}, \AJ{728,2011,8}.
\bibitem{blo07b} J.~M.~{Blondin} and S.~{Shaw}, \AJ{656,2007,366}.
\bibitem{Hempel12} M.~Hempel, T.~Fischer, J.~Schaffner-Bielich and M.~Liebend{\"o}rfer, \AJ{748,2012,70}.
\bibitem{sum08} K.~Sumiyoshi, S.~Yamada and H.~Suzuki \AJ{688,2008,1176}.
\bibitem{sum09} K.~{Sumiyoshi}, C.~{Ishizuka}, A.~{Ohnishi}, S.~{Yamada} and H.~{Suzuki}, \AJ{690,2009,L43}.
\bibitem{nak11} K.~{Nakazato}, S.~{Furusawa}, K.~{Sumiyoshi}, A.~{Ohnishi}, S.~{Yamada} and H.~{Suzuki}, \AJ{745,2011,197}.
\bibitem{swe94} F. D.~Swesty, J. M.~Lattimer and E. S.~Myra, \AJ{425,1994,195}.
\bibitem{Blin11} S.~I.~Blinnikov, I.~V.~Panov, M.~A.~Rudzsky and K.~Sumiyoshi, \aap~\textbf{535} (2011), A37.
\bibitem{fryer99} C.~L.~{Fryer}, \AJ{522,1999,413}.
\bibitem{Maeda03} K.~Maeda and K.~Nomoto, \AJ{598,2003,1163}.
\bibitem{nom05} K.~Nomoto, ASP Conf. Ser.~\textbf{332} (2005), 374; arXiv:astro-ph/0506597.
\bibitem{nak08b} K.~{Nakazato}, K.~{Sumiyoshi}, H.~{Suzuki} and S.~{Yamada}, \PRD{78,2008,083014}.
\bibitem{abe11} K.~Abe et al. Letter of Intent (2011), arXiv:1109.3262.
\bibitem{kistler11} M. D.~Kistler, H.~Y\"uksel, S.~Ando, J. F.~Beacom and Y.~Suzuki, \PRD{83,2011,123008}.
\bibitem{Sumiyoshi98} K.~Sumiyoshi and T.~Ebisuzaki, Parallel Computing~\textbf{24} (1998), 287.
\bibitem{Mihalas99} D.~Mihalas and B.~W.~Mihalas, Foundations of Radiation Hydrodynamics  (Dover Publications, 1999).

\bibitem{abdi} Abdikamalov, E., 
Burrows, A., Ott, C.~D., et al.\ 2012, arXiv:1203.2915 

\bibitem{nagakura} H.~Nagakura, K.~Sumiyoshi and S.~Yamada, in preparation.
\bibitem{Woosley2011} S.~E.~{Woosley}, in Gamma-ray Bursts (eds. C. Kouveliotou, S. E. Woosley and R. A. M. J. Wijers, Cambridge University Press, 2012); arXiv:1105.4193.
\bibitem{macfadyen99} A.~I.~{MacFadyen}, S.~E.~{Woosley} and A.~{Heger}, \AJ{550,1999,410}.
\bibitem{paz90}
{Paczynski}, B. 1990, \apj, 363, 218
\bibitem{mezree}
{Meszaros}, P. \& {Rees}, M.~J. 1992, \mnras, 257, 29P
\bibitem{Harikae2010} S.~Harikae, K,~Kotake and T.~Takiwaki, \AJ{713,2010,304}, arXiv:0912.2590.
\bibitem{Zalamea2011} I.~{Zalamea} and A.~M.~{Beloborodov}, \mnras~\textbf{410} (2011), 2302.

\bibitem{fuji1} Fujimoto, S.-i., 
Hashimoto, M.-a., Kotake, K., \& Yamada, S.\ 2007, \apj, 656, 382 

\bibitem{ono1} Ono, M., Hashimoto, M., 
Fujimoto, S., Kotake, K., 
\& Yamada, S.\ 2009, Progress of Theoretical Physics, 122, 755 

\bibitem{ono2} Ono, M., Hashimoto, M.-a., 
Fujimoto, S.-i., Kotake, K., \& Yamada, S.\ 2012, arXiv:1203.6488 

\bibitem{basel} Winteler, C., 
Kaeppeli, R., Perego, A., et al.\ 2012, arXiv:1203.0616 
\bibitem{ruffert}
{Ruffert}, M., {Janka}, H.-T., {Takahashi}, K., \& {Schaefer}, G. 1997, \aap,
  319, 122
\bibitem{ruff98}
{Ruffert}, M. \& {Janka}, H.-T. 1998, \aap, 338, 535
\bibitem{birkl}
{Birkl}, R., {Aloy}, M.~A., {Janka}, H.-T., \& {M{\"u}ller}, E. 2007, \aap,
  463, 51
\bibitem{harikae} Harikae, S., Kotake, 
K., Takiwaki, T., \& Sekiguchi, Y.-i.\ 2010, \apj, 720, 614 

\bibitem{kotake12_2} Kotake, K., Takiwaki, 
T., Harikae, S, accepted to \apj~(2012); arXiv:1205.6061
\bibitem{dess09}
{Dessart}, L., {Ott}, C.~D., {Burrows}, A., {Rosswog}, S., \& {Livne}, E. 2009,
  \apj, 690, 1681
\bibitem{Sekiguchi11} Y.~Sekiguchi and M.~Shibata, \AJ{737,2011,6}.
\bibitem{saad} Y.~Saad, Iterative Methods for Sparse Linear Systems, 2nd Edition  (Philadelphia, PA: SIAM, 2003).
\bibitem{barrett} R.~Barrett et al., Templates for the Solution of Linear Systems:  Building Blocks for Iterative Methods, 2nd Edition (Philadelphia, PA: SIAM, 1994)
\bibitem{imakura} A.~Imakura, T.~Sakurai, H.~Matsufuru and K.~Sumiyoshi, (2012) in preparation.
%



\end{thebibliography}
\end{document}